\documentclass[]{aastex62}
\usepackage{amsmath}     
\usepackage{wasysym}           
\usepackage{graphicx}
\usepackage{amssymb}
\usepackage{epstopdf}
\usepackage{mathrsfs}
\usepackage{anyfontsize}
\usepackage{natbib}
\usepackage{color}
\usepackage{lipsum}
\usepackage{diagbox}

\shorttitle{Delayed Rebrightenings in Shock-interacting Supernovae}
\shortauthors{Coughlin \& Zrake}
\begin{document}
\title{A Physical Model of Delayed Rebrightenings in Shock-interacting Supernovae without Narrow Line Emission}
\author[0000-0003-3765-6401]{Eric R.~Coughlin}
\affiliation{Department of Physics, Syracuse University, Syracuse, NY 13244, USA}
\author[0000-0002-1895-6516]{Jonathan Zrake}
\affiliation{Department of Physics and Astronomy, Clemson University, Clemson, SC 29634, USA}

\email{ecoughli@syr.edu}

\begin{abstract}
Core-collapse supernovae can display evidence of interaction with pre-existing, circumstellar shells of material by rebrightening and forming spectral lines, and can even change types as Hydrogen appears in previously Hydrogen-poor spectra. However, a recently observed core-collapse supernova -- SN 2019tsf -- was found to brighten after roughly 100 days after it was first observed, suggesting that the supernova ejecta was interacting with surrounding material, but it lacked any observable emission lines and thereby challenged the standard supernova-interaction picture. We show through linear perturbation theory that delayed rebrightenings without the formation of spectral lines are generated as a consequence of the finite sound crossing time of the post-shock gas left in the wake of a supernova explosion. In particular, we demonstrate that sound waves -- generated in the post-shock flow as a consequence of the interaction between a shock and a density enhancement -- traverse the shocked ejecta and impinge upon the shock from behind in a finite time, generating sudden changes in the shock properties in the absence of ambient density enhancements. We also show that a blastwave dominated by gas pressure and propagating in a wind-fed medium is unstable from the standpoint that small perturbations lead to the formation of reverse shocks within the post-shock flow, implying that the gas within the inner regions of these blastwaves should be highly turbulent. 
\end{abstract}


\section{Introduction}
Astrophysical explosions are typically produced by the injection of energy into a medium over a timescale that is much shorter than the dynamical timescale of the medium itself. For example, core-collapse supernovae -- observed with ever-increasing frequency as high-cadence and wide-field surveys (e.g., PTF, \citealt{law09}; ZTF, \citealt{bellm14}; Pan-STARRS, \citealt{chambers16}, ATLAS, \citealt{tonry18}; DES, \citealt{to21}; LSST, \citealt{ivezic19}) have come online in recent years -- are generated by the ``bounce'' of the neutron star that forms from the collapse of the core of a massive star, and the collapse and bounce occur on timescales of milliseconds (though the diffusion of neutrinos out of the protoneutron star, likely responsible for revitalizing the shock launched from the bounce, occurs over seconds; \citealt{burrows86, burrows88}). By comparison, the freefall time from the surface of a supergiant is on the order of months. As a consequence of this disparity in timescales, it is a very good approximation to treat the initial conditions for a supernova explosion as driven by an impulsive energy injection. When the total energy injected is large compared to the binding energy of the star, the characteristic velocity of the shockwave that advances into the stellar envelope (as a byproduct of the energy injection) is much larger than the sound speed and -- because the star was in hydrostatic balance prior to the explosive energy input -- the freefall speed. The Mach number of the shockwave is therefore very large, the thermal and gravitational energy of the star are negligible, and the total energy behind the blast is a conserved quantity\footnote{Energy losses due to radiation only become substantial in terms of the total energy budget of the outflow at significantly later times; e.g., \citet{ostriker88}.}. 

When the density profile of the stellar envelope is well described by a power-law in radial distance from the core, $r$, the well-known Sedov-Taylor blastwave \citep{sedov59, taylor50} simultaneously describes the propagation of the shockwave itself and the fluid variables -- the density $\rho$, pressure $p$, and radial velocity $v$ -- behind the shock. Such a power-law density profile is expected for the convective envelope of a supergiant when the helium core of the star dominates the gravitational potential (e.g., \citealt{coughlin18}). Alternatively, if the progenitor of the supernova possessed a substantial stellar wind throughout the final stages of its nuclear-burning life, the density profile of such a wind (under the assumption that the mass-loss rate from the wind was a constant) scales as $\rho \propto r^{-2}$. In this case, the propagation of the shockwave from the supernova through the wind would approach the energy-conserving regime, and therefore be described by the Sedov-Taylor solution. 

Importantly, however, in these scenarios the density profile of the ambient medium is only {approximately} given by a power-law in radius, and there will be both radial and angle-dependent deviations from a pure power-law. For example, the convective nature of the hydrogen envelopes of supergiants implies that there should be perturbations to the density of the envelope that are induced by the presence of convective eddies, and these perturbations can occur over a range of angular scales (corresponding to a range of spherical harmonic numbers $\ell$; e.g., \citealt{quataert19}). Similarly, the winds from Wolf-Rayet stars are known to possess some degree of ``clumpiness'' (e.g., \citealt{crowther07}), and highly variable stars can eject ``shells'' of material periodically alongside a steady wind (e.g., \citealt{smith08}). High-density (relative to the background wind) clumps would generate anisotropic perturbations, while the presence of a shell would respresent an isotropic, but radially dependent, overdensity within the wind.

As a shockwave encounters overdensities within its surroundings, it responds dynamically and in a way that violates the scaling between the shock position and velocity that results from the Sedov-Taylor solution (see Equation \ref{Vss} below). The initial interaction also ``re-energizes'' the blast, as the adiabatically expanding and cooling material from a supernova explosion slams into a (for example) high-density shell and converts its kinetic energy into thermal energy \citep{chevalier82, chevalier94}. 
In addition to ordinary type-IIn supernovae (e.g., \citealt{filippenko89, schlegel90, turatto93, chugai94}), such interactions have been proposed to explain extreme explosions such as superluminous supernovae \citep{quimby07, smith07, ofek07,gal-yam09, quimby11}, and there have been a number of recent theoretical investigations of the interaction between the ejecta from an explosion and dense gas (e.g., 
\citealt{blondin96, sanz11, chatzoploulos16, andrews18, suzuki19, kurfurst20}; see also, e.g., \citealt{wang00, nakar03, nakar07} for analyses in the ultra-relativistic regime). A number of ``changing-look'' supernovae have also been observed as the material from a Hydrogen-poor supernova interacts with a dense shell of Hydrogen that was (presumably) ejected from the progenitor star at an earlier epoch (e.g., \citealt{milisavljevic15, margutti17, kuncarayakti18, chen18, sollerman20}).

The characteristic signatures of the initial interaction between an expanding blastwave and surrounding, dense gas are the rebrightening of the source and the formation of spectral lines. The latter is thought to occur due to the photoionization of hydrogen-rich circumstellar material by X-ray and UV emission from the shock-heated gas. However, this scenario does not seem to fit the recent supernova SN 2019tsf \citep{sollerman20}, the light curve of which rebrightens from around 80 to 100 days following the initial detection, but does not exhibit the characteristic hydrogen emission lines of supernova interaction. The absence of narrow lines suggests that the shock is no longer propagating in the pre-supernova, hydrogen-rich circumstellar gas (where a history of unsteady mass loss could naturally account for the presence of dense shells). This challenges the standard interaction scenario for powering late-time supernova evolution.

We propose that some late-time rebrightenings of energetic explosions are the signature of internal acoustic waves interacting with the shock front. An internal wave is launched by the collision of the shock front with a dense shell that was at a relatively small radius. This wave propagates in the post-shock gas and ultimately returns to the shock front at a larger radius. The secondary interaction between that wave and the shock (as the shock is ``hit from behind'' by the wave) can induce an abrupt enhancement in the shock power, manifesting as a rebrightened light curve. This qualitative picture motivates an analysis of how a shockwave and the post-shock gas respond to perturbations in the density of the ambient medium. In particular, it would be important to predict the relative amplitude of, and time between, the power enhancements induced by the initial and secondary shock interactions.

Here we analyze the response of a strong, energy-conserving blastwave, described by the Sedov-Taylor self-similar solution, to spherically symmetric density inhomogeneities with linear perturbation theory. Using this approach we show that there is a sound-crossing timescale, which depends on the adiabatic index of the gas and the density profile of the ambient medium, that characterizes the propagation of disturbances in the post-shock flow. Consequently, we demonstrate that a shockwave can exhibit abrupt changes in its (e.g.) velocity at a significantly later time from when it encounters changes in the density profile of the ambient medium, and can therefore exhibit sudden rebrightenings in complete absence of narrow line emission. We also show that a gas-pressure-dominated, energy-conserving shock propagating into a wind-like medium\footnote{This has been described as the ``Primakoff blastwave'' in some references.} is unstable to the formation of reverse shocks -- any small disturbance that the shock encounters creates a wave that increases in amplitude behind the blast -- but is nevertheless stable from the standpoint that the shock speed and position eventually conform to the energy-conserving (i.e., Sedov-Taylor) prescription after encountering the density inhomogeneity.

In Section \ref{sec:equations} we derive the perturbation equations that describe the linear response of a strong shockwave to the presence of ambient density perturbations; this has been done in, e.g., \citet{ryu87}, but the approach and set of coordinates we adopt simplifies the equations considerably, and they manifestly respect the invariance of the solutions to renormalizations of the shock velocity and position (see also the Appendix of \citealt{coughlin19b}). In Section \ref{sec:equations} we take the Laplace transform of the linearized equations constructed in Section \ref{sec:eigenmodes}, we analyze the properties of the eigenvalues and eigenfunctions, and we discuss how to numerically integrate the Laplace-transformed equations without using the eigenfunctions. In Section \ref{sec:solutions} we analyze the general solutions to the set of Laplace-transformed equations and describe the generic properties of the physical response of a blastwave to the presence of ambient density perturbations; we also consider specific examples, and we compare the results to numerical simulations. The astrophysical and observational implications of our results are discussed and presented in Section \ref{sec:implications}, and we summarize and conclude in Section \ref{sec:summary}.

\section{Radial Perturbation Equations}
\label{sec:equations}
We let there be a spherically symmetric, astrophysical explosion that launches a shockwave into an ambient gas. As the shockwave propagates outward from the explosion site through an ambient gas, the post-shock fluid it leaves in its wake is assumed to be adiabatic and characterized by an adiabatic index $\gamma$; $\gamma = 4/3$ corresponds to a radiation-pressure dominated gas (or one composed of relativistic particles that act effectively like radiation), while a monatomic, ideal gas has $\gamma = 5/3$. We denote the time-dependent position of the shockwave by $R(t)$ and the shock velocity by $V(t) = dR/dt$. The shock speed is assumed to be much larger than the ambient sound speed, and any velocities maintained by the gas into which the shock advances are also assumed to be much smaller than that of the shock. We also assume that the density profile of the ambient medium can be approximated as a power-law with spherical radius $r$ from the injection site of the explosion, and we denote the radial power-law index by $n$; the ambient density $\rho$ therefore scales as $\rho \propto r^{-n}$. 

With this set of assumptions, the only relevant scales that characterize the fluid are the shock radius and velocity, which correspondingly sets the dimensionless timescale $\tau$:

\begin{equation}
d\tau = \frac{V\,dt}{R} \quad \Rightarrow \quad \tau = \ln\left(\frac{R}{R_0}\right).
\end{equation}
Here $R_0$ is the position of the shock at some initial time. In this paper we will restrict the perturbations of the ambient medium to be purely radial, and the spherically symmetric fluid equations are

\begin{equation}
\frac{\partial v}{\partial t}+v\frac{\partial v}{\partial r}+\frac{1}{\rho}\frac{\partial p}{\partial r} = 0, \,\,\, \frac{\partial\rho}{\partial t}+\frac{1}{r^2}\frac{\partial}{\partial r}\left[\rho r^2v\right] = 0, \,\,\, \frac{\partial s}{\partial t}+v\frac{\partial s} {\partial r} = 0. \label{eom}
\end{equation}
Here $r$ is spherical radius, $v$ is the radial velocity, $\rho$ is the gas density, $p$ is the pressure, and $s = \ln\left(p/\rho^{\gamma}\right)$ is the specific entropy.

Equations \eqref{eom} govern the time-dependent evolution of the fluid everywhere behind the shock front, at which the entropy jumps discontinuously while the mass, momentum, and energy fluxes are continuous. The continuity of these fluxes at the shock demands that the post-shock velocity, density, and pressure satisfy the jump conditions:

\begin{equation}
v(R) = \frac{2}{\gamma+1}V, \,\,\, \rho(R) = \frac{\gamma+1}{\gamma-1}\rho_{\rm a}(R), \,\,\, p(R) = \frac{2}{\gamma+1}\rho_{\rm a}(R)V^2,
\end{equation}
where $\rho_{\rm a}(R)$ is the ambient density at the shock. As noted above, we assume that the ambient density is well-characterized by a power-law with radius, so $\rho_{\rm a}(R) \simeq \rho_0\left(R/R_0\right)^{-n}$. However, we will also let there be (spherically symmetric) inhomogeneities in the density profile that, in the limit that they are small relative to the power-law profile, can be treated as perturbations that modify the shock propagation. We therefore write

\begin{equation}
\rho_{\rm a} = \rho_0\left(\frac{R(t)}{R_0}\right)^{-n}\left\{1+\delta\rho(R)\right\} = \rho_0 e^{-n\tau}\left\{1+\delta \rho(\tau)\right\}.
\end{equation}
In the last line we simply wrote the quantities in terms of the dimensionless, time-like variable $\tau$ for clarity. Given the boundary conditions on the fluid quantities at the shock front \eqref{eom}, we now parameterize the fluid velocity, density, and pressure by

\begin{equation}
v = Vf(\xi,\tau), \,\,\, \rho = \rho_0 e^{-n\tau}g(\xi,\tau), \,\,\, p = \rho_0 e^{-n\tau}V^2h(\xi,\tau), \label{fluiddefs}
\end{equation}
where 

\begin{equation}
\xi = r/R
\end{equation}
is the spherical, Eulerian radius $r$ normalized by the time-dependent radius of the shock front, $R(t)$. Note that these definitions make no explicit assumptions about the temporal variation of the shock radius or the velocity; they are simply redefinitions of the fluid variables that are motivated by the length and velocity scales provided by the shock position and velocity and the boundary conditions at the shock. 

We can now insert the definitions \eqref{fluiddefs} into the fluid equations \eqref{eom}; doing so and making a few rearrangements gives

\begin{equation}
\frac{\partial g}{\partial \tau}-\xi\frac{\partial g}{\partial \xi}-n g+\frac{1}{\xi^2}\frac{\partial}{\partial \xi}\left[\xi^2fg\right] = 0, \label{cont}
\end{equation}
\begin{equation}
\frac{\partial f}{\partial \tau}+\left(\frac{\partial}{\partial \tau}\ln V\right)f+\left(f-\xi\right)\frac{\partial f}{\partial \xi}+\frac{1}{g}\frac{\partial h}{\partial \xi} = 0, \label{rmom}
\end{equation}
\begin{equation}
\frac{\partial}{\partial \tau}\ln\left(\frac{h}{g^{\gamma}}\right)+2\frac{\partial}{\partial \tau}\ln V+n\left(\gamma-1\right)+\left(f-\xi\right)\frac{\partial}{\partial \xi}\ln\left(\frac{h}{g^{\gamma}}\right) = 0, \label{ent}
\end{equation}
while the boundary conditions at the shock front are

\begin{equation}
f(1,\tau) = \frac{2}{\gamma+1}, \,\,\, g(1,\tau) = \frac{\gamma+1}{\gamma-1}\left\{1+\delta\rho(\tau)\right\}, \,\,\, h(1,\tau) = \frac{2}{\gamma+1}\left\{1+\delta\rho(\tau)\right\}. \label{bcsgen}
\end{equation}
We emphasize that Equations \eqref{cont} -- \eqref{ent} are just the continuity, radial momentum, and entropy equations written in terms of the dimensionless variables $\xi$ and $\tau$, and the only underlying assumption is that the fluid variables retain spherical symmetry. By making this coordinate transformation, however, we see that if the perturbation to the ambient density profile is exactly zero and if the shock velocity behaves in such a way that

\begin{equation}
\frac{\partial}{\partial \tau}\ln V = \alpha, \label{Vss}
\end{equation}
where $\alpha$ a constant but otherwise-unspecified number, then these equations and the boundary conditions can be solved exactly with $f(\xi,\tau) = f(\xi)$, $g(\xi,\tau) = g(\xi)$, and $h(\xi,\tau) = h(\xi)$. These ``time-independent'' solutions are the self-similar solutions and are stationary in the variable $\xi$. 

When the perturbation to the ambient density is not zero, then it is clear that omitting the $\tau$-dependence from the functions $f$, $g$, and $h$ will not simultaneously satisfy the boundary conditions at the shock front and the fluid equations, and the shock velocity will not behave precisely as given by Equation \eqref{Vss}. However, if the perturbations to the ambient density are small, then these constraints -- the $\tau$-independence and the variation of the shock velocity as in Equation \eqref{Vss} -- will be \emph{approximately} upheld up to some correspondingly small corrections that scale linearly with the density perturbations. To accommodate the existence of these perturbations, we therefore write the functions $f$, $g$, and $h$ and the shock velocity as

\begin{equation}
f(\xi,\tau) = f_0(\xi)+f_1(\xi,\tau), \,\,\, g(\xi,\tau) = g_0(\xi)+g_1(\xi,\tau), \,\,\, h(\xi,\tau) = h_0(\xi)+h_1(\xi,\tau), \,\,\, \frac{\partial}{\partial \tau}\ln V = \alpha_0+\alpha_1(\tau). \label{pertquants}
\end{equation}
Subscript-1 quantities are ``small'' in the sense that they are driven by the assumed-small inhomogeneities of the background density relative to the power-law decline, and we can now construct the linear response of the shock to these inhomogeneities by inserting these definitions into the fluid equations and keeping only linear terms. The zeroth-order terms give the following three equations for the ``unperturbed,'' i.e., self-similar functions $f_0$, $g_0$, and $h_0$:

\begin{equation}
-\xi\frac{\partial g_0}{\partial \xi}-n g_0+\frac{1}{\xi^2}\frac{\partial}{\partial \xi}\left[\xi^2f_0g_0\right] = 0, \,\,\, \alpha_0 f_0+\left(f_0-\xi\right)\frac{\partial f_0}{\partial \xi}+\frac{1}{g_0}\frac{\partial h_0}{\partial \xi} = 0, \,\,\, 2\alpha_0+n\left(\gamma-1\right)+\left(f_0-\xi\right)\frac{\partial}{\partial \xi}\ln\left(\frac{h_0}{g_0^{\gamma}}\right) = 0, \label{unperts}
\end{equation}
with the boundary conditions

\begin{equation}
f_0(1) = h_0(1) = \frac{2}{\gamma+1}, \,\,\, g_0(1) = \frac{\gamma+1}{\gamma-1},
\end{equation}
while collecting the first-order terms yields

\begin{equation}
\frac{\partial g_1}{\partial \tau}-ng_1-\xi\frac{\partial g_1}{\partial \xi}+\frac{1}{\xi^2}\frac{\partial}{\partial \xi}\left[\xi^2\left(f_0g_1+g_0f_1\right)\right] = 0, \label{contpert}
\end{equation}
\begin{equation}
\frac{\partial f_1}{\partial \tau}+\alpha_0 f_1+\alpha_1 f_0+\left(f_0-\xi\right)\frac{\partial f_1}{\partial \xi}+f_1\frac{\partial f_0}{\partial \xi}-\frac{g_1}{g_0^2}\frac{\partial h_0}{\partial \xi}+\frac{1}{g_0}\frac{\partial h_1}{\partial \xi} = 0,
\end{equation}
\begin{equation}
\frac{\partial}{\partial \tau}\left[\frac{h_1}{h_0}-\frac{\gamma g_1}{g_0}\right]+2\alpha_1+\left(f_0-\xi\right)\frac{\partial}{\partial \xi}\left[\frac{h_1}{h_0}-\frac{\gamma g_1}{g_0}\right]+f_1\frac{\partial}{\partial \xi}\ln\left(\frac{h_0}{g_0^{\gamma}}\right) = 0. \label{entpert}
\end{equation}
The boundary conditions on the functions $f_1$, $g_1$ and $h_1$ at the shock front can be read off from Equation \eqref{bcsgen}, which gives (after using the boundary conditions on the unperturbed functions)

\begin{equation}
f_1(1,\tau) = 0, \,\,\, g_1(1,\tau) = \frac{\gamma+1}{\gamma-1}\delta \rho(\tau), \,\,\, h_1(1,\tau) = \frac{2}{\gamma+1}\delta \rho(\tau). \label{bcs1}
\end{equation}

\subsection{Global boundary conditions and constraining $\alpha$}
\label{sec:global}

\begin{figure}[htbp] 
   \centering
   \includegraphics[width=0.33\textwidth]{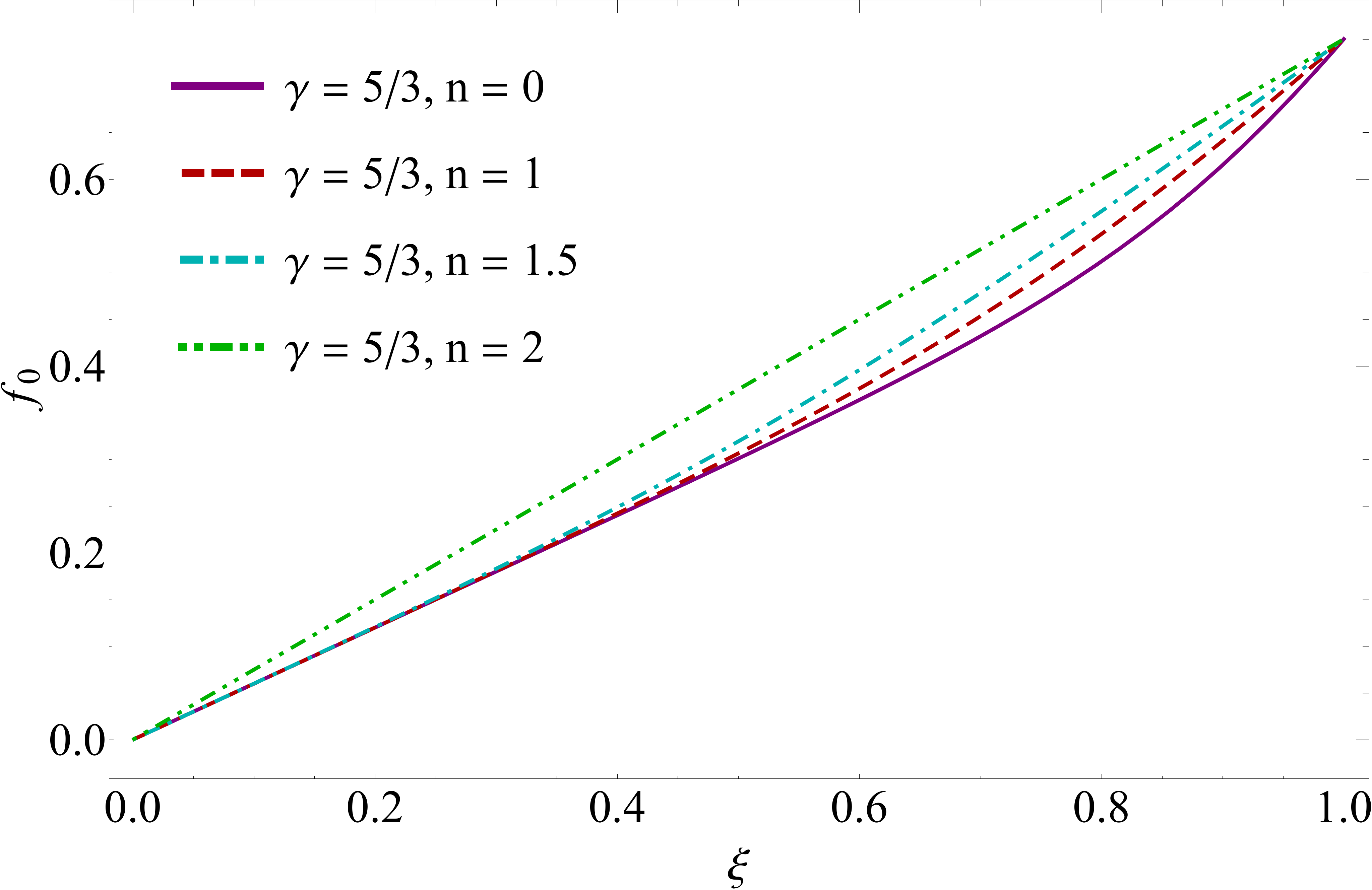} 
   \includegraphics[width=0.32\textwidth]{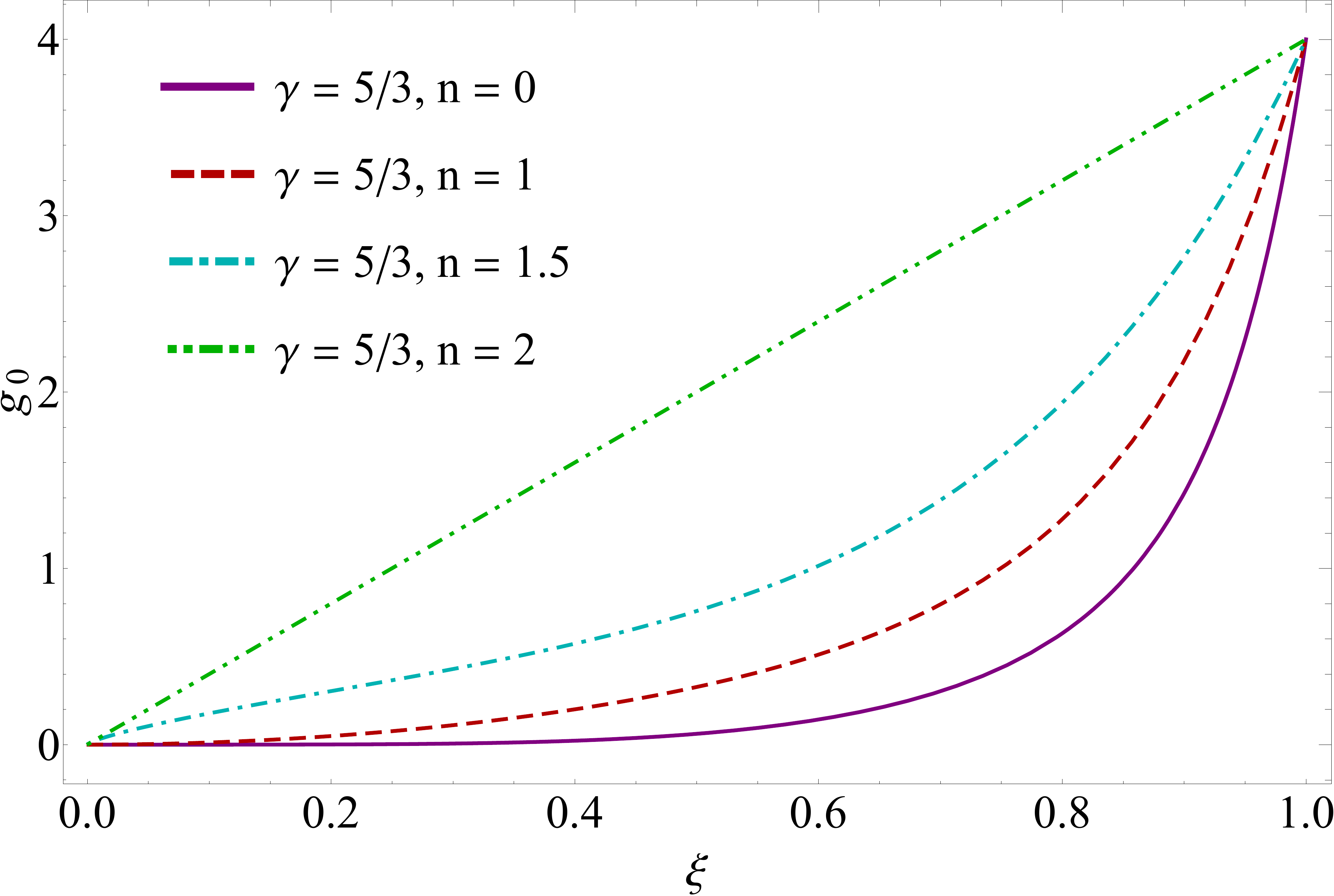}
   \includegraphics[width=0.33\textwidth]{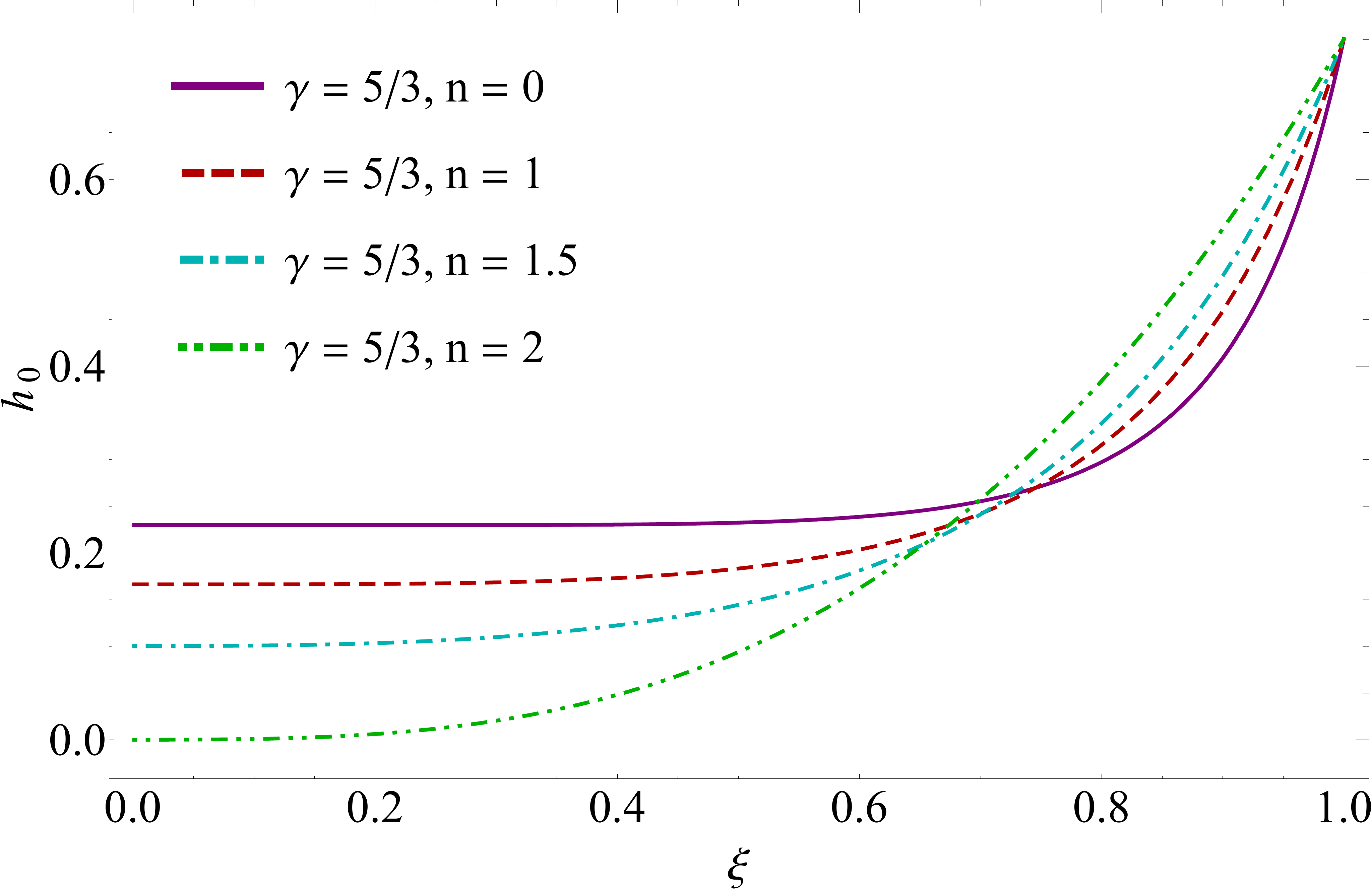}
    \includegraphics[width=0.33\textwidth]{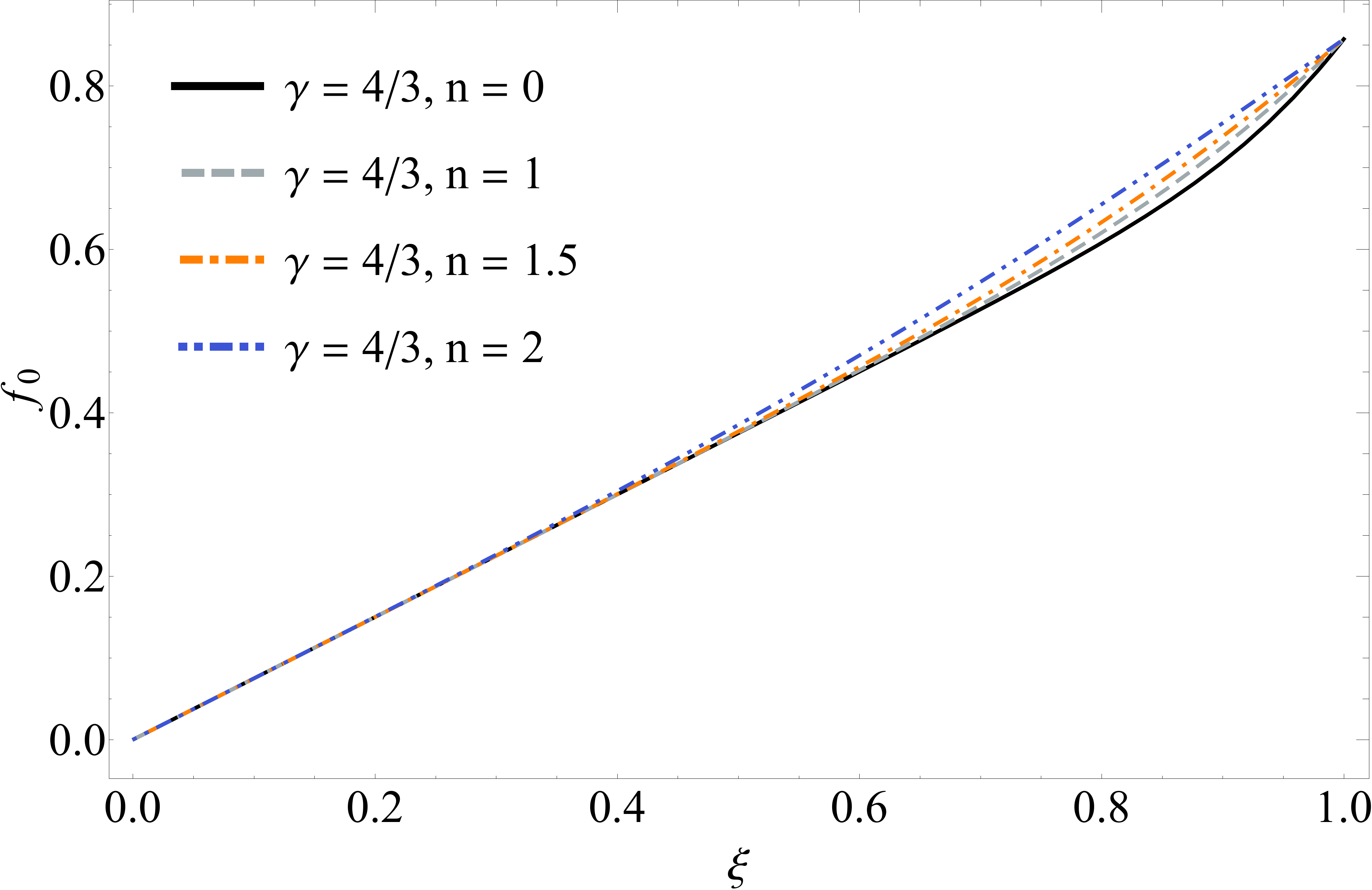} 
   \includegraphics[width=0.32\textwidth]{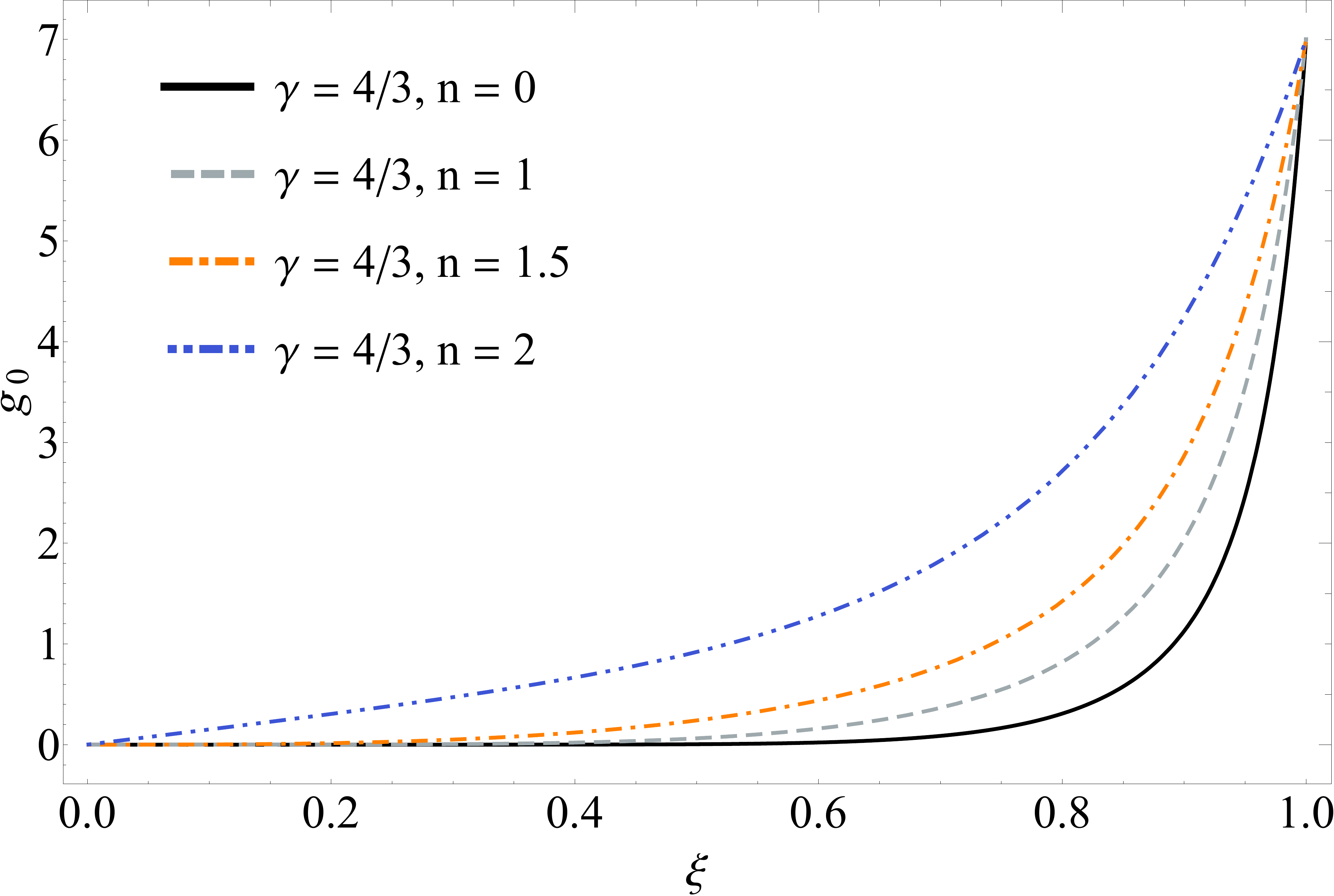}
   \includegraphics[width=0.33\textwidth]{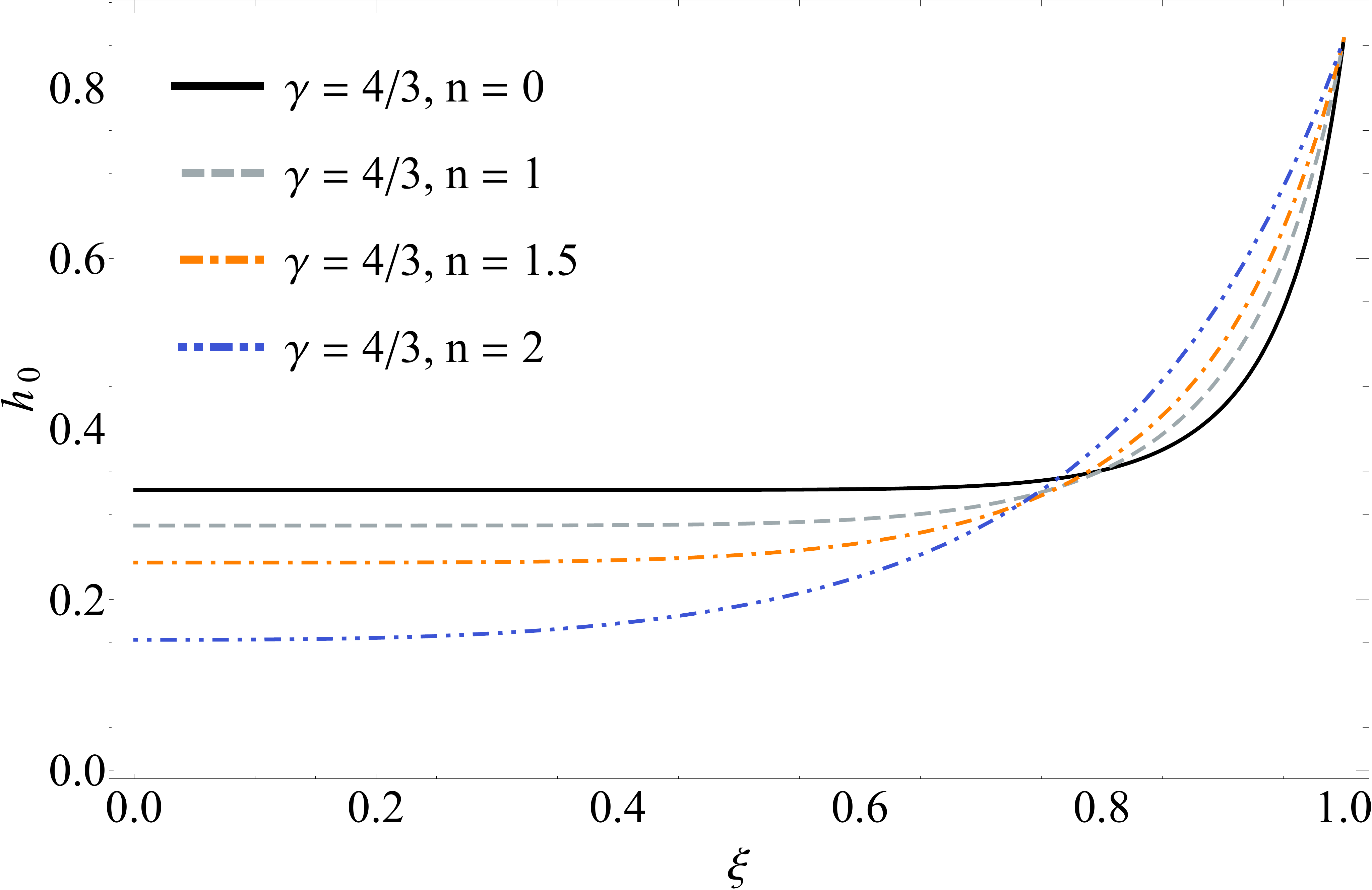}
   \caption{The unperturbed, self-similar solution describing an energy-conserving blastwave (i.e., the Sedov-Taylor solution) with $\gamma = 5/3$ (top row) and $\gamma = 4/3$ (bottom row) and the range of $n$ shown in the legend of each panel. Here $n$ characterizes the falloff of the ambient density with radius, so that $n = 0$ corresponds to a constant-density ambient medium, while $n = 2$ is appropriate to a wind-fed medium. The left, middle, and right panel show the dimensionless post-shock velocity, density, and pressure as functions of $\xi$, which at a given time is just spherical radius normalized by the shock position. }
   \label{fig:sedov}
\end{figure}

Equations \eqref{unperts} constitute the three, ordinary differential equations that can be integrated numerically from the shock front ($\xi = 1$) inward to solve for the self-similar structure of the blastwave, the velocity profile of which is given by $f_0$, the density profile by $g_0$, and the pressure profile by $h_0$. It would seem, however, that the system is under-constrained owing to the existence of the fourth parameter $\alpha_0$ that characterizes the variation of the shock velocity with position. This parameter is governed by a fourth boundary condition that does not take place at the shock front and establishes the global nature of the post-shock fluid. For example, if the self-similar solution describes the entirety of the post-shock flow and the center of the blastwave is physically well-behaved, then the spherical symmetry of the solution demands that $v(r = 0) = v(\xi = 0) = 0$ and hence $f_0(0) = 0$. Only a special value of $\alpha_0$ will satisfy this fourth boundary condition for a given $n$ and $\gamma$. Alternatively, in the absence of any dissipation behind the shock, we require that the total energy behind the blastwave $E \propto R^{3-n}V^2$ be conserved; differentiating this expression with respect to time and straightforward algebra then shows that $\alpha_0 = (n-3)/2$ if the energy is exactly conserved. For this value of $\alpha_0$ the velocity is zero at the origin, and this energy-conserving, global solution is the well-known, Sedov-Taylor blastwave. Figure \ref{fig:sedov} illustrates these solutions for $\gamma = 5/3$ (top row) and $\gamma = 4/3$ (bottom row) and a range of $n$; note that, for $n = 2$, the solution is homologous with $f_0 = 3\xi/4$, $g_0 = 4\xi$, and $h_0 = 3\xi^3/4$, which is also known as the ``Primakoff blastwave.'' It is clear that, as $\gamma$ declines, more of the mass becomes compressed into a thin shell immediately behind the shock. In the limit that $\gamma \rightarrow 1$, a redefinition of the self-similar variable shows that the density falls off exponentially rapidly behind the shock over the characteristic distance $\simeq (\gamma-1)R$ \citep{sanz16,coughlin20}.

However, it may not necessarily be the case that the self-similar solution describes the entirety of the post-shock flow. \citet{waxman93} used this restricted applicability of the self-similarity of the flow to search for self-similar solutions that describe \emph{accelerating} blastwaves, which are characterized by ambient density profiles with power-law indices that satisfy\footnote{There is a narrow range of $n \gtrsim 3$ that constitutes a ``gap'' between the Sedov-Taylor decelerating and Waxman-Shvarts accelerating regimes, where the upper limit on $n$ within this gap depends on the adiabatic index of the fluid; see \citet{waxman93, kushnir10}} $n \gtrsim 3$ and would -- if the self-similar solution were required to describe the entire post-shock flow -- possess infinite energy. In this case, the value of $\alpha_0$ is constrained by the smooth passage of the fluid variables through a sonic point that exists within the post-shock flow and causally disconnects the self-similar region of the flow (between the sonic point and the shock) and the origin. Alternatively, the post-shock flow may contain a contact discontinuity that separates the shocked ambient medium from shocked ejecta, which occurs during the early stages of core-collapse supernovae when the ejecta interacts with its surroundings. In this case, $\alpha_0$ is constrained by the power-law decline of the ejecta density with radius and the ambient density profile \citep{chevalier82}; if the ejecta density declines with radius $r$ as $\propto r^{-m}$, then it follows that $\alpha_0 = (n-3)/(m-3)$. Finally\footnote{If the gravitational field of a point mass (or at least a spatially confined and spherically symmetric mass distribution that behaves effectively like a point) modifies the behavior of the fluid, then the only choice of $\alpha$ that can satisfy self-similarity is $\alpha_0 = -1/2$ (e.g., \citealt{chevalier89, coughlin18b, coughlin19b}).}, it may also be the case that the shock that expands into the ambient medium is advanced by the presence of an ongoing supply of energy in the form of a wind, in which case a contact discontinuity separates the shocked ambient gas from the shocked wind \citep{weaver77}. In this scenario, a constant energy supply from the wind implies that $\alpha_0 = \left(n-2\right)/3$.  

As we did not specify the type of self-similar solution -- global without sonic points, causally disconnected with a sonic point, or causally connected with a contact discontinuity -- Equations \eqref{unperts} characterize the Sedov-Taylor (energy-conserving, decelerating), the Waxman-Shvarts (accelerating), and the decelerating interaction and wind-driven regimes, provided that one uses the appropriate value of $\alpha_0$. For concreteness, for the remainder of this paper we deal almost exclusively with the Sedov-Taylor, decelerating case and we set $\alpha_0 = (n-3)/2$. 

Similarly, Equations \eqref{cont} -- \eqref{ent} appear to represent three equations for the four unknowns $f_1$, $g_1$, $h_1$, and $\alpha_1$ and therefore seem to be under-constrained (given the boundary conditions on the functions $f_1$, $g_1$, and $h_1$ in Equation \ref{bcs1}). However, if we choose a specific $\alpha_1$, then integrating the three differential equations inward will not necessarily yield a well-behaved solution in the interior of the shock. Similar to the additional criterion that selects the value of $\alpha_0$, the correction to the shock acceleration parameter $\alpha_1$ is constrained neither by the differential equations nor the boundary conditions at the shock, but by a fourth, global boundary condition. One way to understand the nature of this fourth condition is to write Equations \eqref{contpert} -- \eqref{entpert} as a single matrix equation of the form $\partial \mathbf{f}/\partial \tau+\mathbf{M}\partial \mathbf{f}/\partial \xi+\mathbf{P}\mathbf{f}+\mathbf{D} = 0$, where $\mathbf{f}$ is the vector containing the functions $f_1$, $g_1$, and $h_1$, and $\mathbf{M}$, $\mathbf{P}$, and $\mathbf{D}$ are matrices. The characteristics $\lambda_0$, $\lambda_+$, and $\lambda_-$ are the three eigenvalues of the matrix $\mathbf{M}$, where

\begin{equation}
    \mathbf{M} = \left(
    \begin{array}{ccc}
    f_0-\xi & 0 & g_0^{-1} \\
    g_0 & f_0-\xi & 0 \\
    \gamma h_0 & 0 & f_0-\xi
    \end{array} \right)
\end{equation}
and we have let $\mathbf{f} = \left\{f_1,g_1,h_1\right\}^{T}$. From this the characteristics (eigenvalues) can be shown to be

\begin{equation}
\lambda_0 = f_0-\xi, \quad \lambda_{\pm} =f_0-\xi\pm\sqrt{\gamma h_0/g_0}. \label{lambdas}
\end{equation} 
These three eigenvalues represent the dimensionless speeds at which information propagates through the fluid. In particular, information from the shock travels back into the shocked gas at a speed $\lambda_{-}$. Any sudden changes in the density of the ambient medium are then communicated from the shock to the origin in a time $T(0)$, where the function $T(\xi)$ is given by

\begin{equation}
T(\xi) = \int_1^{\xi}\frac{d\tilde{\xi}}{\lambda_{-}(\tilde{\xi})} = \int_1^{\xi}\frac{d\tilde{\xi}}{f_0(\tilde{\xi})-\tilde{\xi}-\sqrt{\gamma h_0(\tilde{\xi})/g_0(\tilde{\xi})}}
\end{equation}
and is the dimensionless time taken for a discontinuous change in the density at the shock to travel from $\xi = 1$ to some distance $\xi$ interior to the shock. 

\begin{figure}[htbp] 
   \centering
   \includegraphics[width=0.495\textwidth]{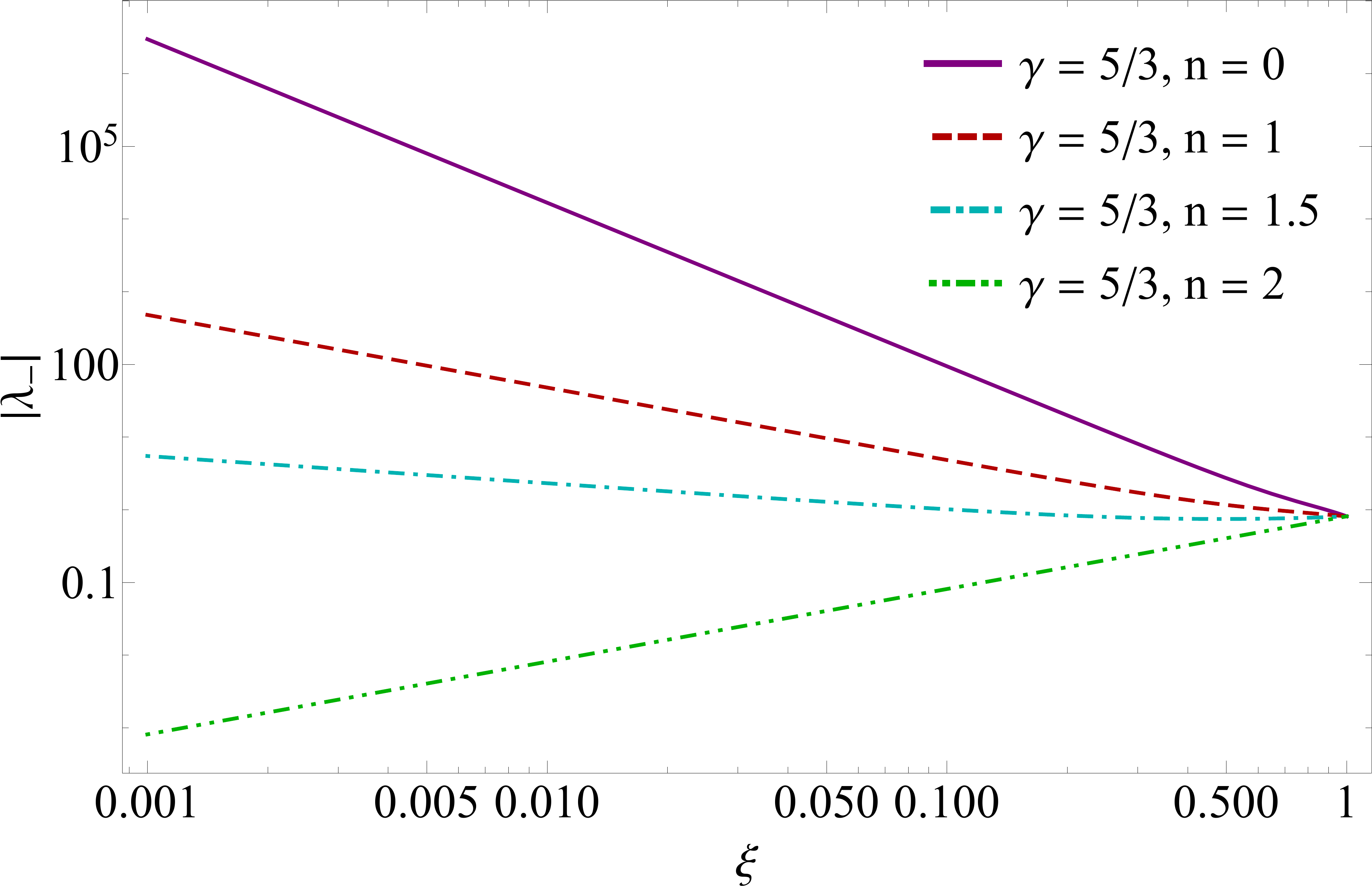}
   \includegraphics[width=0.485\textwidth]{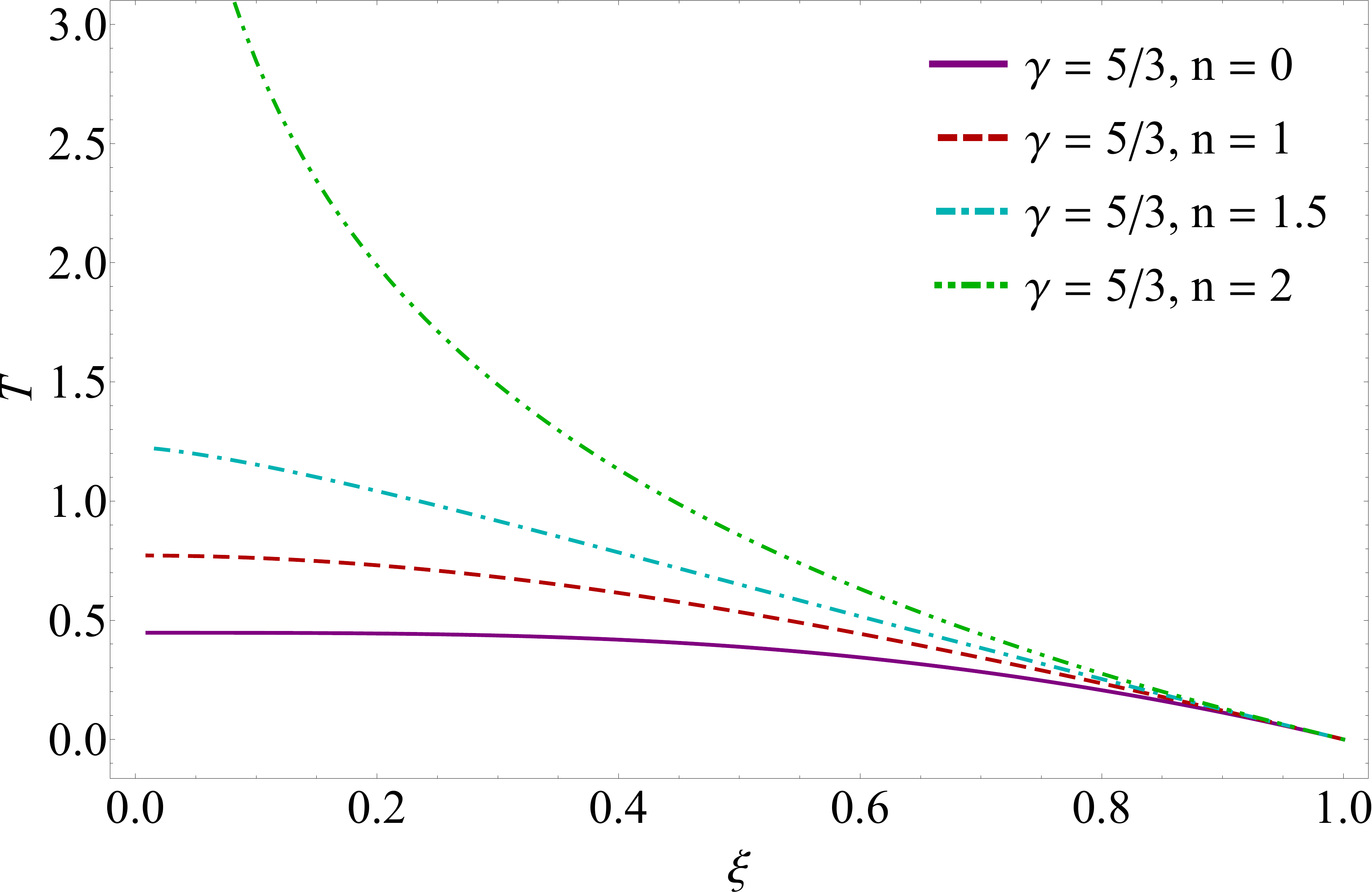} 
     \includegraphics[width=0.495\textwidth]{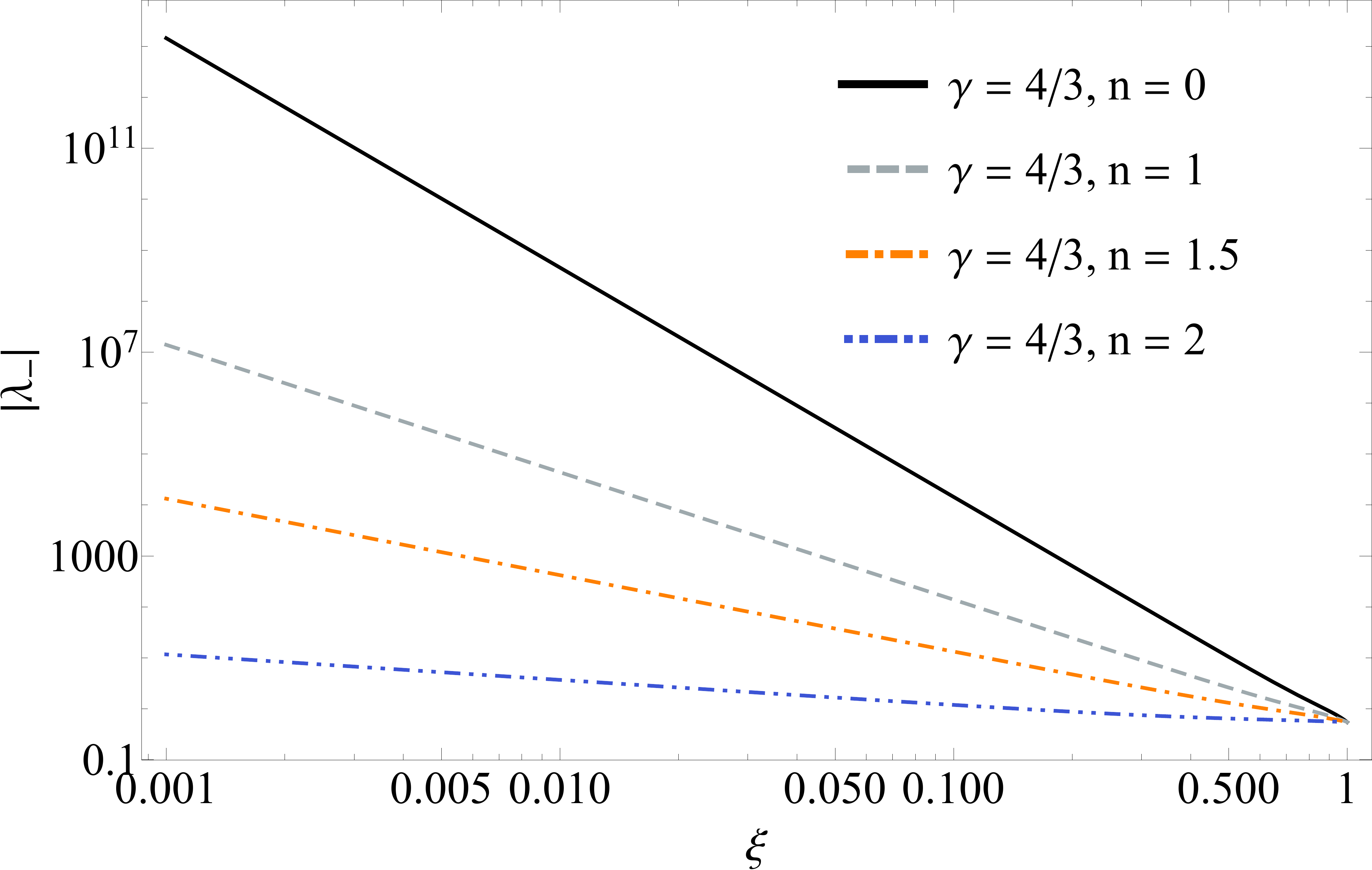}
   \includegraphics[width=0.485\textwidth]{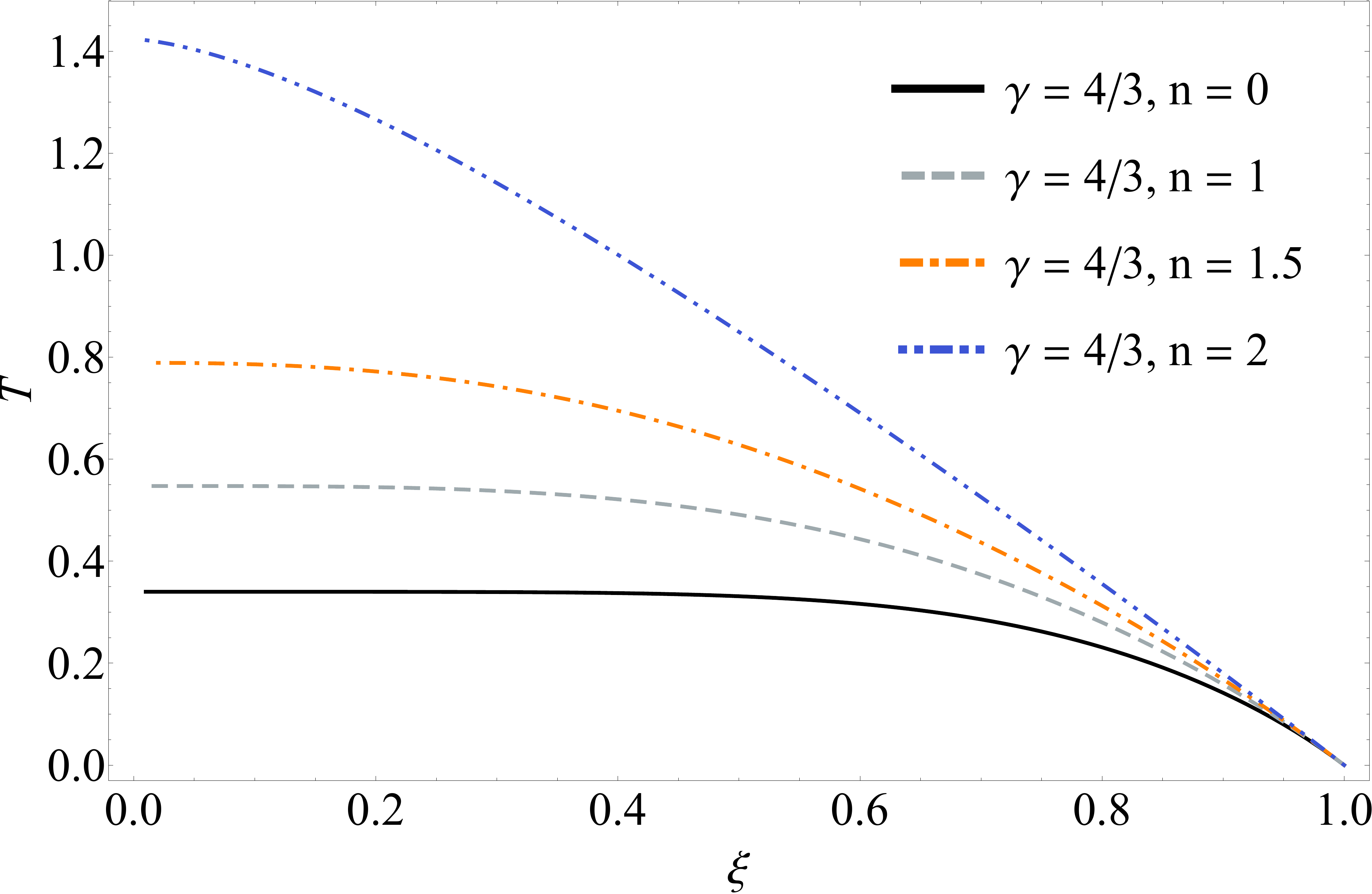} 
   \caption{Left: On a log-log scale, the absolute value of the characteristic $\lambda_{-}$, which yields the instantaneous speed at which information propagates backward from the shock front into the interior of the shocked fluid, for $\gamma = 5/3$ (top) and $\gamma = 4/3$ (bottom). For a constant-density ambient medium ($n = 0$), the sound speed increases rapidly toward and diverges at the origin, with the divergence being much more pronounced for $\gamma = 4/3$. As $n$ increases the divergence becomes less pronounced, and for $n = 2$ (a wind-like ambient medium) and $\gamma = 5/3$ the speed declines linearly with distance behind the shock. Right: The time taken for the presence of density changes at the shock to communicate to a distance $\xi$ behind the shock for $\gamma = 5/3$ (top) and $\gamma = 4/3$ (bottom). For $n < 2$ and $\gamma = 5/3$, the entire post-shock region is causally connected, and the rapid divergence of $|\lambda_-|$ means that the time flattens effectively to a constant at a finite $\xi$; for $\gamma = 5/3$, the shock front communicates to the origin in a time $T(0) = 0.447$, 0.771, and 1.23 for $n = 0$, 1, and 1.5, respectively, while the same values for $\gamma = 4/3$ are $T(0) = 0.340$, 0.547, and 0.789. For $n = 2$ and $\gamma = 5/3$, the decline of the sound speed as $|\lambda_-| \propto 1/\xi$ implies that $T \propto -\ln \xi$, and the origin is causally disconnected from the shock front, implying that the time for the backward-propagating characteristic to reach the origin is infinitely long. On the other hand, $\gamma = 4/3$ and $n = 2$ remains causally connected, with $T(0) = 1.42$}
   \label{fig:lambdam}
\end{figure}

The top-left panel of Figure \ref{fig:lambdam} shows the absolute value of the $\lambda_{-}$ characteristic as a function of $\xi$ for $\gamma = 5/3$ and the $n$ shown in the legend, while the bottom-left panel gives the same curves for $\gamma = 4/3$. For small $n$, the rapidly declining density and nearly constant pressure at small $\xi$ (see Figure \ref{fig:sedov}) imply that the sound speed increases rapidly near the origin, which is seen in this figure. However, the analytic solution for $\gamma = 5/3$ and $n = 2$ demonstrates that $|\lambda_-| \propto \xi$, and the sound speed declines exactly linearly behind the shock front. The right panels of this figure give the sound crossing time $T(\xi)$ to a distance $\xi$ behind the shock for the same $\gamma$ and $n$ as in the left panels. The rapid increase in the sound speed near the origin for small $n$ implies that the function $T$ quickly asymptotes to a constant behind the shock, and the sound crossing time from the shock to the origin for $\gamma = 5/3$ is $T(0) \simeq 0.447$, 0.771, and 1.23 for $n = 0$, 1, and 1.5, respectively, while the same values for $\gamma = 4/3$ are $T(0) = 0.340$, 0.547, and 0.789. For $n = 2$ and $\gamma = 5/3$, the linear decline in the sound speed near the origin implies that $T \propto -\ln \xi$, and the shock front is causally disconnected from the origin, while for $\gamma = 4/3$ we find $T(0) \simeq 1.42$ for $n = 2$.

The causal connectedness (or lack thereof) of the post-shock solutions yields insight into the additional, global boundary condition that we impose near the origin and the corresponding constraint on $\alpha_1$. When the solution is causally connected, inward-propagating sound waves from the surface reach the origin in a finite time and must be totally reflected back into the shocked fluid (i.e., sound waves cannot penetrate to $\xi < 0)$, and hence the perturbation to the fluid velocity must be zero at the origin (e.g., \citealt{leveque02}). 

This total reflection condition then gives $f_1(0) = 0$, and this is the additional boundary condition that will only be satisfied for a special choice of $\alpha_1(\tau)$; this boundary condition also ensures that the Lagrangian displacement of the origin is zero, which is the standard boundary condition used in the study of the radial oscillations of stars (e.g., \citealt{cox80}). This similarity between these two systems is also not coincidental, as the Sedov-Taylor blastwave is essentially a hydrostatic atmosphere near the origin (albeit one with a diverging sound speed) for small $n$ and $\gamma$. 

When the solution is causally disconnected, the origin is never reached by waves that propagate inward from the surface. In this case, no information can emanate outward from the origin into the shocked fluid, and the eigenfunction appropriate to the wave speed $\lambda_+$ must approach zero near the origin. Investigating the eigenfunctions then shows that this can only be achieved if $h_1(0) = 0$, which is the additional boundary condition that -- in this case where the origin is causally disconnected from the shock\footnote{\citet{ryu87} used this boundary condition -- the vanishing of the perturbation to the pressure at the origin -- for Sedov-Taylor blastwaves that maintain causal connectedness. As a consequence, their solutions lead to non-physical divergences of the fluid velocity near the origin and violate the scale invariance of the shock velocity and position; see Figure 14 and the discussion in the Appendix of \citet{coughlin19b}, and see \citet{kushnir05}.} -- specifies the parameter $\alpha_1$. Moreover, since the shock front does not ``know'' about the physical conditions of the fluid at the origin, it may not be the case that the other fluid variables (the velocity and the density) remain well-behaved near this point; in fact, owing to the declining sound speed, we expect inward-propagating waves to steepen in a manner similar to what occurs as waves propagate into the outer regions of a stellar envelope (e.g., \citealt{shiode14, fernandez18,coughlin18}). We see below that this is indeed the case.

Investigating the nature of the characteristics yields insight into the causal connectedness of the post-shock fluid and, correspondingly, the boundary conditions on the functions near the origin that will only be satisfied for a specific $\alpha_1$. However, there is an additional, global boundary condition that we can use to constrain $\alpha_1$: as we noted above, the Sedov-Taylor blastwave maintains the regularity of the solutions near the origin \emph{and} conserves the total energy behind the blastwave. Since changing the density of the ambient medium does not contribute any additional energy flux at the shock, the perturbations must also satisfy this global boundary condition, which will only occur for a special value of $\alpha_1$. This condition also does not depend on the causal connectedness of the solution, as the blastwave is always bounded by the finite region $0 \le \xi \le 1$. We will show in the next section how to construct this energy conservation condition in terms of the Fourier transforms of the solutions and, for all the solutions that satisfy this energy-conserving condition, they also satisfy the corresponding boundary conditions on the fluid variables near the center of the blast.

Before continuing with the analysis of the perturbation equations, we note one important feature of our approach to describing the perturbations to the shock front: in deriving the linearized equations from the general fluid equations, we remained agnostic to the precise variation of the shock position and velocity with time until the unperturbed (i.e., self-similar) solutions were derived. The self-similar solutions require not that the shock position or velocity behave in any specific way\footnote{Note that if gravitational effects modify the solution, then there is a well-defined velocity scale, but the spatial scale (if the gravitating body is compact and acts effectively as a point) remains arbitrary; thus there is meaning to rescaling the shock velocity in this case, but not the shock radius \citealt{coughlin19b}.}, but that the shock \emph{acceleration} satisfy the constraint given by Equation \eqref{Vss}. Correspondingly, the perturbations to the fluid quantities fundamentally affect the shock acceleration, and as such the last expression in Equation \eqref{pertquants} can be regarded as a differential equation for the shock position that is valid for any initial position and initial velocity. This feature of the solutions -- that the acceleration is the quantity that is fundamentally affected by the presence of perturbations -- must be present owing to the scale invariance of both the shock position and the velocity as described by the unperturbed solutions, and our approach is consistent with both of these fundamental invariants.

\section{Eigenmode Analysis and Laplace-transformed Equations}
\label{sec:eigenmodes}
The linearized equations \eqref{contpert} -- \eqref{entpert} are separable in the variables $\xi$ and $\tau$. We now take the Laplace transform of these equations in $\tau$, where the Laplace transform of (e.g.) $f_1$ is

\begin{equation}
{\rm LT}\left[{f}_1\right](\xi,\sigma) = \int_{\tau_0}^{\infty}f_1(\xi,\tau)e^{-\sigma\tau}d\tau. \label{LT}
\end{equation}
Here the lower limit on the integral, $\tau_0$, is the earliest time at which there are no perturbations to the post-shock fluid quantities or the shock acceleration; when we have a localized perturbation to the density, such that the perturbation is exactly zero for all radii less than some value, then $\tau_0$ is equal to that value and can be defined to be zero without loss of generality (i.e., the scale radius $R_0$ can always be chosen to coincide with the location at which the perturbation is introduced). When the perturbation is non-zero everywhere but decays sufficiently rapidly at large $|\tau|$, such as a Gaussian ``bump'' in radius, then we can let $\tau_0 \rightarrow -\infty$. Note that, upon taking the Laplace transform, we can further divide all of the equations by the Laplace transform of the perturbation to the density, and we then remove all explicit dependence on this quantity. Defining $\tilde{f}_1$ as the ratio of the Laplace transform of $f_1$ to the Laplace transform of $\delta\rho$, i.e.,

\begin{equation}
    \tilde{f}_1 = \frac{{\rm LT}[f_1]}{{\rm LT}[\delta \rho]},
\end{equation}
and similarly for the perturbation to the density and the pressure, the Laplace-transformed set of equations is

\begin{equation}
\sigma \tilde{g}_1-n\tilde{g}_1-\xi\frac{\partial \tilde{g}_1}{\partial \xi}+\frac{1}{\xi^2}\frac{\partial}{\partial \xi}\left[\xi^2\left(f_0 \tilde{g}_1+g_0\tilde{f}_1\right)\right] = 0, \label{L1}
\end{equation}
\begin{equation}
\sigma \tilde{f}_1+\alpha_0 \tilde{f}_1+\tilde{\alpha}_1f_0+\left(f_0-\xi\right)\frac{\partial \tilde{f}_1}{\partial \xi}+\tilde{f}_1\frac{\partial f_0}{\partial \xi}-\frac{\tilde{g}_1}{g_0^2}\frac{\partial h_0}{\partial \xi}+\frac{1}{g_0}\frac{\partial \tilde{h}_1}{\partial \xi} = 0, \label{L2}
\end{equation}
\begin{equation}
\sigma\left(\frac{\tilde{h}_1}{h_0}-\frac{\gamma \tilde{g}_1}{g_0}\right)+2\tilde{\alpha}_1+\left(f_0-\xi\right)\frac{\partial}{\partial \xi}\left[\frac{\tilde{h}_1}{h_0}-\frac{\gamma \tilde{g}_1}{g_0}\right]+\tilde{f}_1\frac{\partial}{\partial \xi}\ln\left(\frac{h_0}{g_0^{\gamma}}\right) = 0, \label{L3}
\end{equation}
with the boundary conditions 

\begin{equation}
\tilde{f}_1(1) = 0, \,\,\, \tilde{g}_1(1) = \frac{\gamma+1}{\gamma-1}, \,\,\, \tilde{h}_1(1) = \frac{2}{\gamma+1}.
\end{equation}
We recover the solutions for the fluid quantities by taking the inverse Laplace transform:

\begin{equation}
f_1 = \frac{1}{2\pi i}\int_{C}\tilde{f}_1\times \delta\tilde{\rho}\,e^{\sigma\tau}d\sigma. \label{Laplaceinv}
\end{equation}
Here the contour $C$ is in the complex plane and is a line that extends from $-i\infty$ to $i\infty$. The location at which the line intersects the real line must be more positive than all of the poles of the Laplace-transformed functions. These poles are the \emph{eigenvalues} $\sigma_{\rm i}$, and are the locations in the complex plane at which the perturbation to the shock acceleration and the other fluid variables diverge as simple poles (i.e., the parameter $\tilde{\alpha}_1$ varies as $\tilde{\alpha}_1 \propto \left(\sigma-\sigma_{\rm i}\right)^{-1}$ in the vicinity of $\sigma_{\rm i}$); if we divide Equations \eqref{L1} -- \eqref{L3} by $\tilde{\alpha}_1$, let $\tilde{f}_1 \rightarrow \tilde{f}_1/\tilde{\alpha}_1$ and similarly for the other variables, and take the limit as $\sigma \rightarrow \sigma_{\rm i}$, then the equations for the eigenmodes are identical to Equations \eqref{L1} -- \eqref{L3} but with $\tilde{\alpha}_1 = 1$, and the functions $\tilde{g}_1$ and $\tilde{h}_1$ satisfy homogeneous boundary conditions at the shock front (i.e., $\tilde{f}_1(1) = \tilde{g}_1(1) = \tilde{h}_1(1) = 0$). The eigenvalues are then constrained by requiring that the functions satisfy the appropriate boundary conditions at the origin ($\tilde{f}_1(0) = 0$ for $n < 2$, $\tilde{h}_1(0) = 0$ for $n = 2$). 

Ordinarily the eigenmodes are useful for numerically calculating the inverse Laplace transforms of the functions and for determining the asymptotic, temporal behavior of the solutions, because the eigenvalues typically form an infinite set of discrete points, implying that we can close the contour integral in Equation \eqref{Laplaceinv} in the left half of the complex plane and use the residue theorem to write the continuous integral as a discrete sum over the poles. This is the approach taken in stellar oscillation theory, and can also be used to understand the unstable nature of some shocks (e.g., \citealt{ryu87, goodman90, coughlin19b, ro19}). The pole with the largest real part dominates the late-time behavior of the perturbations, and if there are any eigenvalues with a positive real part, the solution is dynamically unstable. On the other hand, if all of the poles are imaginary, then the solution is neutrally stable and oscillates in response to the presence of perturbations.

\begin{figure}
    \centering
    \includegraphics[width=0.495\textwidth]{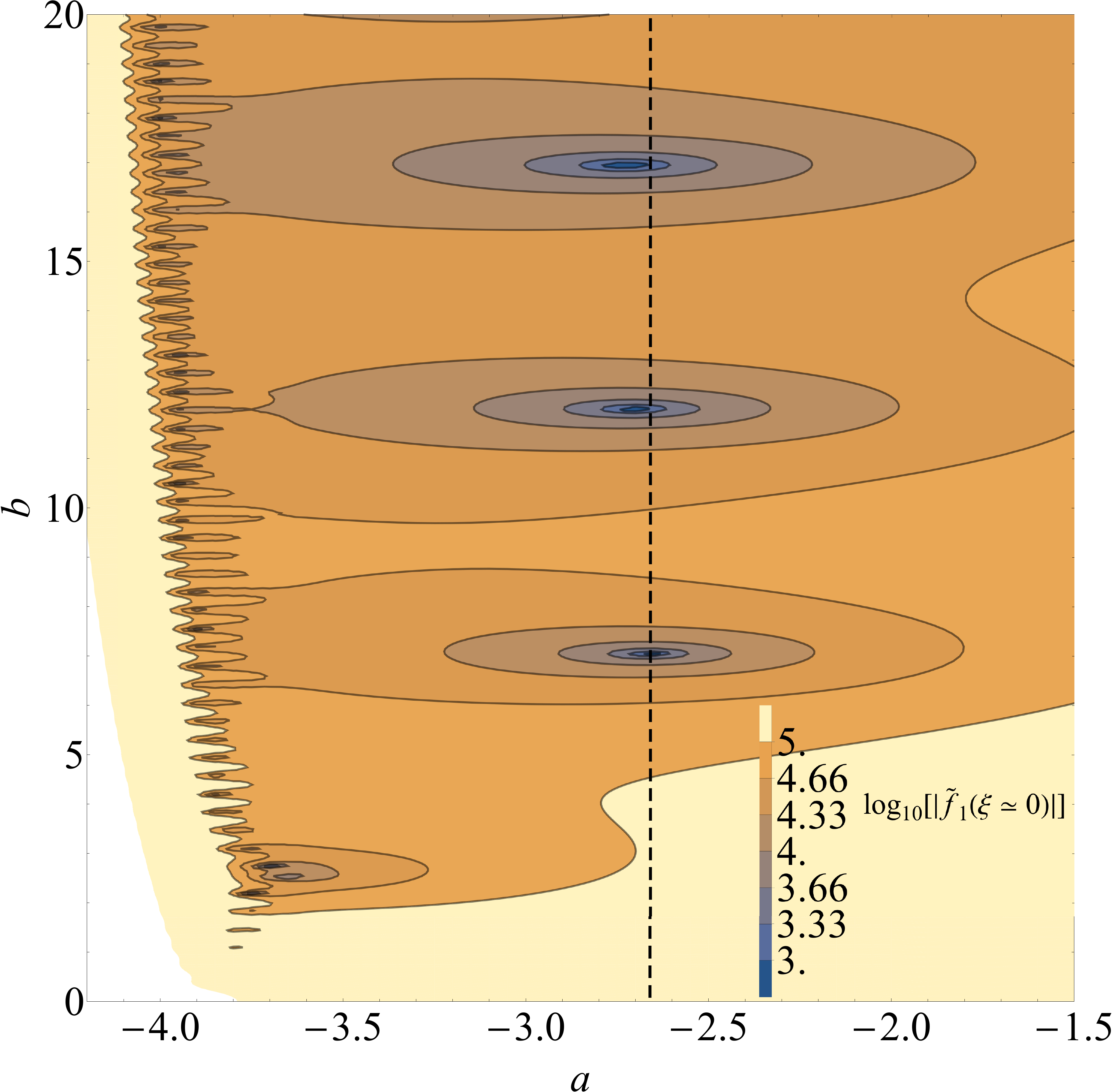}
     \includegraphics[width=0.465\textwidth]{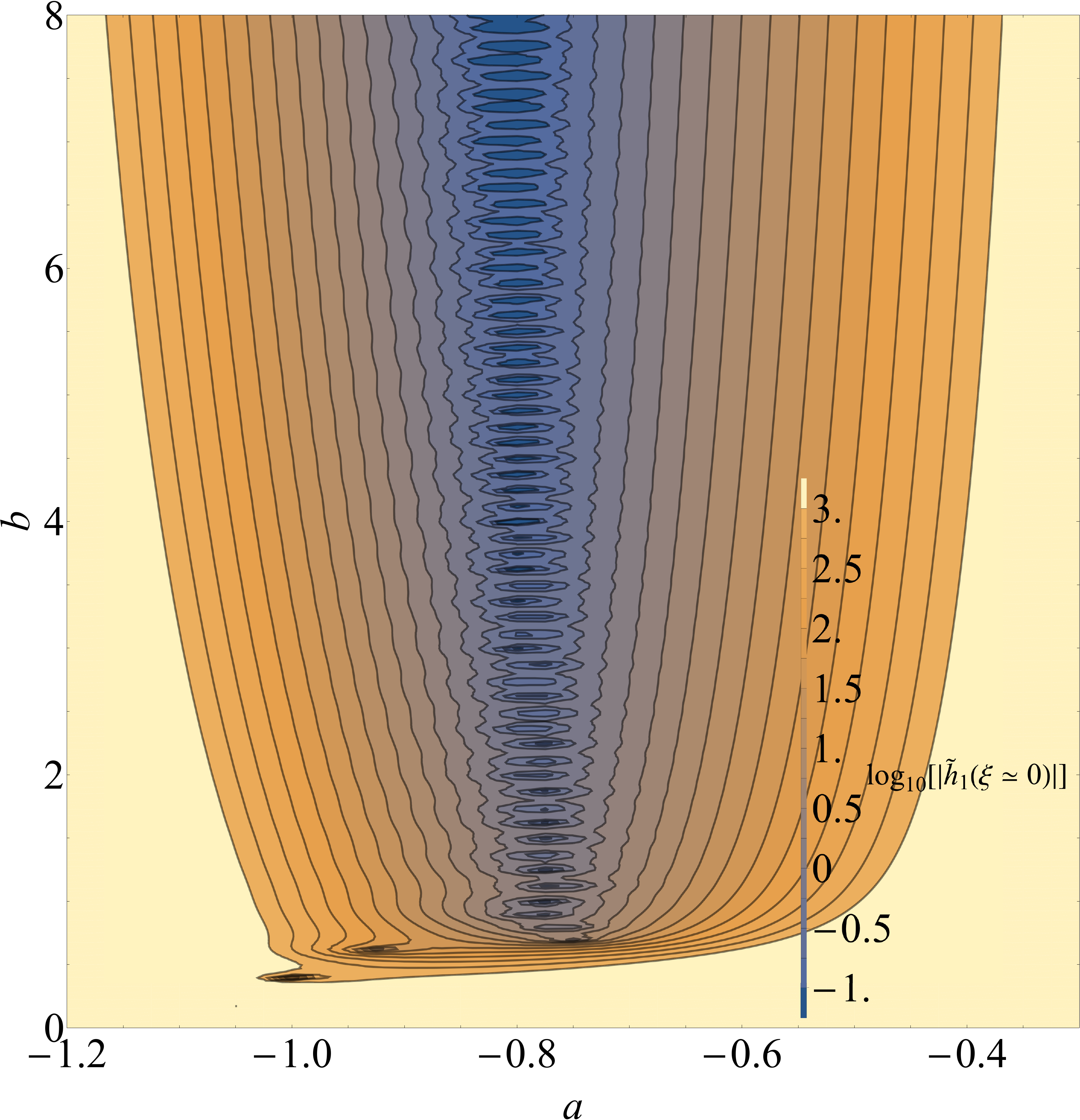}
    \caption{Left: A contour plot of the base-10 logarithm of $|\tilde{f}_1(\xi \simeq 0)|$, which is the absolute value of the correction to the velocity near the origin, as a function of the real ($a$) and imaginary ($b$) parts of the Fourier variable $\sigma$ for a constant-density medium ($n = 0$) and $\gamma = 5/3$. The eigenmodes are locations where the perturbation to the velocity equals zero, and are therefore characterized by blue regions in this panel (i.e., where the base-10 logarithm of $|\tilde{f}_1(\xi \simeq 0)|$ becomes small). The eigenmodes are characterized by discrete points, which occur at $a \simeq -2.7$ and $b \simeq 7$ and $b \simeq 12$ and are the localized islands in this figure, and also a continuum that has $a \simeq -4$ and extends over all $b$. Right: A contour plot of the base-10 logarithm of $|\tilde{h}_1(\xi \simeq 0)|$, being the absolute value of the correction to the pressure near the origin, as a function of the real ($a$) and imaginary ($b$) parts of the Fourier variable $\sigma$ for a wind-like medium ($n = 2$) and $\gamma = 5/3$. The eigenmodes are locations where the perturbation to the pressure equals zero near the origin, and are therefore characterized by blue regions in this panel. Unlike the constant-density case (left panel), here there are no discrete poles, and the eigenmodes are characterized by a continuum that lies in a ``valley'' near $a \simeq -0.8$. }
    \label{fig:poles}
\end{figure}

Many investigations have dealt exclusively with the eigenmodes for these reasons (e.g., \citealt{bernstein80, ryu87, chevalier90, goodman90, ryu91, coughlin19b}). Unfortunately, however, the eigenmodes describing the perturbations to the Sedov-Taylor blastwave are represented by a discrete of set poles \emph{and a continuum of poles} with approximately the same real part, as determined by \citet{sanz16}. By integrating the eigenmode equations numerically for a number of $n$ and $\gamma$, we find results that agree with \citet{sanz16}: there is a set of discrete, localized poles and a continuum of poles that all have roughly the same real component and, therefore, lie approximately on a vertical line in the complex plane. All the modes we find also have a negative real part, implying that the Sedov-Taylor blastwave is stable to radial perturbations. To illustrate these features, Figure \ref{fig:poles} shows a contour map of the base-10 logarithm of $|\tilde{f}_1(\xi = 0.001)|$ as a function of the real and imaginary parts of $\sigma$ (where $a$ and $b$ are the real and imaginary parts of the complex eigenvalue $\sigma$, respectively) calculated from the eigenmode equations when $n = 0$ and $\gamma = 5/3$ (left panel), and $|\tilde{h}_1(\xi = 0.001)|$ when $n = 2$ and $\gamma = 5/3$ (right panel). The locations at which $|\tilde{f}_1(\xi \simeq 0)| = 0$ (left) or $|\tilde{h}_1(\xi \simeq 0)| = 0$ (right) represent the eigenmodes, and are the blue regions in this figure. In the left panel we see that there are two discrete poles with $a \simeq -2.7$ and $b \simeq 7$ and $12$, but there are also many, finely spaced poles that lie along the line $a \simeq -4$ that approach a continuum in the limit that the radius at which we evaluate the inner boundary condition goes to zero (i.e., as $\xi$ becomes smaller, the function $\tilde{f}_1$ becomes highly oscillatory in this region and there are more solutions that satisfy $\tilde{f}_1(\xi \simeq 0) = 0$). In the right panel, there are no discrete eigenmodes, and instead the poles all lie along a continuum that has a real part $a \simeq -0.8$. Table \ref{tab:eigenvalues} gives the largest, discrete eigenvalue for both $\gamma = 4/3$ and $\gamma = 5/3$ and a range of $n$. Notice that, because our approach perturbs only the \emph{acceleration} of the shockwave (i.e., the self-similarity of the solutions only necessitates that $\partial \ln V/\partial \tau$ is a constant) and defines the self-similar variable in terms of the true shock position, the trivial solutions that correspond to renormalizations of the shock velocity and position do not appear; these solutions are manifestly obtained by integrating the differential equation for the shock, the right-most in Equation \eqref{pertquants}, and including the constants of integration. 

\begin{table}
\begin{center}
\begin{tabular}{|c|c|c|c|c|}
\hline
  $\sigma$  & $n = 0$ & $n = 1$ & $n = 3/2$ & $n = 2$ \\
  \hline
   $ \gamma = 4/3 $ & $\sigma_1 = -2.36 + 9.58i$ & $-1.68+6.23i$ & $-1.34+4.52i$ & $-1.01+4.48i$ \\
    \hline
     $ \gamma = 5/3 $ & $ -2.66 + 7.05i$ & $-1.78+4.32i$ & $-1.32+2.86i$ & \ldots \\
     \hline
\end{tabular}
       \caption{The largest, discrete eigenvalue for the Sedov-Taylor blastwave for $\gamma$ and $n$ given in the left column and top row. For $\gamma = 5/3$, $n = 2$, there is only a continuum of poles with a real value around $\simeq -0.8$ (see Figure \ref{fig:poles}). }
           \label{tab:eigenvalues}
\end{center}
\end{table}

For $n < 2$, these modes satisfy the constraint that the radial velocity equals zero at $\xi \simeq 0$. However, the perturbation to the density fluctuates violently about the origin, and -- while the sign of the density perturbation oscillates increasingly rapidly as we near $\xi \simeq 0$ -- the magnitude of the perturbation diverges (see Figure \ref{fig:funsofn} below). We also must include an extremely large number of modes in our calculation of the inverse transform in this case, and the inverse transform would still be in the form of an integral (i.e., using the eigenmodes no longer transforms the continuous sum into a discrete one). For these reasons, the eigenmodes do not appear to offer a useful means for recovering the linear response of a strong shock to density inhomogeneities in an ambient medium. 

Note, however, that \emph{we do not need to use the eigenmodes} to evaluate the integral in Equation \eqref{Laplaceinv}: since all of the poles lie to the left of the imaginary axis, we can let $\sigma$ be a purely imaginary number in Equations \eqref{L1} -- \eqref{L3}, integrate the equations numerically inward from the shock front for an arbitrarily chosen $\tilde{\alpha}_1$ and calculate the value of $\tilde{f}_1(\xi \simeq 0)$ or $\tilde{h}_1(\xi \simeq 0)$. For the same $\sigma$, we can then perturb the value of $\tilde{\alpha}_1$ and calculate the new solution for $\tilde{f}_1(\xi \simeq 0)$ or $\tilde{h}_1(\xi \simeq 0)$, and use the difference between the new and old values to inform the new guess for $\tilde{\alpha}_1$ that will better satisfy the boundary condition near the origin. We can continue to iterate on the value of $\tilde{\alpha}_1$ until the boundary condition near the origin is satisfied to a high level of tolerance. By repeating this process for a densely sampled range of imaginary $\sigma$ that extends to a large value, we can interpolate the solution for $\tilde{\alpha}_1$ over that range and thereby construct the function $\tilde{\alpha}_1(\sigma)$, and numerically calculate the integral appearing in Equation \eqref{Laplaceinv} for any given density perturbation (assuming the integral remains finite and decays sufficiently rapidly at large $|\sigma|$; see below for a discussion of when this is true).

In the next section we employ this procedure to calculate the response of an adiabatic, strong shock to an ambient density perturbation. Before doing so, we note two features of the solutions, the first being that if $\tilde{\alpha}_1$ is a solution to the equations for a given $\sigma$, then $\tilde{\alpha}_1^{*}$ is the solution with $\sigma \rightarrow \sigma^{*}$. Thus, if we restrict $\sigma$ to purely imaginary values, then we only need to calculate $\tilde{\alpha}_1$ for ${\rm Im[\sigma]} \ge 0$. Also, with $\sigma$ purely imaginary the Laplace transform (and the inverse) reduces to the Fourier transform, as can be seen from Equation \eqref{LT}. We can therefore use the features of the Fourier transform, and the Fourier transforms of specific functions, to understand the behavior of the shock to various perturbations.

The second is that, as we mentioned at the end of Section \ref{sec:global}, the Sedov-Taylor solution maintains the regularity of the fluid variables at $\xi = 0$ \emph{and} conserves the total energy behind the blastwave. The perturbations to the blastwave must also possess this energy-conserving feature, as the pressure of the ambient medium is formally zero (i.e., the shock Mach number is infinite) and adding a density perturbation does not change the enthalpy of the ambient fluid; we can use this property of the solutions to construct a distinct, integral constraint on the value of $\tilde{\alpha}_1$ by noting that the total energy behind the blastwave is

\begin{equation}
E = 4\pi \int_0^{R} \left(\frac{1}{2}\rho v^2+\frac{1}{\gamma-1}p\right)r^2\,dr = 4\pi\rho_0 R_0^3\left(\frac{R}{R_0}\right)^{3-n}V^2\int_0^{1}\left(\frac{1}{2}g f^2+\frac{1}{\gamma-1}h\right)\xi^2d\xi, \label{enint}
\end{equation}
where in the last line we introduced our definitions of the fluid variables in terms of the dimensionless quantities. Differentiating both sides of this expression with respect to $t$, using our definitions for the perturbations to the fluid variables \eqref{pertquants}, keeping only linear terms, setting the result equal to zero, rearranging and taking the Laplace transform then gives

\begin{equation}
2\tilde{\alpha}_1\int_0^{1}\left(\frac{1}{2}g_0f_0^2+\frac{1}{\gamma-1}h_0\right)\xi^2d\xi + \sigma\int_0^{1}\left(\frac{1}{2}\tilde{g}_1f_0^2+g_0f_0\tilde{f}_1+\frac{1}{\gamma-1}\tilde{h}_1\right)\xi^2d\xi = 0. \label{alphaen}
\end{equation}
This condition for $\tilde{\alpha}_1$ must hold independently of the causal connectedness of the fluid and the boundary condition on the fluid variables imposed at the origin. We have checked that this additional, energy-conserving condition holds for all solutions found in the next section, where we find solutions to the Fourier-transformed set of Equations \eqref{L1} -- \eqref{L3}. Equation \eqref{alphaen} also shows that $\tilde{\alpha}_1(\sigma = 0) = 0$; we return to the significance of this feature below.

\section{Solutions}
\label{sec:solutions}
\subsection{General solutions}

\begin{figure}[htbp] 
   \centering
   \includegraphics[width=0.495\textwidth]{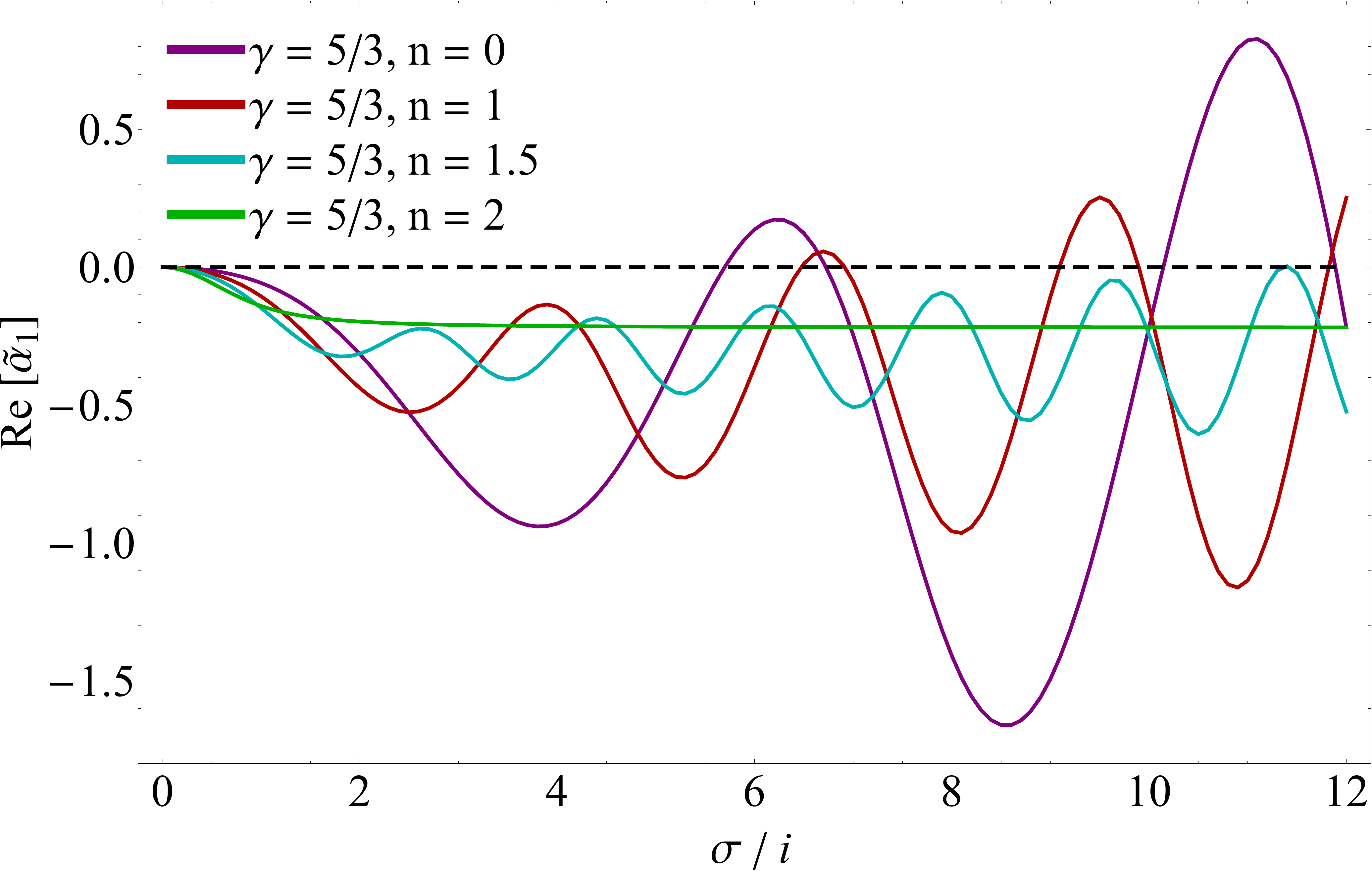} 
   \includegraphics[width=0.485\textwidth]{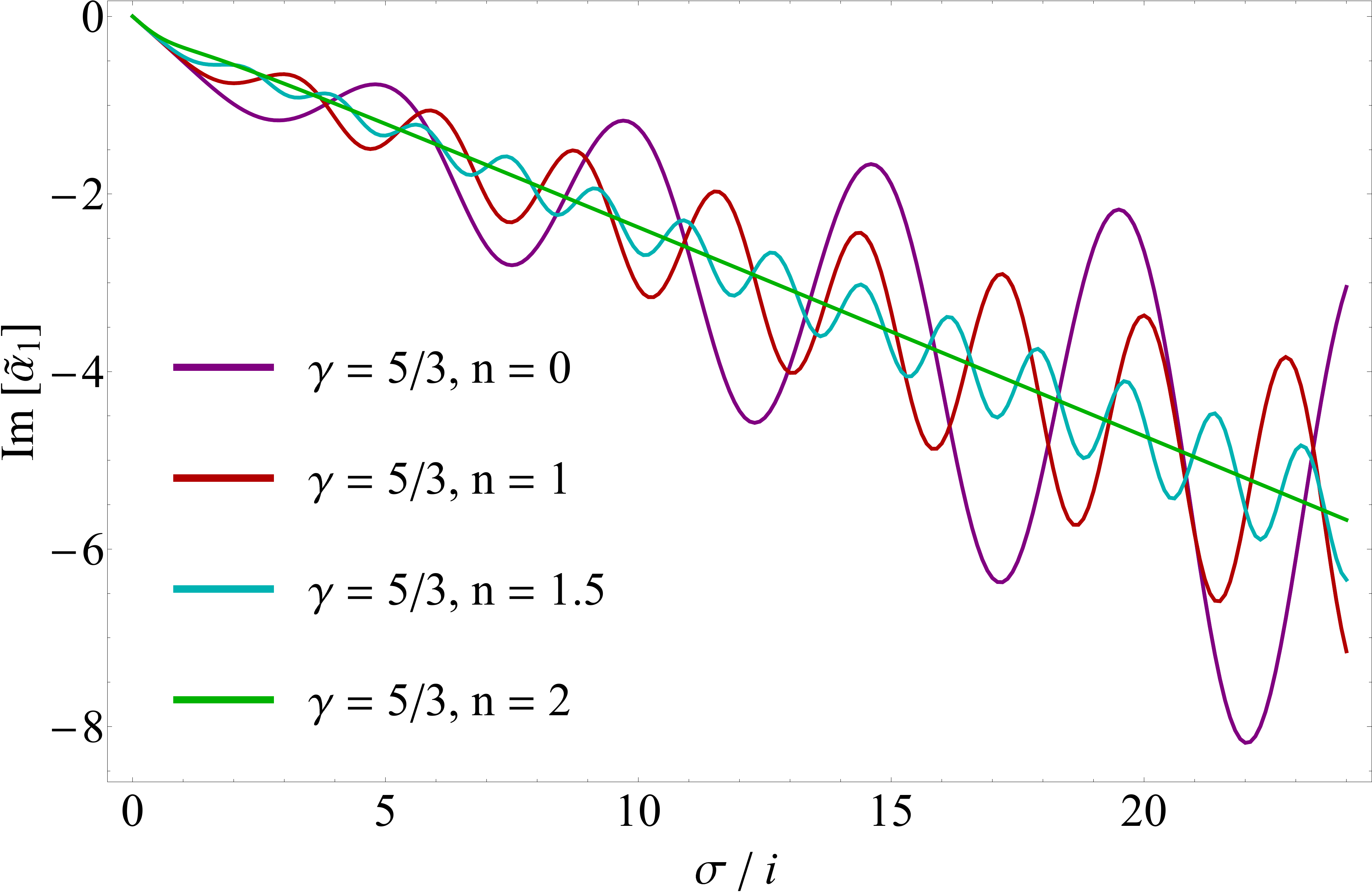} 
   \caption{The real (left panel) and imaginary (right panel) components of the Fourier transform of the perturbation to the shock acceleration. Here the solutions were calculated with $\gamma = 5/3$, and the value of $n$ -- which is the radial power-law index of the density profile of the ambient medium -- is shown in the legend. The black, dashed line in the left panel shows ${\rm Re[\tilde{\alpha}_1]} = 0$ for reference. Solutions with $n = 0$, 1, and 1.5, which remain causally connected with the origin, exhibit oscillatory behavior and grow as $\propto \sigma$. For $n = 2$ (a wind-fed medium), for which the solution is causally disconnected from the origin, the real part of the perturbation to the shock acceleration levels to a constant value ($\simeq -0.2$) as $\sigma$ increases, the imaginary part declines linearly with increasing $\sigma$, and no oscillatory behavior is exhibited.}
   \label{fig:Realpha}
\end{figure}

\begin{figure}[htbp] 
   \centering
   \includegraphics[width=0.32\textwidth]{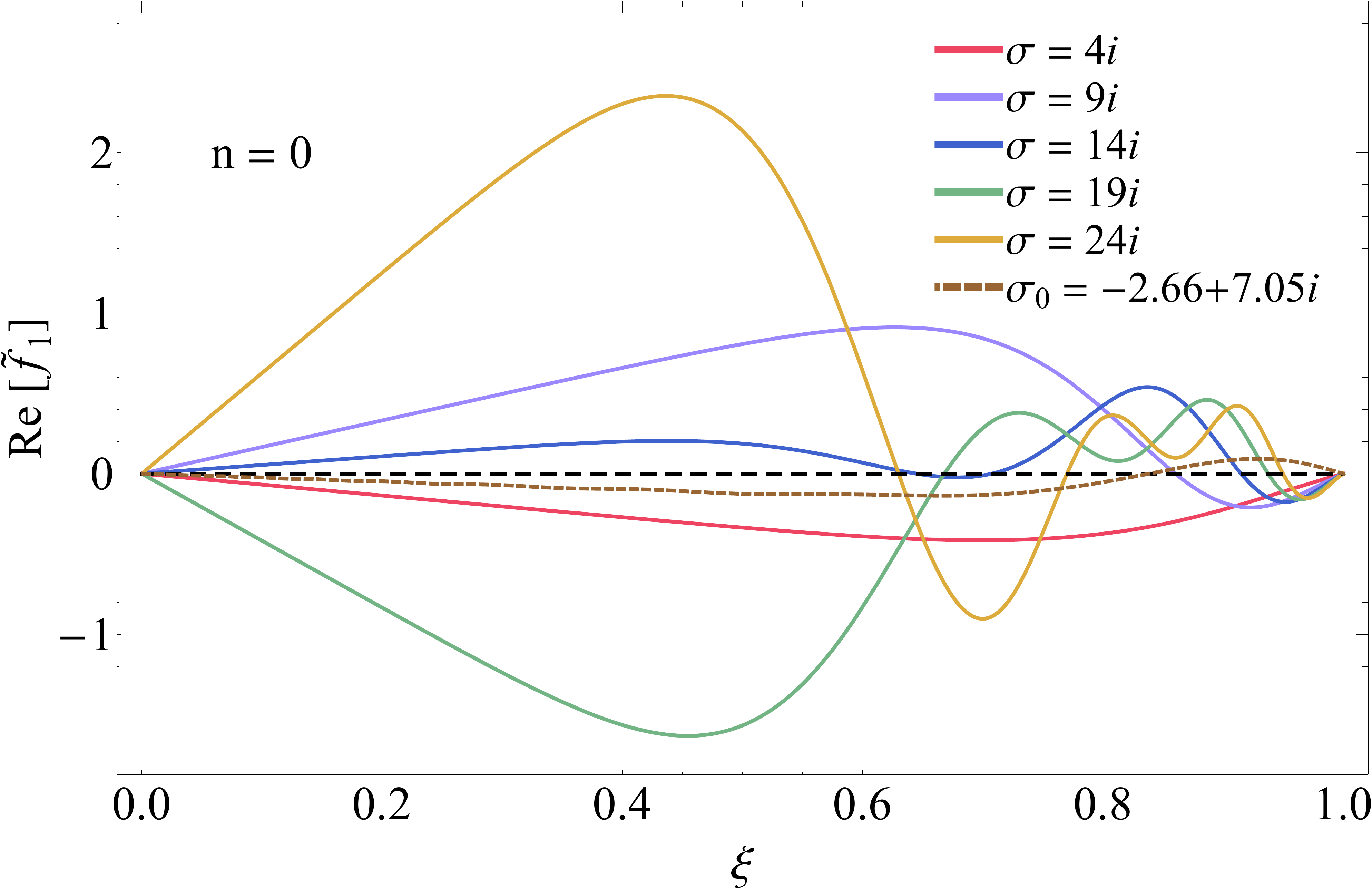} 
   \includegraphics[width=0.32\textwidth]{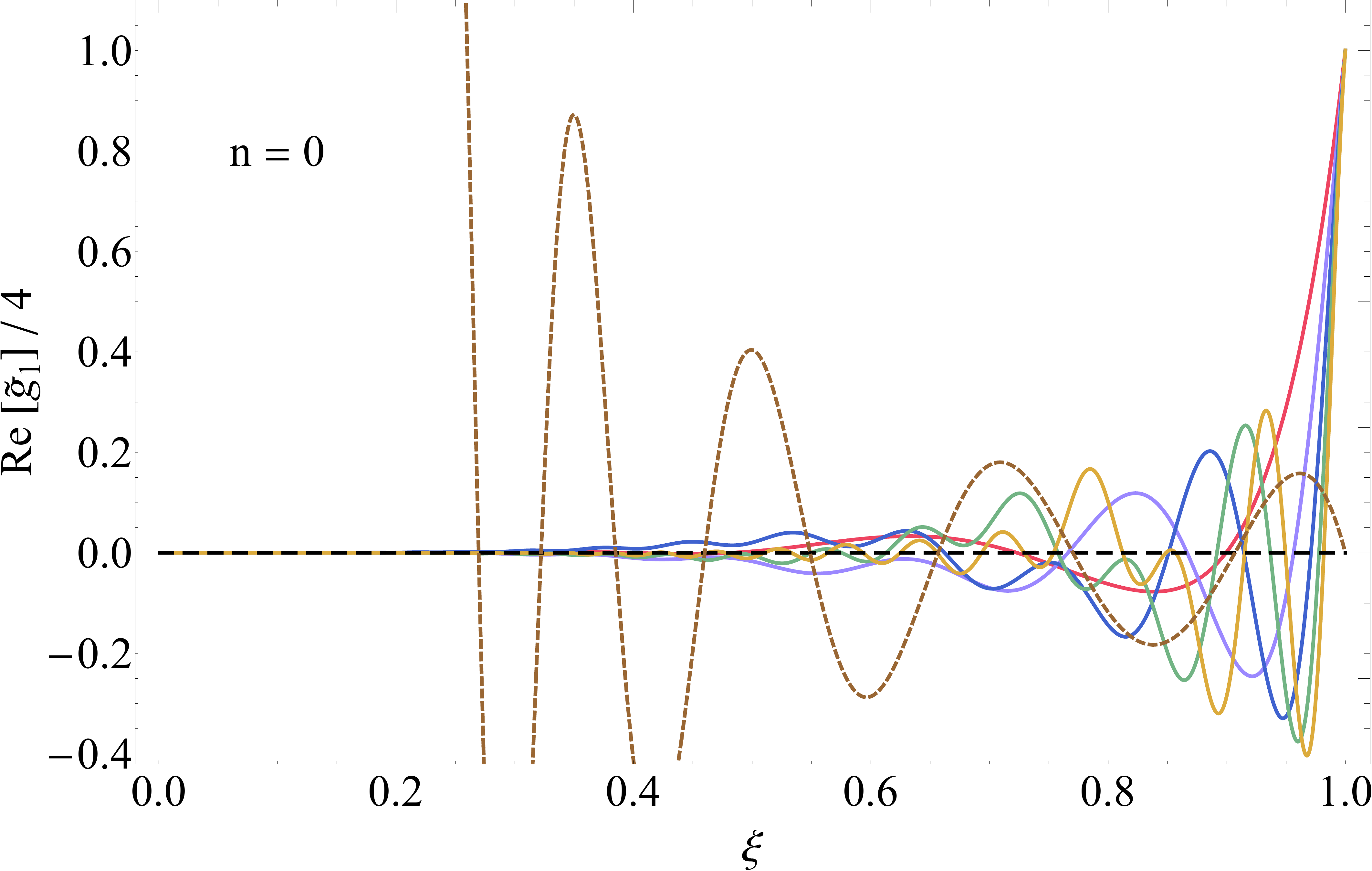} 
   \includegraphics[width=0.32\textwidth]{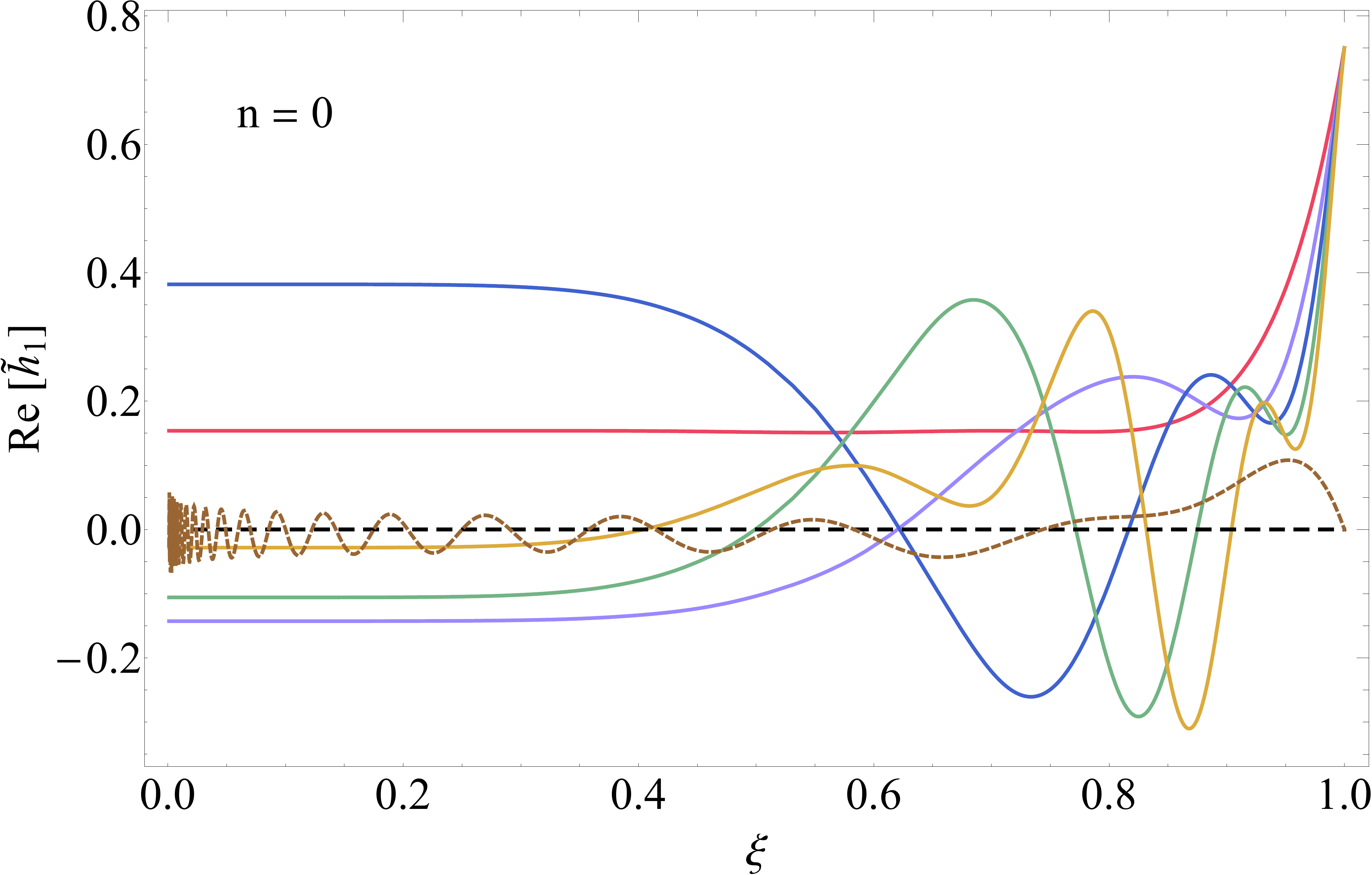} 
     \includegraphics[width=0.32\textwidth]{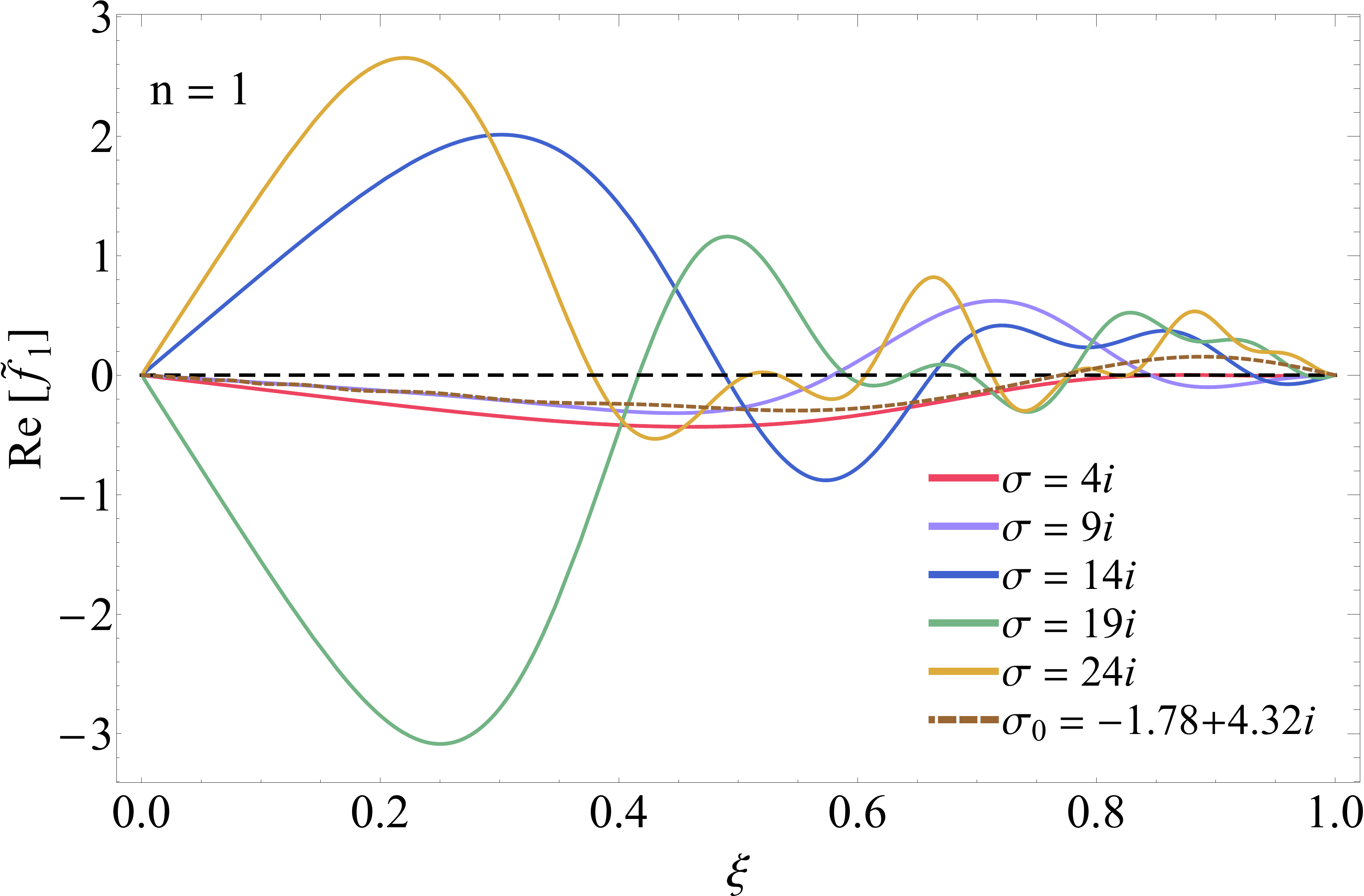} 
   \includegraphics[width=0.32\textwidth]{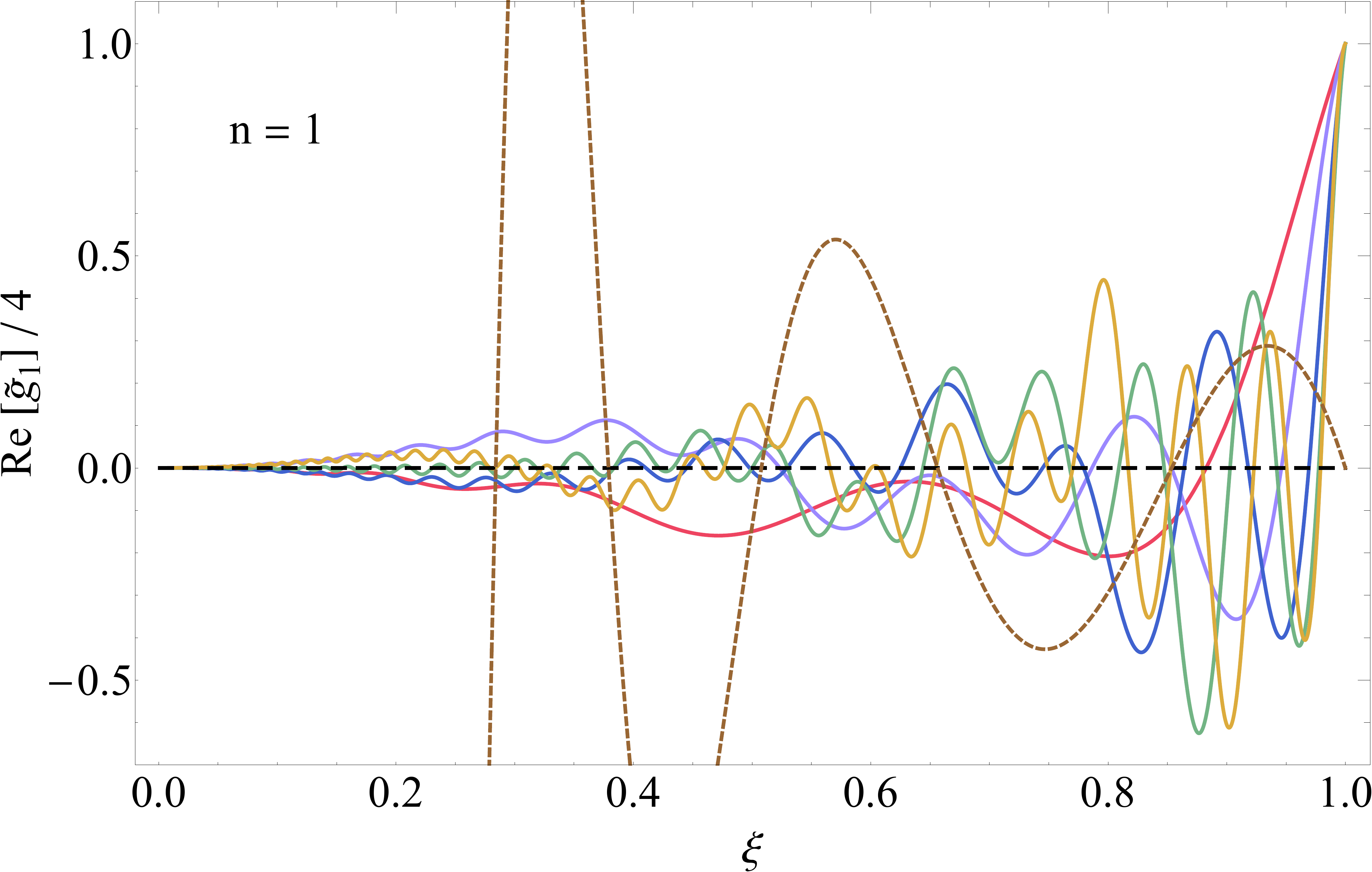} 
   \includegraphics[width=0.32\textwidth]{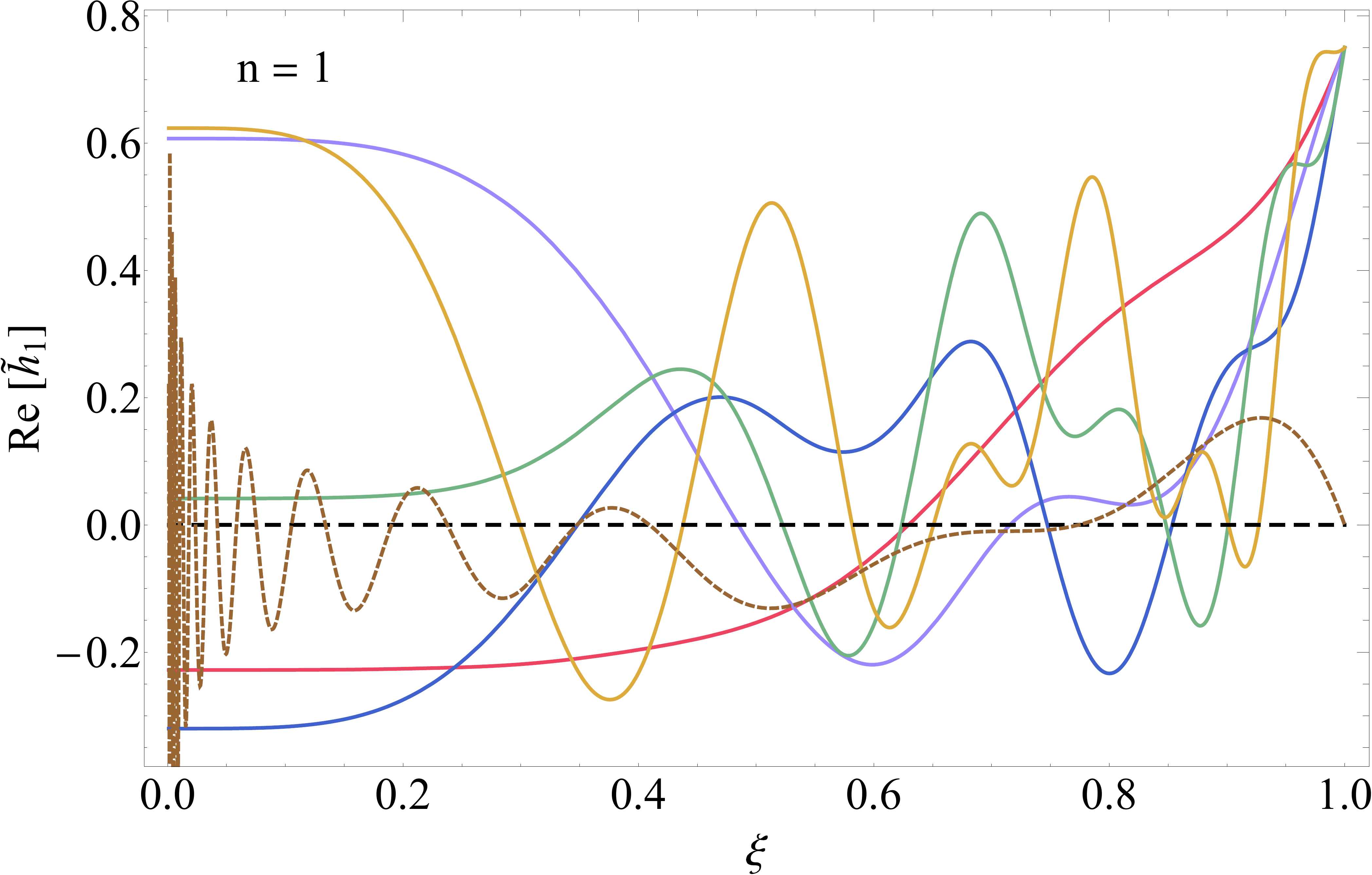} 
    \includegraphics[width=0.32\textwidth]{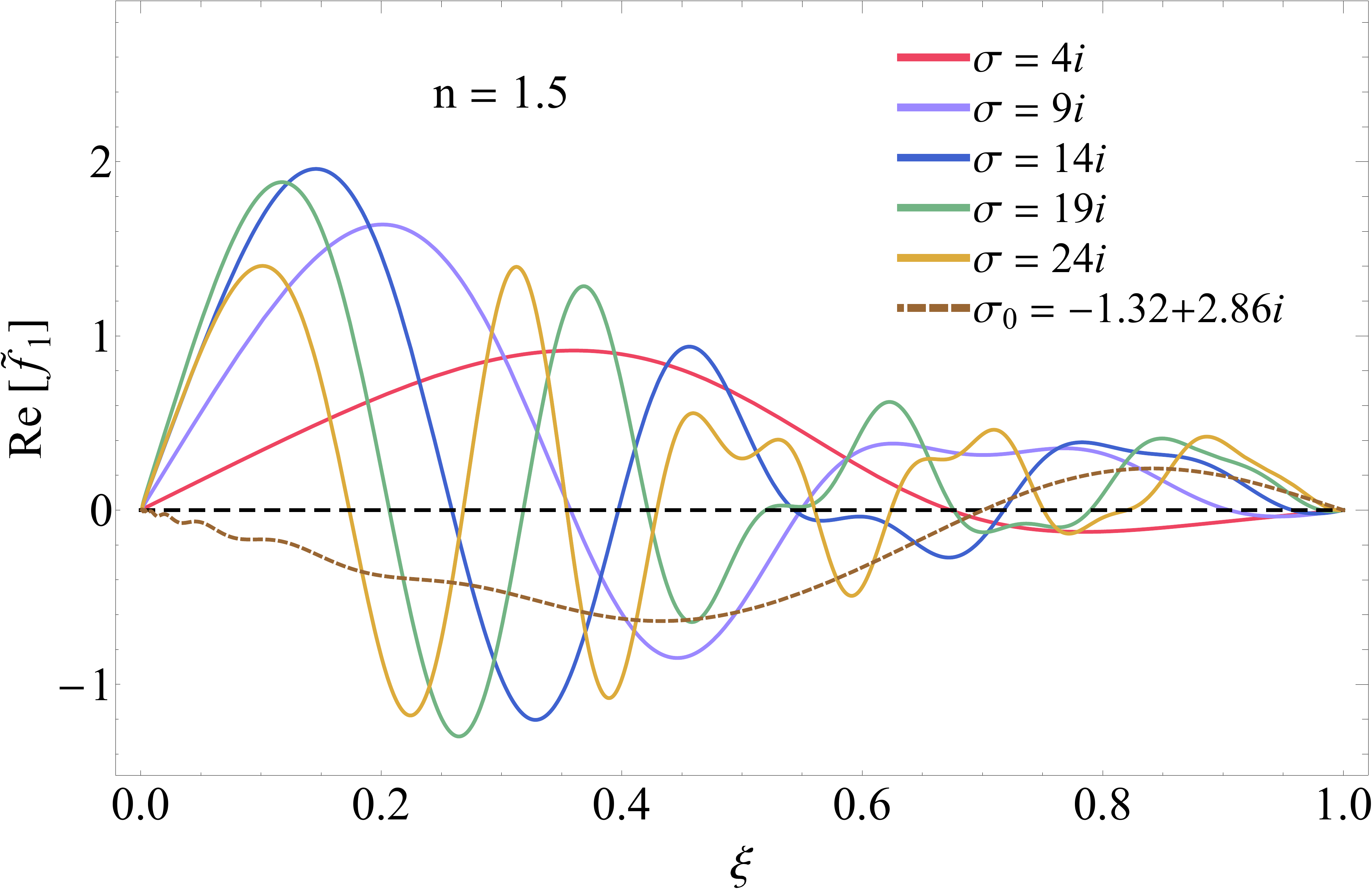} 
   \includegraphics[width=0.32\textwidth]{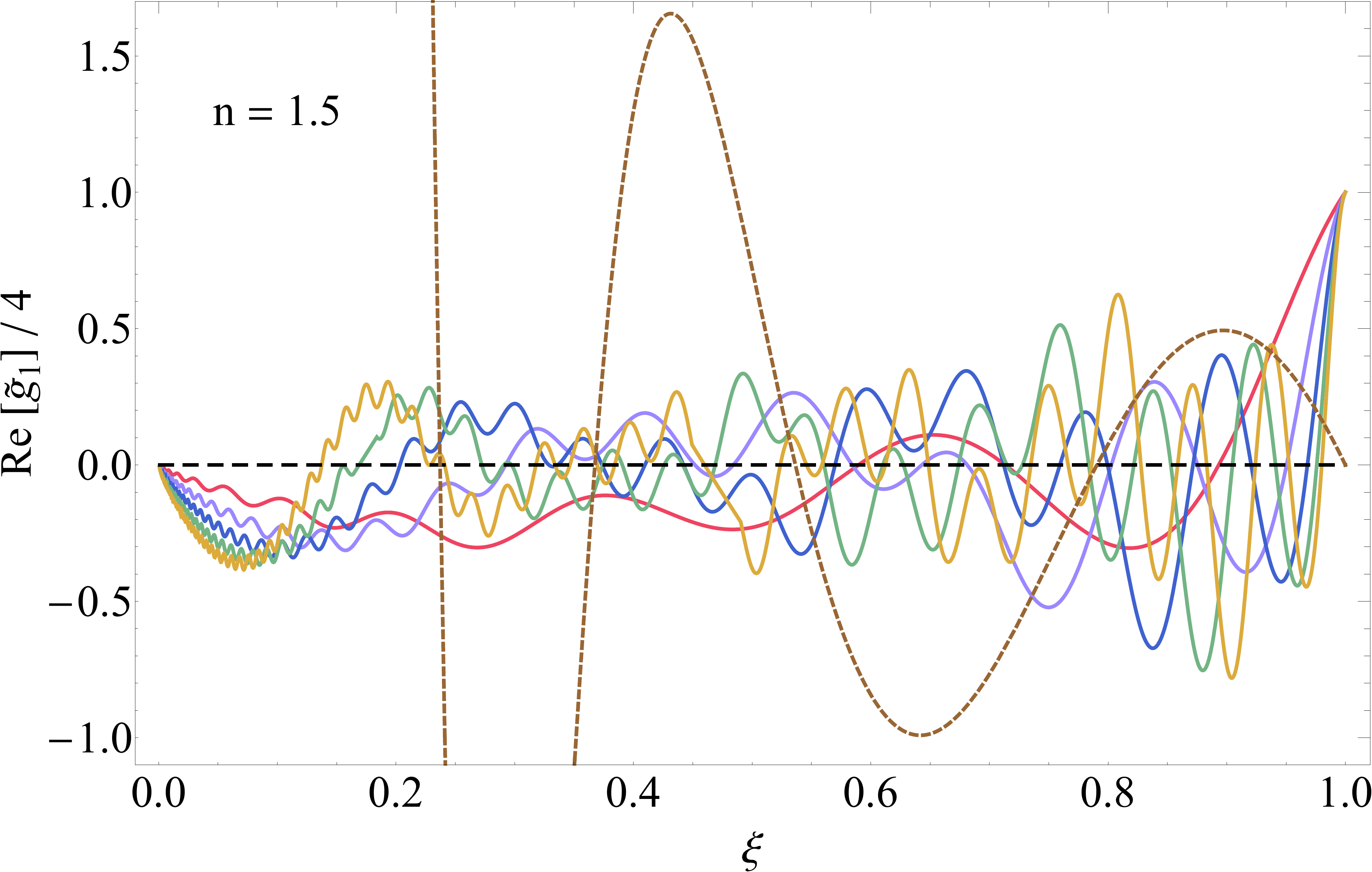} 
   \includegraphics[width=0.32\textwidth]{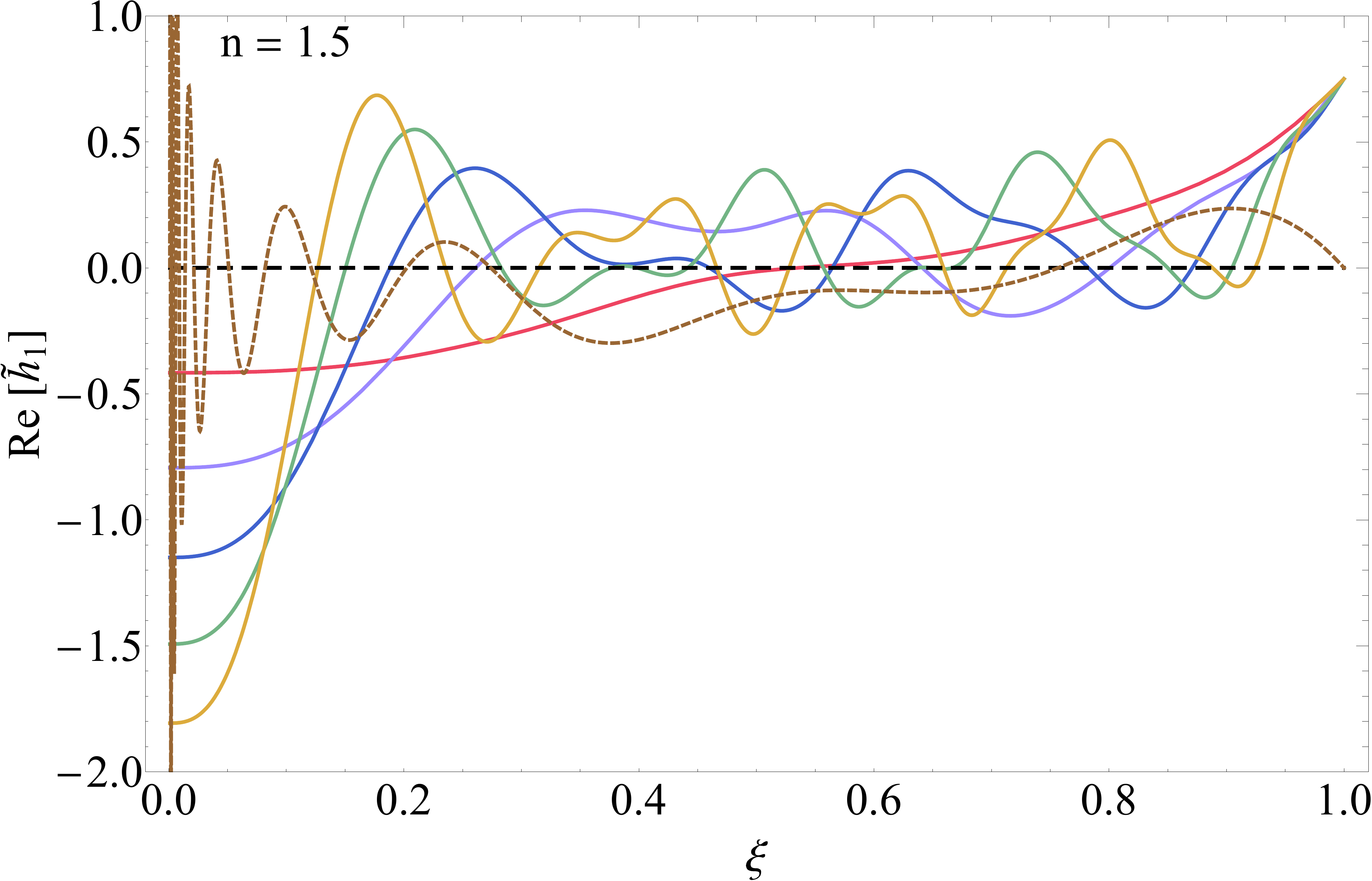} 
     \includegraphics[width=0.32\textwidth]{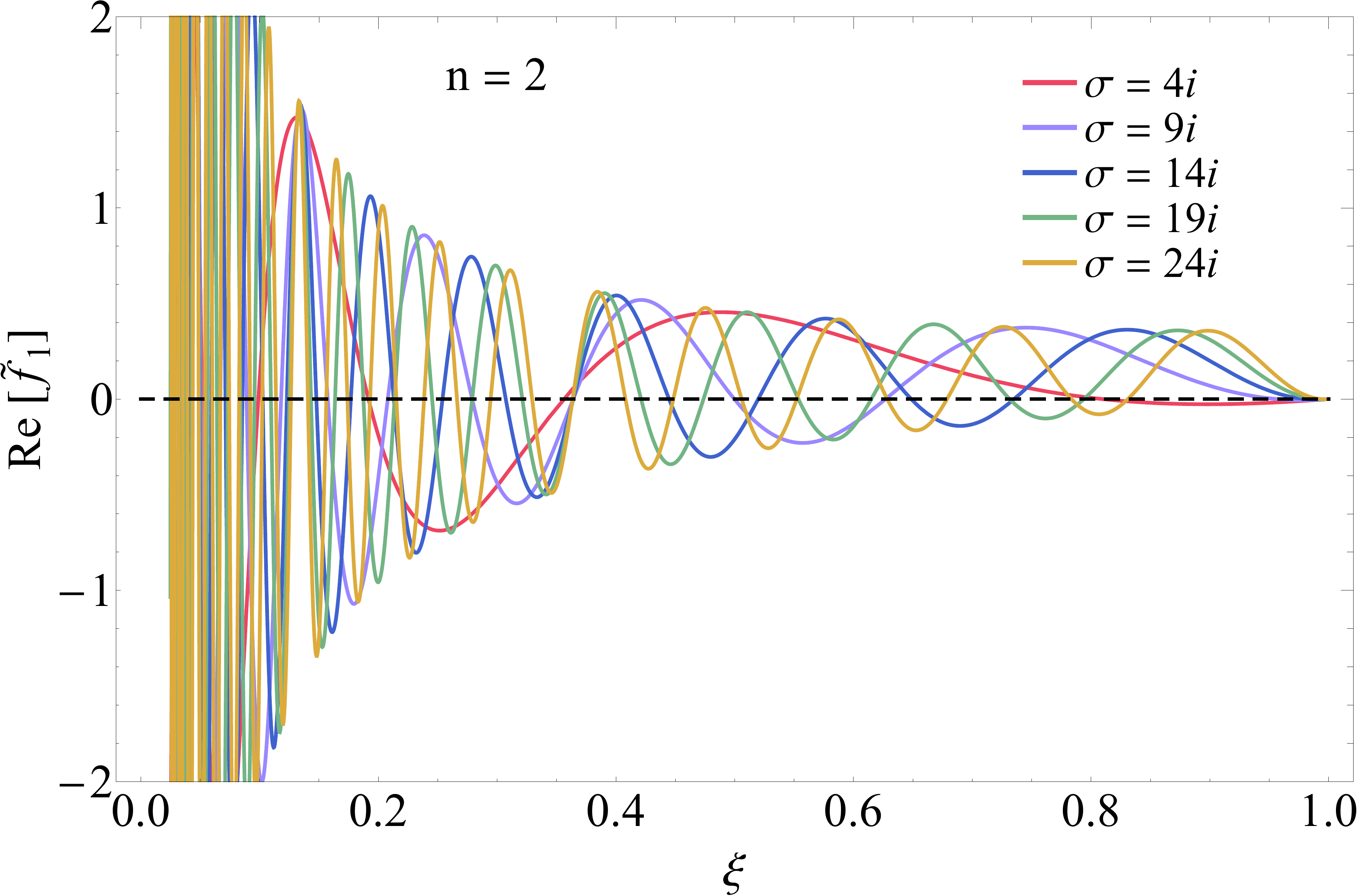} 
   \includegraphics[width=0.32\textwidth]{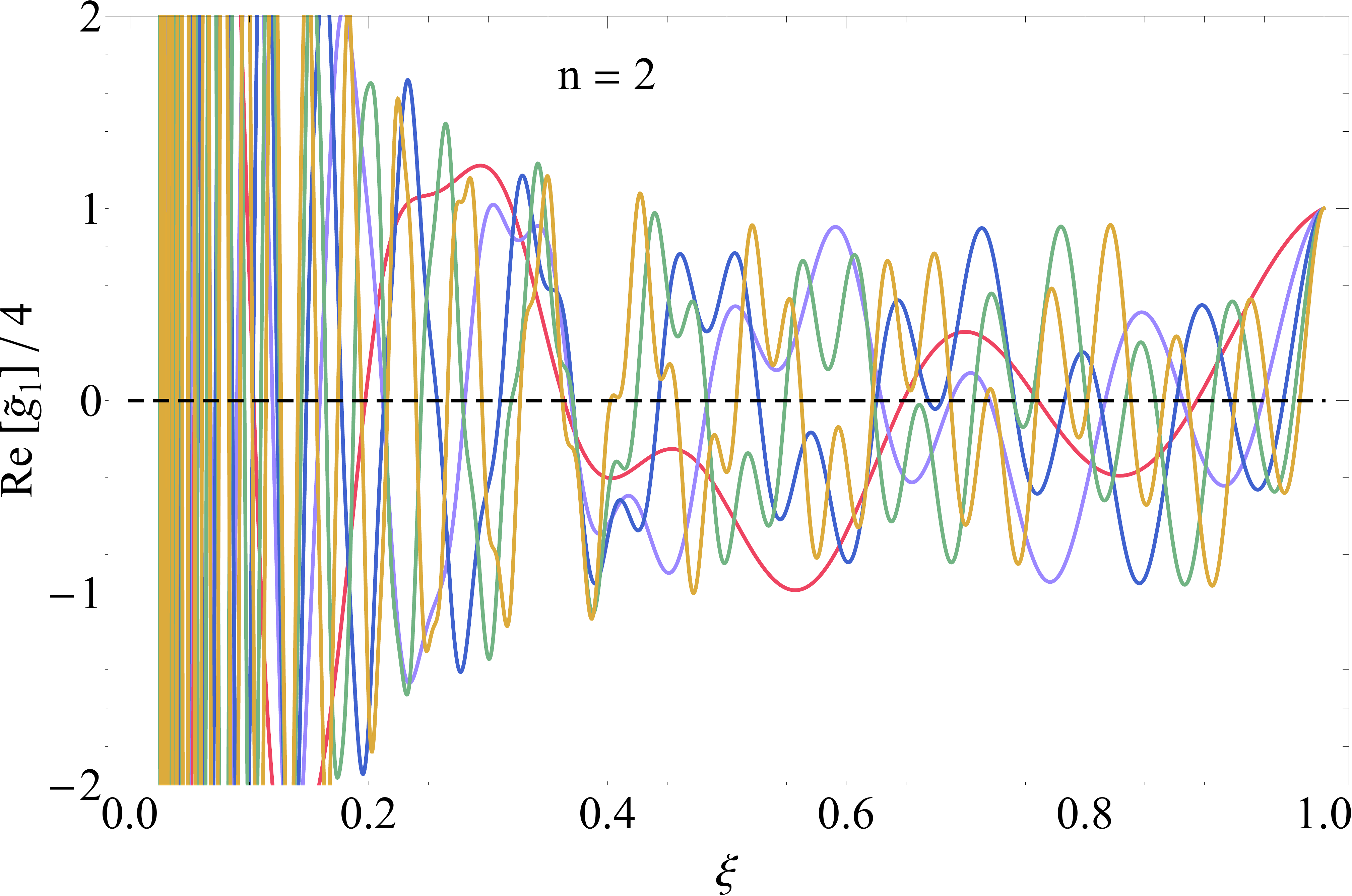} 
   \includegraphics[width=0.32\textwidth]{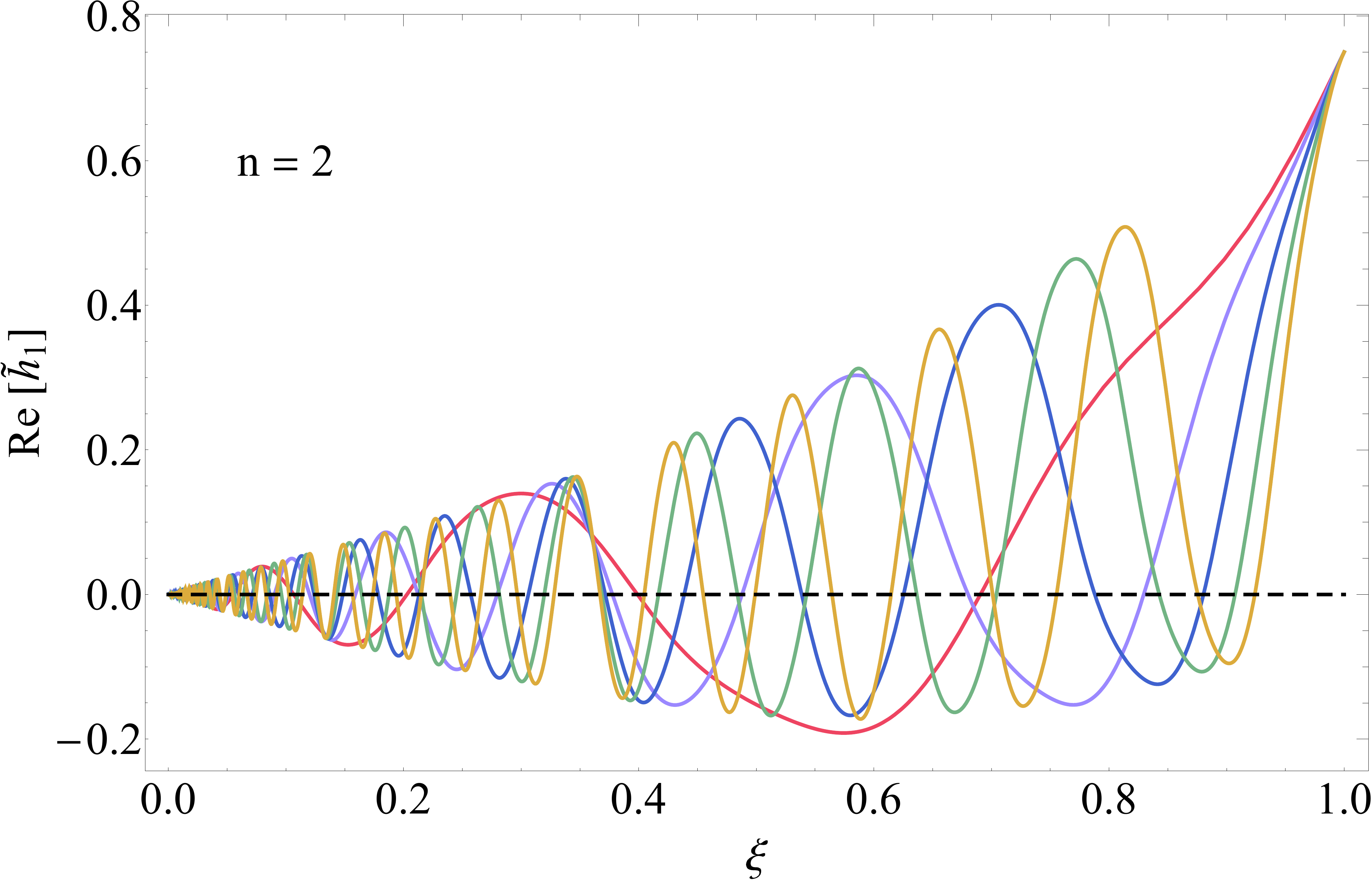} 
   \caption{The real part of the Fourier-transformed correction to the velocity (left column), the density (middle column), and the pressure (right column) for $n = 0$ (top row), $n = 1$ (second row from the top), $n = 3/2$ (second row from the bottom), and $n = 2$ (bottom row), all for a post-shock adiabatic index of $\gamma = 5/3$. These solutions satisfy the boundary conditions at the shock front ($\xi = 1$; note that the density is normalized by $(\gamma+1)/(\gamma-1) = 4$) and maintain global energy conservation.  The different colors correspond to the Fourier-transformed frequency in the legend, and the brown, dashed curves in the top and middle rows are the eigenfunctions that correspond to the smallest, discrete ``eigenvalue,'' the eigenvalue itself, denoted by $\sigma_0$, shown in the legend. The legend for each row is given in the left-most panel of the respective row. }
   \label{fig:funsofn}
\end{figure}

Figure \ref{fig:Realpha} shows the solution for the real (left) and imaginary (right) components of the correction to the shock acceleration that satisfies global energy conservation behind the shock (Equation \ref{alphaen}) and the appropriate boundary condition on the fluid variables at the origin. These solutions are plotted as a function of $\sigma/i$ with $\sigma$ a purely imaginary number. For large $\sigma$ and for $n < 2$, the solution behaves approximately as $\tilde{\alpha} \propto \sigma e^{b\sigma}$, where $b$ is a complex number that depends on $n$. For $n = 2$, the real part of the solution levels to a constant value ($\simeq -0.2)$ and has no oscillatory behavior, while the imaginary part varies linearly with $\sigma$ and similarly displays no oscillation. 

Figure \ref{fig:funsofn} gives the real part of the solutions for the functions that satisfy the Fourier-transformed set of equations and energy conservation (and the appropriate boundary condition at the origin), where the value of $n$ is shown in the top-left of each panel, and the left, middle, and right columns illustrate the perturbation to the velocity, density, and pressure, respectively. Each curve in these panels is appropriate to the Fourier frequency shown in the legend, and the legend appropriate to each row is given in the left-most panel of the corresponding row. The brown, dashed curves in the top three rows illustrate the ``eigenfunctions'' that satisfy the eigenvalue equations, being Equations \eqref{L1} -- \eqref{L3} but with $\tilde{\alpha}_1 = 1$ and homogeneous boundary conditions for the perturbed functions, at the lowest-order, discrete eigenvalue; this eigenvalue is shown in the legend in the left-most panel of each row. We do not find any discrete eigenvalues when $n = 2$. The imaginary components of these functions exhibit similar qualitative trends to the real parts shown here.

For $n = 0$, 1, and 1.5, the velocity perturbation approaches zero near the origin as required by the total internal reflection of inward-propagating waves, and as $|\sigma|$ increases, each solution shows qualitatively more oscillatory behavior. The eigenfunction for the velocity is also well behaved and also satisfies the correct boundary condition at the origin. The perturbations to the density and the pressure are also regular everywhere for all $\sigma$. However, the eigenfunctions for the density and the pressure exhibit highly erratic behavior, and the density diverges extremely rapidly near the origin. 

For $n = 2$, the perturbation to the pressure approaches zero near the origin, which ensures that there are no outward-propagating waves that emanate from the center of the blast. In this case, neither the velocity nor the density remains regular near the origin, and each diverges and oscillates increasingly rapidly as we approach $\xi = 0$. However, this behavior does not invalidate the solutions: the appropriate boundary conditions at the shock front are satisfied, as is global energy conservation (note that the integrand in the second term in Equation \ref{alphaen} is convergent because $g_0 \propto f_0 \propto \xi$). The divergence of these quantities near the origin means that nonlinear terms will become important in the inner region of the shocked fluid, and that linear waves launched from the shock -- following the encounter with any density perturbations in the ambient medium -- will eventually steepen nonlinearly as they continue to propagate (which they will do indefinitely in the linear regime). As we noted above, precisely the same behavior is exhibited by sound waves propagating down a steep density gradient in a hydrostatic atmosphere, and that does not invalidate the linear theory while the waves remain subsonic. See the right panel of Figure \ref{fig:vel_n0_n2} for a specific demonstration of this steepening and reverse-shock formation when nonlinear effects are included.

\subsubsection{Physical interpretation and asymptotic scalings}
To understand the physical meaning of the Fourier-transformed shock acceleration shown in Figure \ref{fig:Realpha}, recall the general expression for the inverse transform:

\begin{equation}
\alpha_1(\tau) = \frac{1}{2\pi}\int_{-\infty}^{\infty}\tilde{\alpha}_1(\sigma)\,\delta\tilde{\rho}(\sigma)\,e^{i\sigma\tau}d\sigma \label{invtrans}
\end{equation}
and consider the case where $\delta \rho(\tau) \propto e^{i\sigma_0\tau}$ for some $\sigma_0$, meaning that the perturbation to the density is oscillatory in $\ln R$ for all $R\in(0,\infty)$. Since the Fourier transform of $e^{i\sigma_0\tau}$ is $\propto \delta_{\rm D}(\sigma-\sigma_0)$, where $\delta_{\rm D}$ is the Dirac delta function, the response of the shock acceleration to this density perturbation is to oscillate at the frequency $\sigma_0$ with the coefficient given by $\tilde{\alpha}_1(\sigma_0)$. Further, if $\sigma_0 = 0$, which corresponds to a uniform offset in the density, then the fact that $\tilde{\alpha}_1(0) = 0$ shows that $\alpha_1(\tau) \equiv 0$. This is precisely what we expect, as a uniform offset in the density everywhere simply corresponds to a change in the normalization of the ambient density. As a consequence, the solution is the same Sedov-Taylor solution in the absence of any perturbation but suitably renormalized, and hence the perturbation to the acceleration must be -- and is -- zero. The functions in Figure \ref{fig:funsofn} then represent the post-shock variation of the fluid variables in the presence of a sinusoidally varying density perturbation in $\ln R$. 

Now consider the other limit where $\delta \rho(\tau)$ is localized in space, and let the density perturbation be a square ``bump'' that jumps discontinuously from zero to a finite value over the range $-\Delta\tau$ to $\Delta\tau$, the Fourier transform of which is

\begin{equation}
\delta\tilde{\rho} \propto \frac{\sin\left(\sigma\Delta\tau\right)}{\sigma}.
\end{equation}
Investigating Figure \ref{fig:Realpha}, we see that the imaginary component of every $\tilde{\alpha}_1$ (i.e., independent of the choice of $n$) varies as $\propto -\sigma$ for $|\sigma| \gg 1$. Therefore, a contribution to the limiting behavior of the integrand in Equation \eqref{invtrans} will be $\propto \sin(\sigma\Delta\tau)e^{i\sigma\tau}$, and hence the acceleration of the shock is given by a $\delta$-function at the locations $\pm \Delta \tau$. This therefore demonstrates that the velocity changes discontinuously upon encountering a discontinuous change in the density of the ambient medium, and \emph{decelerates} upon encountering the positive change in the density and \emph{accelerates} when the density declines. 

\begin{figure}[htbp] 
   \centering
   \includegraphics[width=0.495\textwidth]{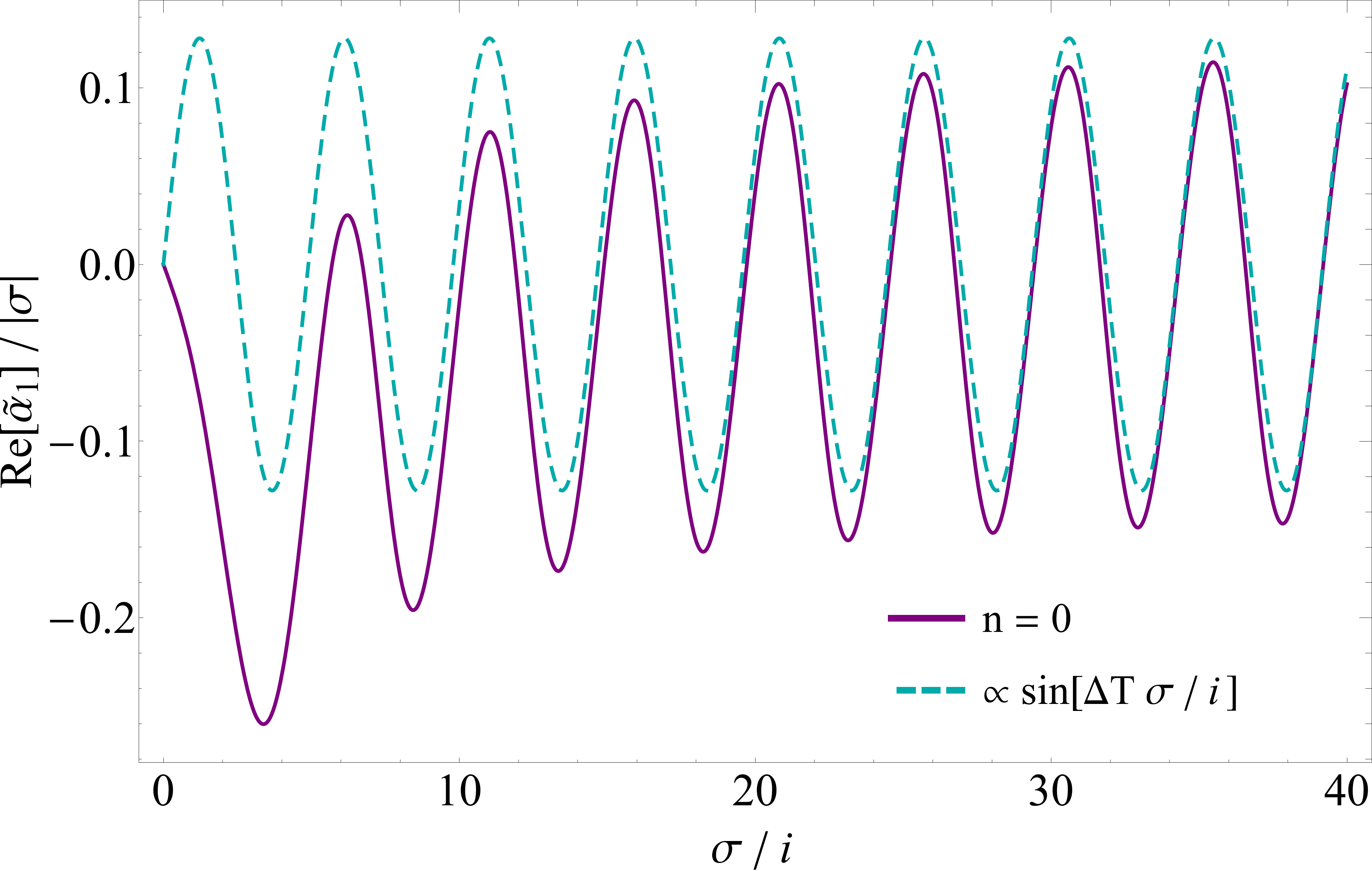} 
   \caption{The real part of the Fourier transform of the perturbation to the shock acceleration (purple curve) as a function of the Fourier frequency $\sigma / i$. The blue, dashed curve highlights the sinusoidal dependence of this function at large $|\sigma|$, and the fact that the period of oscillation is given by $\Delta T \simeq 1.283$, which is the dimensionless time taken for a discontinuity in the fluid properties to traverse the post-shock gas. This periodicity implies that shock should exhibit variation in its (e.g.) velocity not just upon encountering a discontinuous ``bump'' in the density profile of the ambient medium, but also after the shock expands by a factor of $e^{\Delta T}$.} 
   \label{fig:realpha53}
\end{figure}

In addition to declining linearly with $\sigma$, both the real and imaginary components of the Fourier-transformed acceleration for $n = 0$, 1 and 1.5 exhibit an oscillatory dependence that also grows in amplitude in a way that is proportional to $|\sigma|$. The period of the oscillation is also very well matched by the dimensionless timescale $\Delta T$, defined as

\begin{equation}
\Delta T = \int_1^{0}\frac{d\xi}{\lambda_{-}}+\int_0^{1}\frac{d\xi}{\lambda_{+}},
\end{equation}
where $\lambda_{-}$ and $\lambda_{+}$ are the backward and forward-propagating characteristics defined in Equation \eqref{lambdas}. Figure \ref{fig:realpha53} illustrates this agreement for the specific case of $n = 0$, for which $\Delta T \simeq 1.283$, where the solid, purple curve gives $Re[\tilde{\alpha}_1]/|\sigma|$ and the blue, dashed curve is proportional to $\sin\left(\Delta T\sigma / i\right)$. This additional, sinusoidal dependence -- that also increases in magnitude proportionally to $|\sigma|$ -- implies that the shock acceleration exhibits a delta-function-like response not only immediately upon encountering the discontinuous change in the density, but also at a time \emph{delayed by $\Delta T$}. 

We can understand this behavior by noting that the solution to the Riemann problem implies that the discontinuity generated within the fluid as the shock encounters a discontinuous density jump travels at the speed delimited by the characteristics of the flow. The discontinuity therefore travels from the shock to the origin and back to the shock in a time of $\Delta T$, and the velocity of the shock abruptly changes as it is hit by the discontinuity \emph{from behind}. For $n = 0$, 1, and 1.5, we find numerically that $\Delta T = 1.283$, 2.236, and 3.608, while $\Delta T \rightarrow \infty$ for $n = 2$; the fact that $\Delta T \rightarrow \infty$ for $n = 2$, which represents the causal disconnectedness between the shock and the origin for this solution, implies that the discontinuity in the fluid properties cannot reflect off of the origin and return to the shock in a finite time, which is consistent with the fact that the solution for $\tilde{\alpha}_1$ does not display any additional, sinusoidal dependence (see the green curve in the left and right panels of Figure \ref{fig:Realpha}). The values of $\Delta T$ for different $n$ and $\gamma = 4/3$ and $5/3$ are shown in Table \ref{tab:DeltaT}. 

\begin{table}
\begin{center}
\begin{tabular}{|c|c|c|c|c|}
\hline
  $\Delta T$  & $n = 0$ & $n = 1$ & $n = 3/2$ & $n = 2$ \\
  \hline
   $ \gamma = 4/3 $ & $\Delta T = 0.886$ & 1.43 & 2.07 & 3.77 \\
    \hline
     $ \gamma = 5/3 $ & $ 1.283$ & $2.236$ & $3.608$ & $\infty$ \\
     \hline
\end{tabular}
       \caption{The dimensionless time taken for a discontinuity in the ambient medium to propagate through the post-shock fluid, reflect off the origin, and return to the shock front. The shock position expands by a factor of $e^{\Delta T}$ in this time. }
           \label{tab:DeltaT}
\end{center}
\end{table}

This result demonstrates that Sedov-Taylor blastwaves that are completely causally connected should not only exhibit variations in velocity immediately upon encountering discontinuous (or at least rapidly varying over the scale of the shock position) density variations in the ambient medium, but also in temporal increments of the fundamental period, or sound crossing time, of the post-shock fluid. Furthermore, this periodicity varies \emph{logarithmically} with the position of the shock, and hence if the shock encounters a discontinuous change in the ambient density at a position of $R_0$, then we expect additional variations to occur (from the propagation and reflection of the discontinuity within the post-shock flow) when the shock expands in radius by a factor $\Delta R = R_0\times e^{\Delta T}$. Despite the fact that we expect this periodicity, there is no evidence of this from the lowest-order, discrete eigenvalue, which has an imaginary part (which controls the frequency of oscillation) of $\sim 7.05$ for $n = 0$ (i.e., not equal to the expected value of $\sim 1.283$). 

In addition to $\Delta T$, we also expect periodicities in integer increments of this fundamental period (i.e., $2\Delta T$, $3\Delta T$, etc.), as the discontinuity will reflect off of the shock and again traverse the post-shock fluid in a time of $\Delta T$. However, because of the continued expansion of the gas and the velocity gradient in the interior of the blastwave, the magnitude of the perturbation to the shock will lessen over time. This feature is qualitatively evident from the fact that in Figure \ref{fig:realpha53}, the cyclic variation at large $|\sigma|$ can be very well-approximated by a single sine function with frequency $\Delta T$, implying that the contribution from the higher-order frequencies is small. We also see that the amplitude of the sinusoidal variation of the $n = 1$ and $n = 1.5$, $\tilde{\alpha}_1$ curves is reduced compared to that of $n = 0$, which reflects the fact that the sound crossing time increases as $n$ increases, and hence any discontinuity is effectively reduced in amplitude more drastically over the longer duration over which it traverses the post-shock flow.

\subsection{Examples}
\label{sec:examples}
Here we consider specific density perturbations within the ambient medium and calculate the corresponding corrections to the shock propagation, and we make comparisons to numerical simulations. Using the properties of $\tilde{\alpha}_1$ with respect to the transformation $\sigma \rightarrow -\sigma$ and the assumed realness of $\delta\rho(\tau)$, 
the inverse Fourier transform of $\tilde{\alpha}_1$ becomes

\begin{equation}
\alpha_1(\tau) = \frac{1}{\pi}\int_0^{\infty}\left\{\tilde{\alpha}_{1,\rm r}(\sigma)\left(\delta\tilde{\rho}_{\rm r}(\sigma)\cos(\sigma\tau)-\delta\tilde{\rho}_{\rm i}(\sigma)\sin\left(\sigma\tau\right)\right)-\tilde{\alpha}_{1,\rm i}(\sigma)\left(\delta\tilde{\rho}_{\rm i}(\sigma)\cos\left(\sigma\tau\right)+\delta\tilde{\rho}_{\rm r}(\sigma)\sin\left(\sigma\tau\right)\right)\right\}d\sigma.
\end{equation}
Here $\tilde{\alpha}_{1, \rm r}$ ($\tilde{\alpha}_{1, \rm i}$) is the real (imaginary) component of $\tilde{\alpha}_1$, and $\delta\tilde{\rho}_{\rm r}$ ($\delta\tilde{\rho}_{\rm i}$) is the real (imaginary) component of the Fourier transform of $\delta \rho(\tau)$. Provided that $\delta\tilde{\rho}$ falls off sufficiently rapidly at large $\sigma$, we can approximate this integral by truncating the upper limit at a large but finite value. When $\delta\tilde{\rho}$ does not fall off more steeply than $\propto 1/\sigma$, the integrand in this expression will diverge for specific $\tau$, indicating that the perturbation to the shock acceleration is characterized by a $\delta$-function (see the discussion in the preceding subsection). In this case it is numerically convenient to work with the shock velocity, which is related to the integral of the shock acceleration parameter and defined via (cf.~Equation \ref{pertquants})

\begin{equation}
    \frac{\partial\ln V}{\partial \tau} = -\frac{1}{2}\left(3-n\right)+\alpha_1(t),
\end{equation}
and can be piecewise-continuous even when the acceleration diverges. Finally, it is useful to parameterize the deviation of the blastwave evolution from the Sedov-Taylor solution by defining the energy variable $\mathscr{E}$:

\begin{equation}
    \mathscr{E}_1 = \ln\left[V^2e^{\left(3-n\right)\tau}\right] \propto \ln\left[V^2R^{3-n}\right] = 2\int\alpha_1\,d\tau, \label{escript}
\end{equation}
which \emph{would be constant} (and scale with the total energy) in the absence of perturbations\footnote{Introducing the logarithm in the definition of the energy variable is convenient because it scales linearly with $\alpha_1$, and hence we can scale out the dependence on the magnitude of the density perturbation, $\delta\rho_0$.}. Note, however, that this quantity will not be constant in the presence of perturbations, even though the energy is still absolutely conserved, because of the time-dependent evolution of the post-shock fluid quantities (i.e., the total energy behind the blastwave, which is conserved and independent of time, has additional terms that are not contained in the above expression; see Equation \ref{enint}).

Similarly, the solution for the perturbation to the post-shock velocity profile is

\begin{equation}
f_1(\xi,\tau) = \frac{1}{\pi}\int_{0}^{\infty}\left\{\tilde{f}_{1,\rm r}(\xi,\sigma)\left(\delta\tilde{\rho}_{\rm r}(\sigma)\cos(\sigma\tau)-\delta\tilde{\rho}_{\rm i}(\sigma)\sin\left(\sigma\tau\right)\right)-\tilde{f}_{1,\rm i}(\xi,\sigma)\left(\delta\tilde{\rho}_{\rm i}(\sigma)\cos\left(\sigma\tau\right)+\delta\tilde{\rho}_{\rm r}(\sigma)\sin\left(\sigma\tau\right)\right)\right\}d\sigma,
\end{equation}
and analogous expressions hold for the corrections to the density and pressure of the post-shock fluid. Here $\tilde{f}_{1,\rm r}(\xi,\sigma)$ and $\tilde{f}_{1, \rm i}(\xi,\sigma)$ are the Fourier-transformed perturbations to the velocity, evaluated at a given position within the post-shock flow. Thus, to evaluate these integrals, we first select a value of $\xi$ and then interpolate the solution for $\tilde{f}_1$ over $\sigma$, which we then integrate numerically over a large (but finite) range in $\sigma$. We use these expressions to calculate the properties of the shock and the post-shock fluid given the specific density perturbations considered in the ensuing subsections. 

For concreteness and to keep the number of plots to a relative minimum, we focus exclusively on the case where the post-shock adiabatic index is $\gamma = 5/3$, and we consider only the ambient power-law indices $n = 0$ and $n = 2$. Also to keep the number of plots manageable, we only analyze the post-shock fluid properties for the first example (a ``Gaussian bump''), and otherwise we only consider the evolution of the shock itself. In calculating the above integrals, in practice and for the plots shown below we truncate the upper limit at $\sigma=300$; we have checked that changing this upper limit by a modest factor does not change the results in any discernible way, except when the density perturbation is discontinuous, in which case the Gibbs phenomenon possesses a noticeable dependence on the upper limit (see the discussion in Section \ref{sec:Rectangular} below). We also set the inner boundary at which we evaluate the boundary condition on the perturbation to the velocity or pressure at $\xi = 0.001$; modest changes to this value do not affect our results.

\subsubsection{Gaussian bump}
\label{sec:Gaussian}
\begin{figure}[htbp] 
   \centering
   \includegraphics[width=0.325\textwidth]{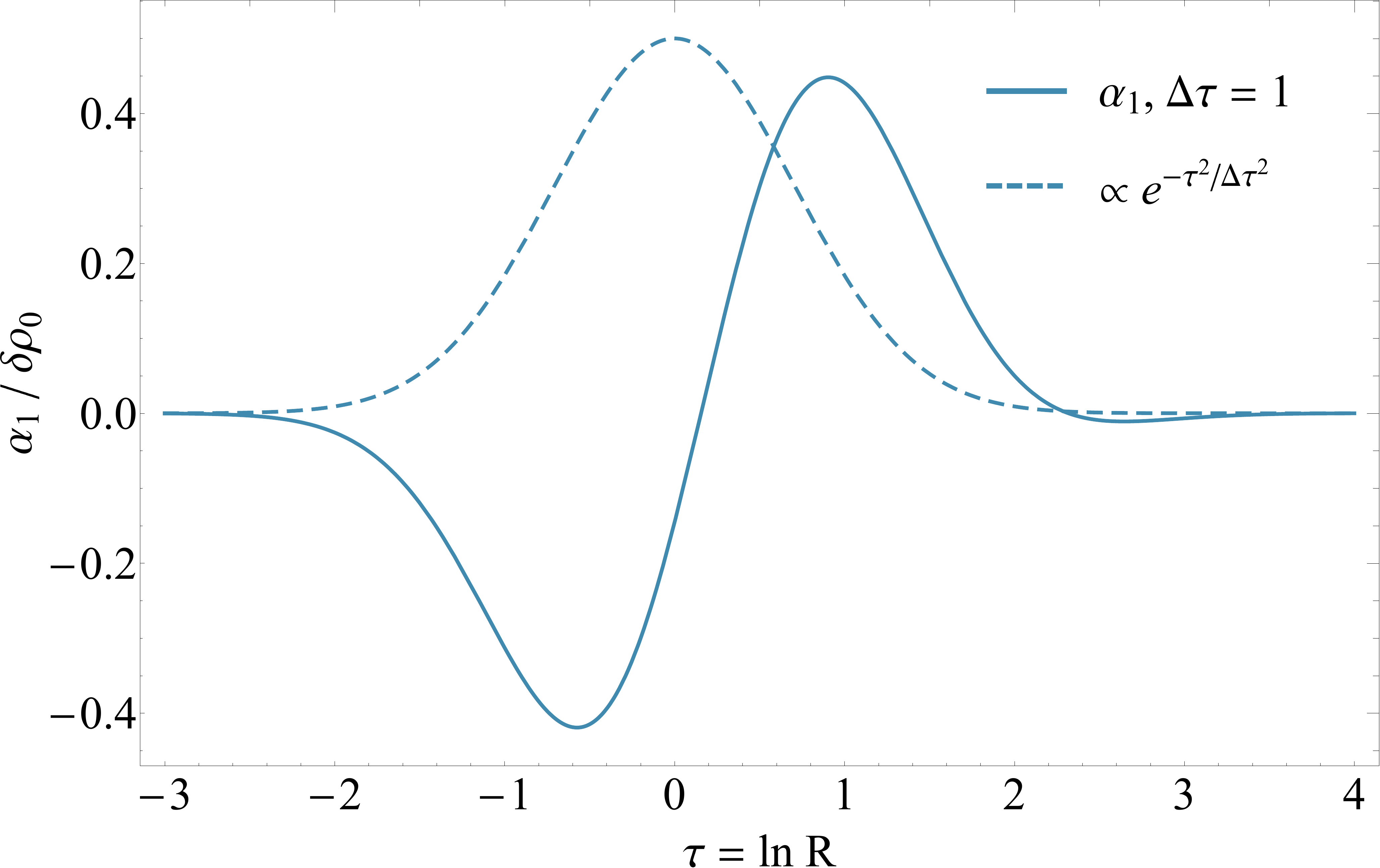} 
   \includegraphics[width=0.325\textwidth]{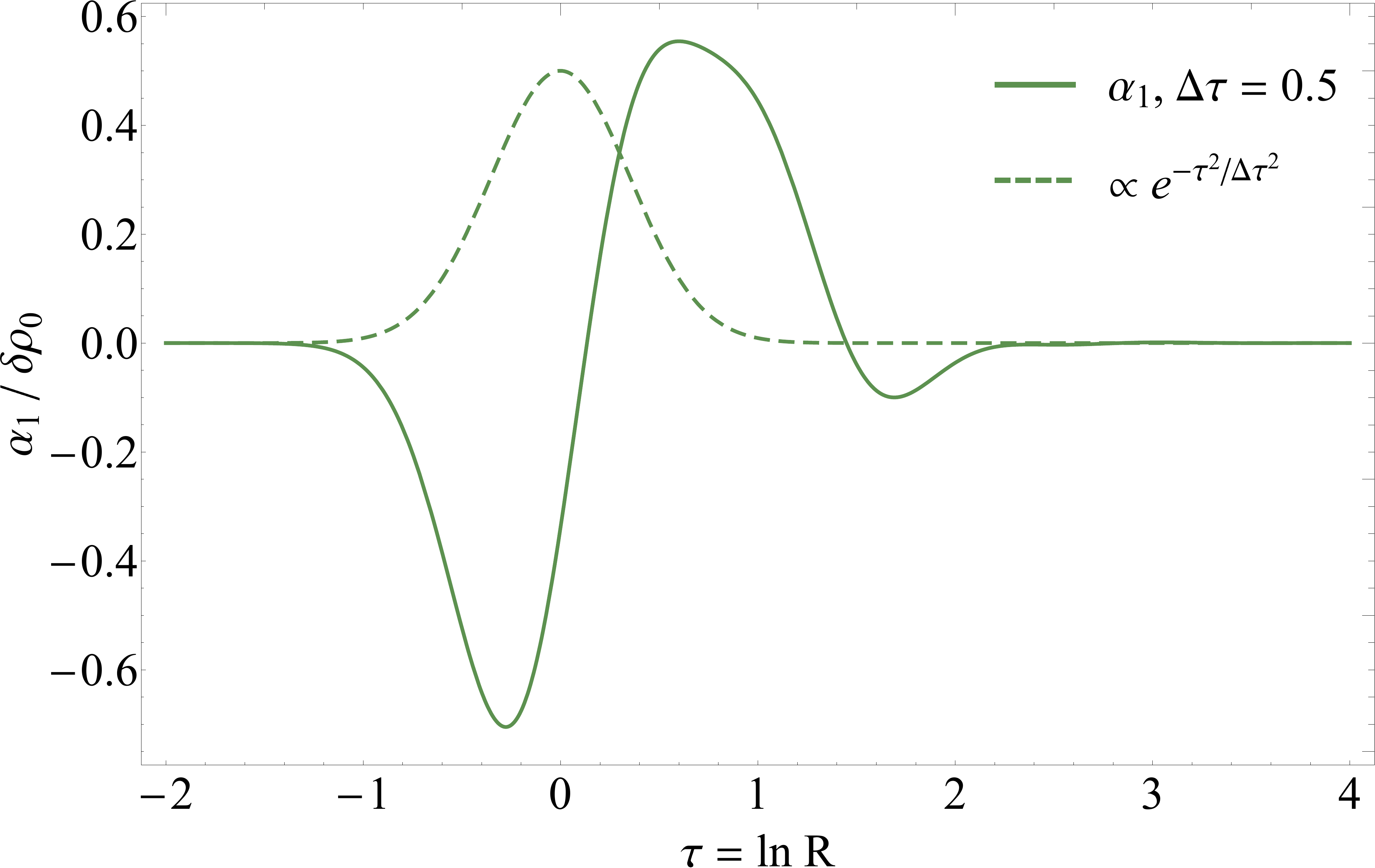} 
   \includegraphics[width=0.325\textwidth]{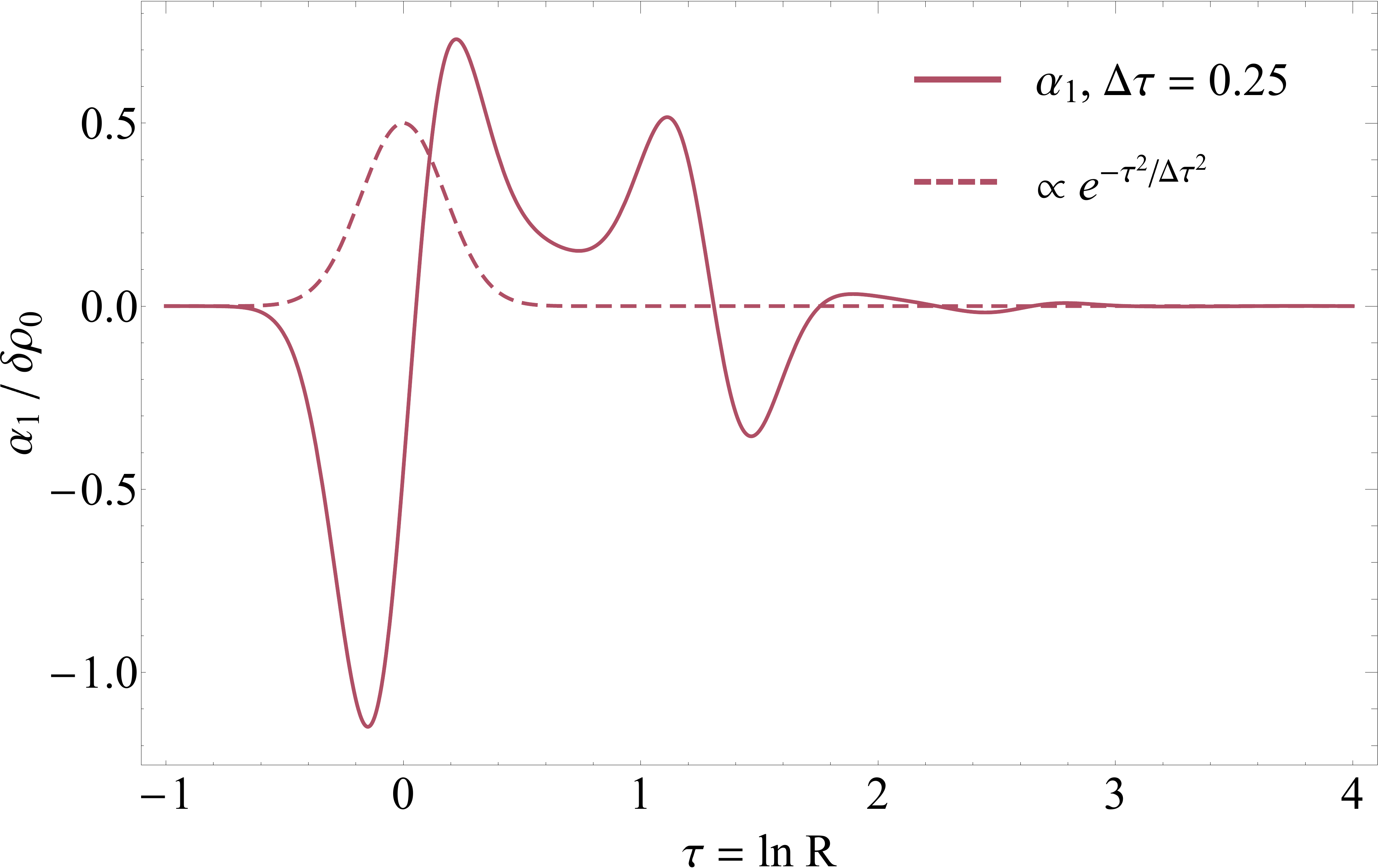} 
     \includegraphics[width=0.325\textwidth]{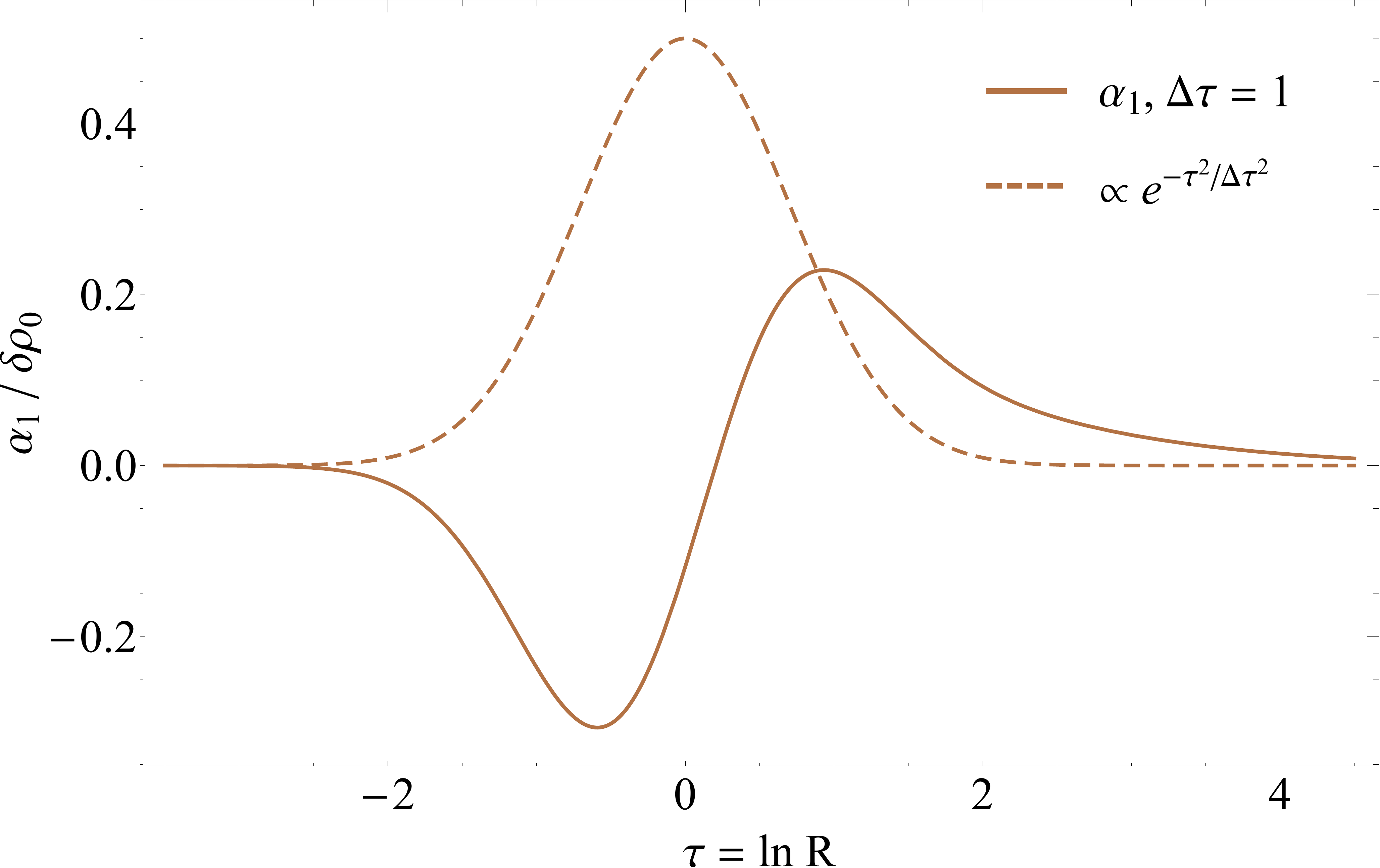} 
   \includegraphics[width=0.325\textwidth]{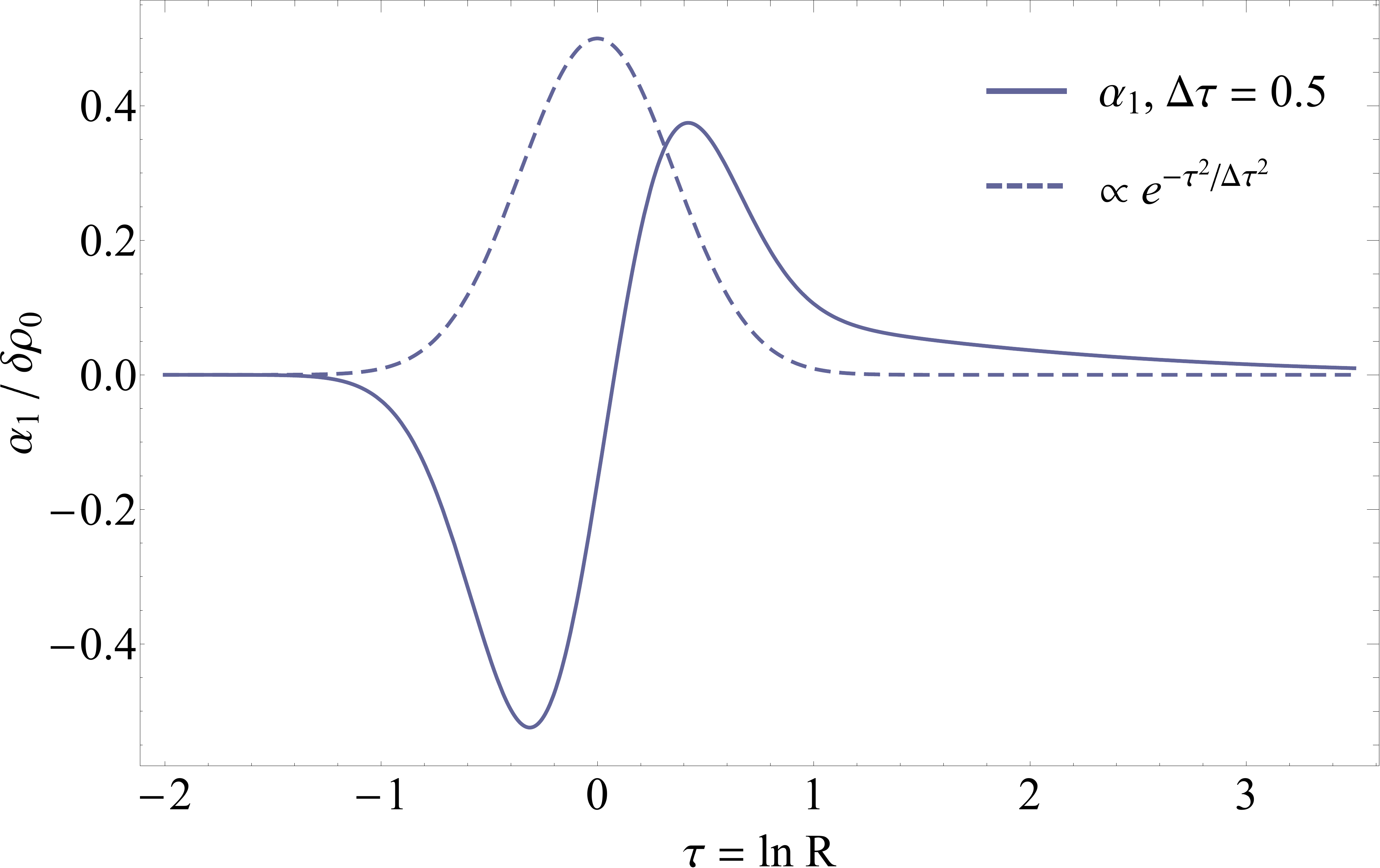} 
   \includegraphics[width=0.325\textwidth]{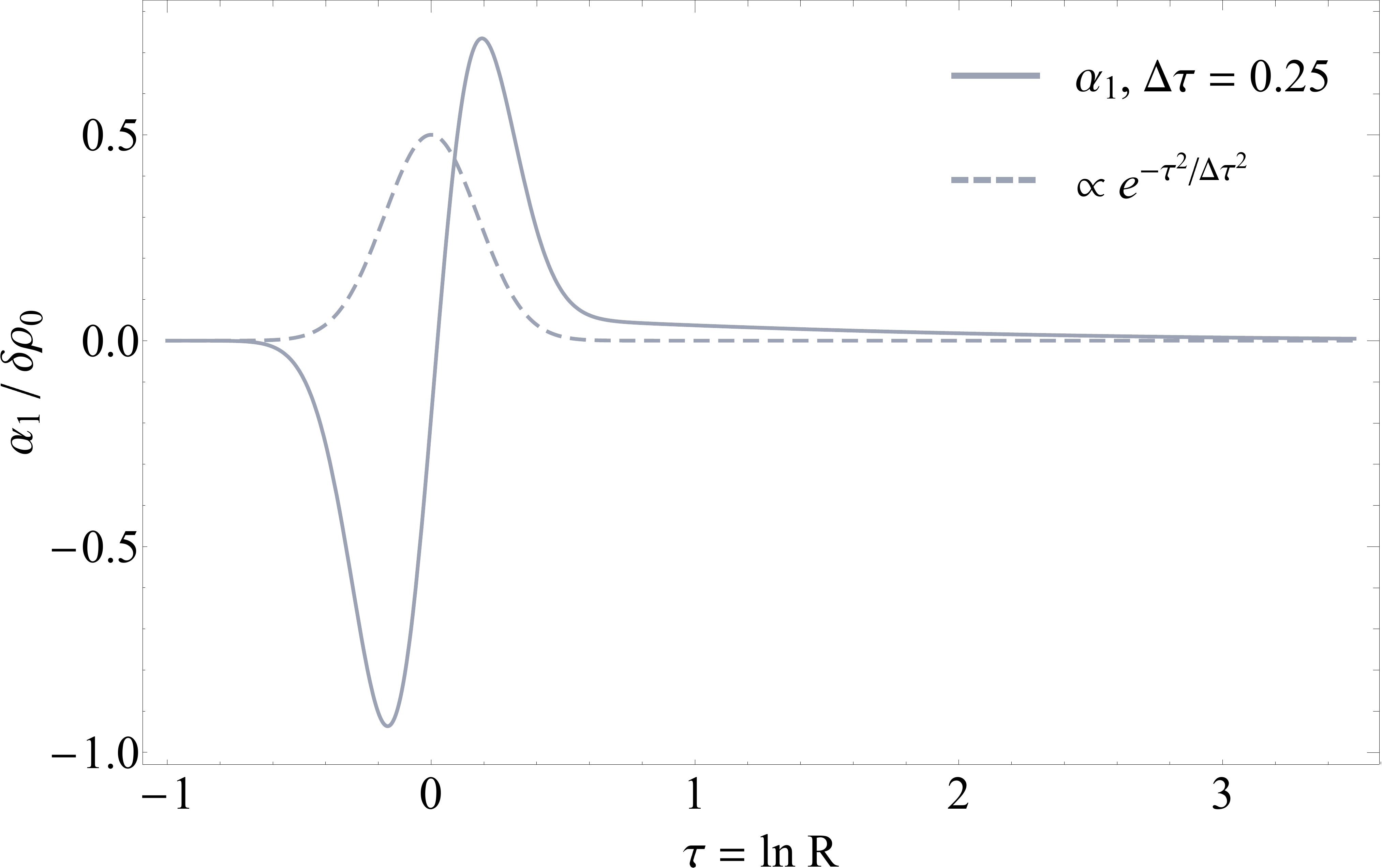}
   \caption{The correction to the shock acceleration, $\alpha_1$, normalized by the magnitude of the density perturbation, $\delta\rho_0$, induced by a Gaussian density ``bump'' in the ambient medium; in the top row the density of the ambient medium, $\rho_{\rm a}$, is homogeneous (except for the Gaussian bump), and hence with $\rho_{\rm a} \propto r^{-n}$, $n = 0$, while the bottom row is for a wind-like medium with $n = 2$. The width of the Gaussian, in units of the shock radius, is $\Delta \tau = 1$ (left column), 0.5 (middle column), and 0.25 (right column). The dashed curves in these panels show the qualitative behavior of the density perturbation.}
   \label{fig:alpha_Gaussian_n0}
\end{figure}

Consider the case where the ambient density perturbation is in the form of a Gaussian ``bump,''

\begin{equation}
    \delta\rho(\tau) = \delta\rho_0 e^{-\frac{\tau^2}{\Delta\tau^2}} \quad \Rightarrow \quad \delta\tilde{\rho} = \delta\rho_0\sqrt{\pi\Delta\tau}e^{-\frac{1}{4}\sigma^2\Delta\tau^2}.
\end{equation}
Here $\delta\rho_0$ is an overall normalization factor that can be set to unity within the linear theory and represents the magnitude of the density perturbation at $\tau = 0$, and $\Delta \tau$ is the width of the Gaussian (also note that, since $\tau = \ln\left(R/R_0\right)$, this is a log-normal distribution in terms of the shock position). Figure \ref{fig:alpha_Gaussian_n0} shows the acceleration parameter $\alpha_1$, where the top row is for $n = 0$ (constant-density ambient medium) and the bottom row is for $n = 2$ (wind-like ambient medium), and here we set $\delta\rho_0 = 1$ (alternatively, here $\alpha_1$ is normalized by $\rho_0$, and hence the perturbation to the acceleration is scaled by this parameter when the linear theory is accurate). We see that the shock decelerates as it encounters the increasing density profile, and then accelerates down the density gradient for $\tau > 0$. For the constant-density medium, the shock also oscillates after encountering the bump, with the oscillations becoming more erratic as the width of the Gaussian decreases. On the other hand, the acceleration parameter of the shock in the wind-fed medium does not exhibit any oscillatory behavior after encountering the bump, and instead appears to monotonically decline after encountering the density perturbation.

\begin{figure}
    \centering
    \includegraphics[width=0.49\textwidth]{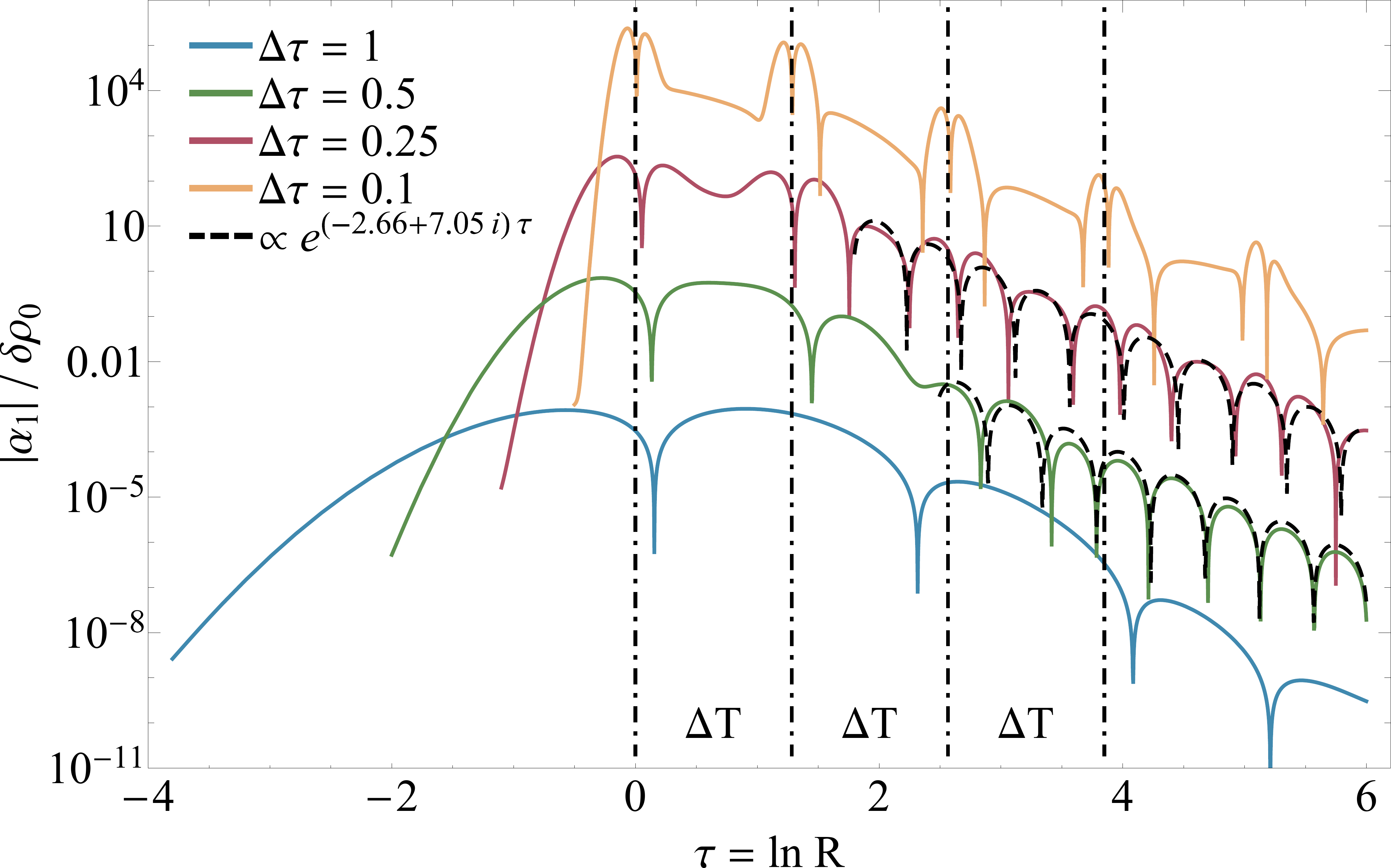}
     \includegraphics[width=0.49\textwidth]{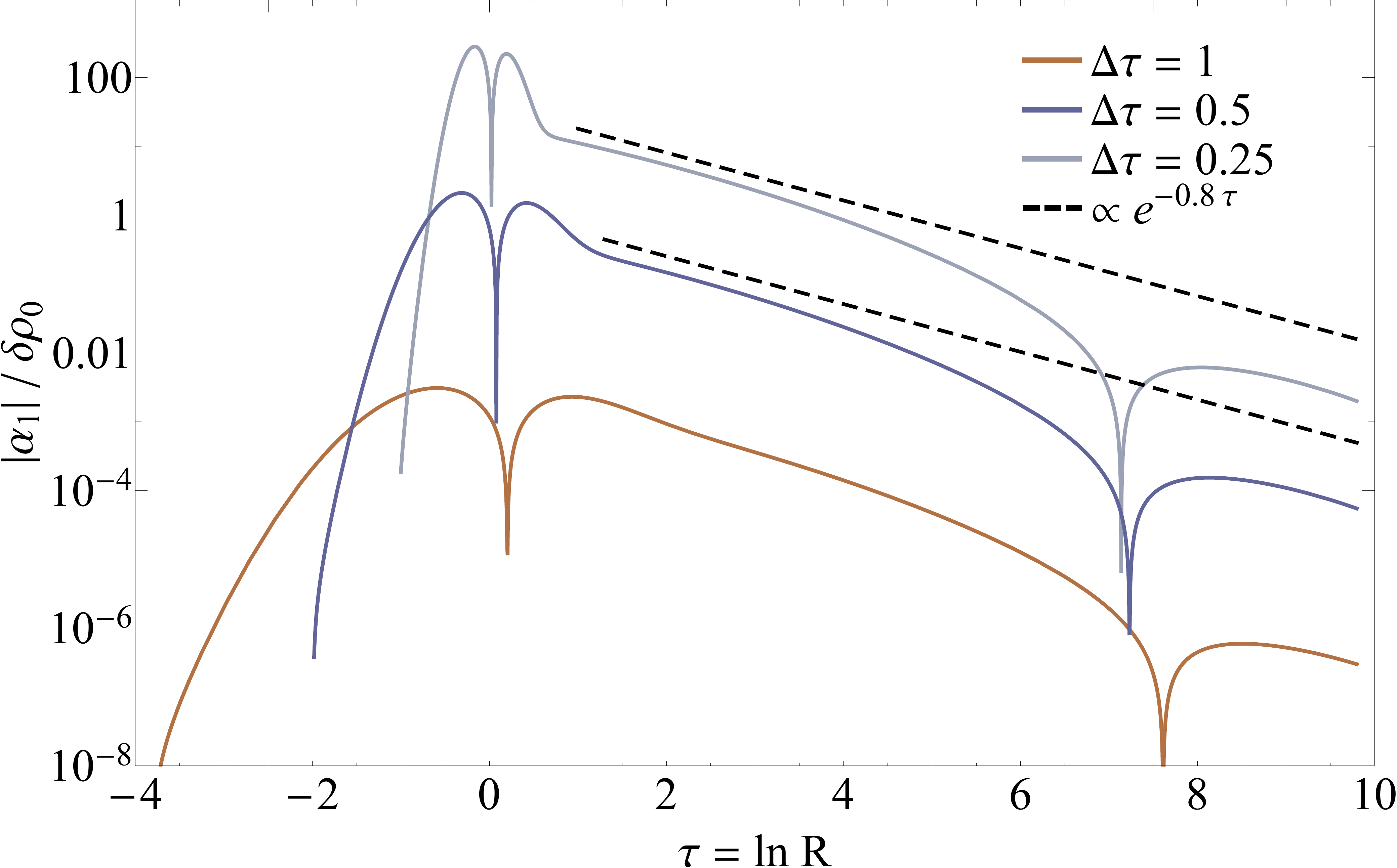}
    \caption{The absolute value of the perturbation to the shock acceleration, $|\alpha_1|$, normalized by the magnitude of the density perturbation $\delta\rho_0$, as a function of the log of the shock position as the shock encounters a Gaussian bump of width $\Delta \tau$, where the different curves are appropriate to the $\Delta \tau$ shown in the legends. Each curve is arbitrarily scaled by a constant factor so that they can be more easily distinguished from one another. The left panel is for a constant-density ambient medium with $n = 0$, while the right-hand panel is for a wind-like medium with $n = 2$. The dashed lines in the left panel give the expected behavior from the largest eigenvalue, which scales with time as shown in the legend, while those in the right panel show the decline expected from the fact that the poles for $n = 2$, $\gamma = 5/3$ lie approximately along a line in the complex plane. The vertical, dot-dashed lines in the left-hand panel show the sound-crossing time over the post-shock fluid, being $\Delta T \simeq 1.283$ for $n = 0$ and $\gamma =5/3$.}
    \label{fig:abs_alpha}
\end{figure}

Figure \ref{fig:abs_alpha} illustrates the absolute value of the perturbation to the shock acceleration for the same three Gaussian ``bumps'' as in Figure \ref{fig:alpha_Gaussian_n0}, but on a log-linear scale; the magnitude of the density perturbation is also scaled arbitrarily for the different solutions so that they can be more easily distinguished from one another, and the large, vertical dips are the location where the perturbation equals zero (i.e., the log of the acceleration goes approaches $-\infty$ at these locations, but the divergences are truncated for clarity). We see that the constant-density blastwave (left panel) oscillates periodically in $\tau$ after encountering the bump, and for $\Delta \tau = 0.5$ and $0.25$, the decay and oscillation frequency are well-matched by the largest eigenvalue for these parameters, as shown by the dashed curves, but only after the shock has expanded by a modest amount ($\tau \simeq 1$ for $\Delta\tau =0.25$, and $\tau \simeq 2$ for $\Delta\tau = 0.5$). For $\Delta\tau = 1$, on the other hand, the oscillation frequency and decline of the perturbation to the shock acceleration do not match those of the eigenvalue even after 6 $e$-foldings of the shock position, with the decline being steeper and the oscillation timescale longer. For the wind-like medium (right panel), the perturbation to the acceleration appears to decline at a power-law rate that is roughly matched by the rate predicted by the continuum of eigenvalues (dashed lines). Eventually, however, the solution starts to exhibit oscillations, and the oscillations appears between roughly 7 and 8 $e$-foldings of the shock position after encountering the initial bump. 

Also shown in the left-hand panel of Figure \ref{fig:alpha_Gaussian_n0} is the solution for the shock acceleration when the width of the Gaussian bump is $\Delta \tau = 0.1$. In this case, the bump is highly confined in space and the width is much smaller than the radius of the shock (i.e., $\Delta \tau \ll \tau$), and the shock acceleration does not follow the oscillatory behavior predicted by the eigenvalue with the largest real part. Instead, we see that the correction to the shock acceleration shows sudden changes when the shock expands by factors of $\Delta T$, which are shown by the vertical, dashed lines in this figure. Thus, in this limit the density perturbation behaves more like a discontinuity, and its presence is communicated back through the post-shock fluid at the sound crossing time. The decay of the perturbations also follows roughly the same power-law decline as the eigenmode.

We thus see that the degree to which the decline and oscillation frequency of the acceleration matches the most-positive eigenvalue (i.e., the one with the largest real part, which in this case is $\sigma \simeq -2.66\pm 7.05i$) depends on the nature of the perturbation to the ambient density, which is in agreement with the suggestion by \citet{sanz16}. However, this result is not surprising: the perturbation to the acceleration is a \emph{sum} over the eigenvalues, where in this case the sum is both discrete and continuous owing to the isolated and continuous spectrum of poles (see Figure \ref{fig:poles}). When the perturbation is highly localized in space, we need to include many higher-order modes to accurately reconstruct the response of the blastwave to the perturbation, meaning that it takes a long time (in $\tau = \ln R$) for the most positive eigenvalue to dominate the evolution. Instead, the oscillations occur on the sound-crossing time of the post-shock fluid, though the decay rate is still comparable to that of the most-positive eigenvalue because all of the discrete poles have nearly the same real part (again, see Figure \ref{fig:poles}). On the other hand, when the variation in the density perturbation is comparable to the size of the shock itself, the low-frequency terms (i.e., those that probe the largest length scales) in the eigenvalue decomposition are most dominant. Thus, in this limit the coefficient multiplying the most-positive eigenvalue is much smaller than those at the low-frequency end of the expansion, and those with longer oscillation timescales and faster decay timescales (again, see Figure \ref{fig:poles}) dominate the solution; the perturbation to the shock acceleration therefore decays faster and is less oscillatory when the ambient density perturbation is spread over scales that are comparable to the shock radius. Nevertheless, since there is a most-positive eigenvalue, at sufficiently late times the shock will exhibit the decay and oscillation frequency of this eigenvalue. However, since the variation occurs as the \emph{log} of the shock position, the time at which the eigenvalue takes over can be many $e$-foldings of the shock position.

\begin{figure}
    \centering
    \includegraphics[width=0.49\textwidth]{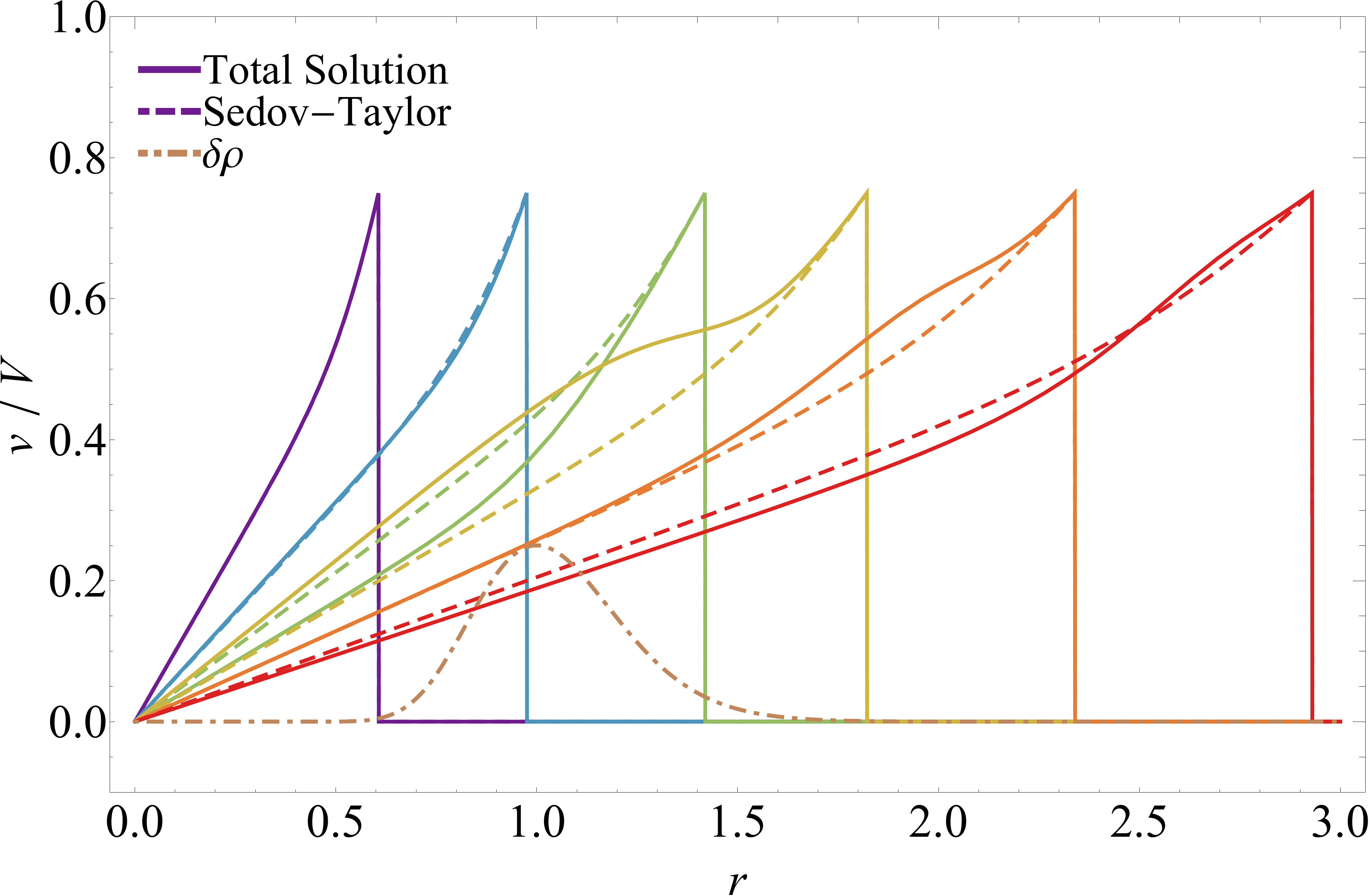}
    \includegraphics[width=0.49\textwidth]{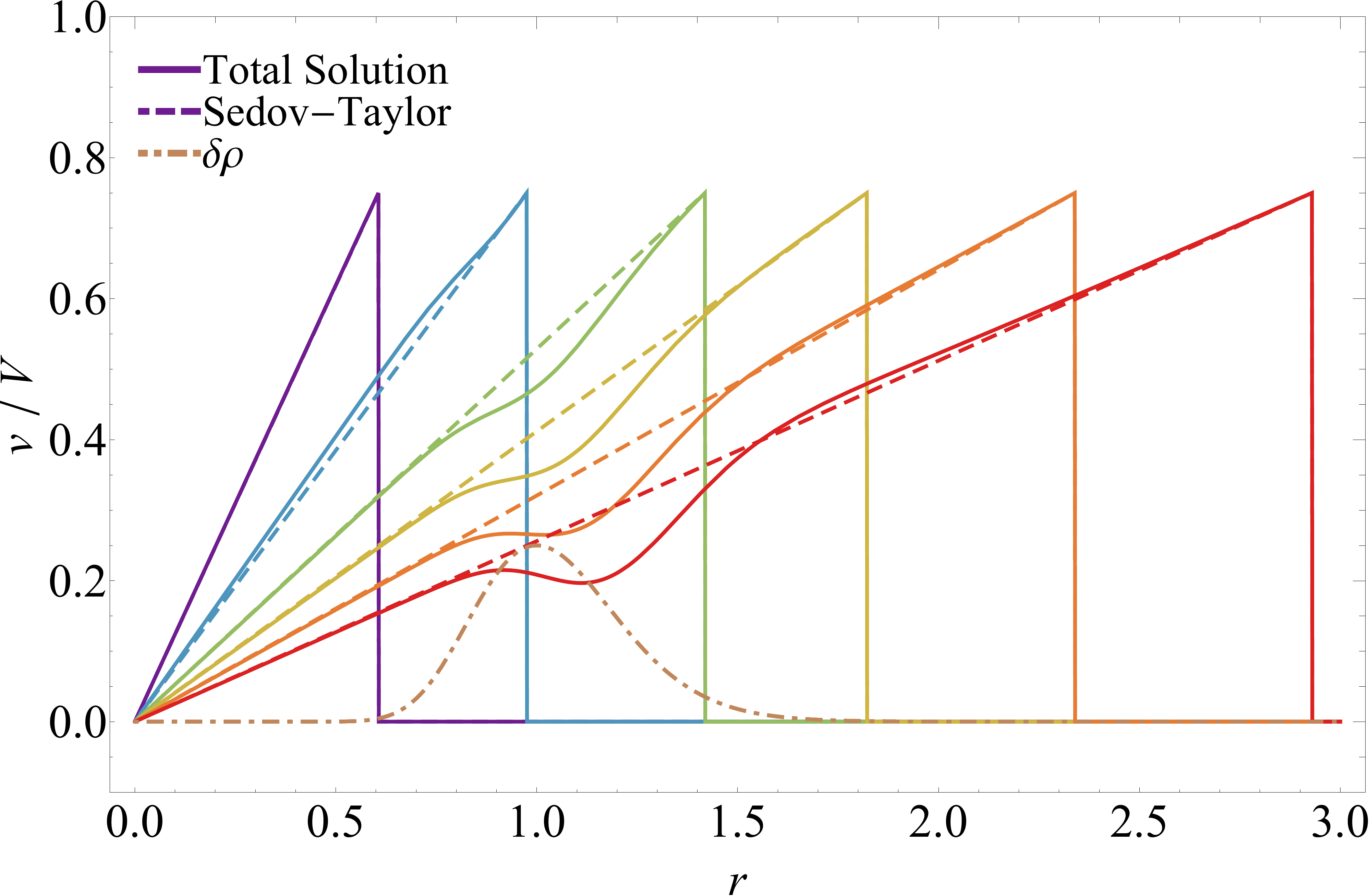}
    \caption{The fluid velocity, $v$, normalized to the shock velocity, $V$, as a function of spherical radius, $r$, for a constant-density ambient medium ($n = 0$, left panel) and a wind-like ambient medium ($n = 2$, right panel) as a shock encounters a Gaussian bump of width $\Delta\tau = 0.25$ and magnitude $\delta\rho_0 = 0.25$; the dot-dashed, brown curve illustrates the perturbation to the density. The solid curves show the total solution for the velocity, while the dashed curves illustrate the Sedov-Taylor solution (i.e., without the correction to the velocity profile) for reference.}
    \label{fig:vel_profile}
\end{figure}

Figure \ref{fig:vel_profile} shows the solution for the post-shock velocity normalized by the shock velocity as the shock encounters a Gaussian bump of width $\Delta \tau = 0.25$, which is shown by the brown, dot-dashed curve for reference; the amplitude of the perturbation was set to $\delta \rho_0 = 0.25$ for concreteness. The left-hand panel shows the solution for a constant-density ambient medium ($n = 0$), while the right-hand panel gives the solution for a wind-fed medium ($n = 2$). Different curves represent different times in the shock evolution, the shock itself shown by the discontinuity in the velocity profile. For the constant-density solution, the encounter with the bump results in a perturbation to the post-shock velocity that oscillates in time and position. Over time, the amplitude of the perturbation to the post-shock velocity weakens, and the velocity eventually returns to the unperturbed (Sedov-Taylor) solution. 

\begin{figure}
    \centering
    \includegraphics[width=0.495\textwidth]{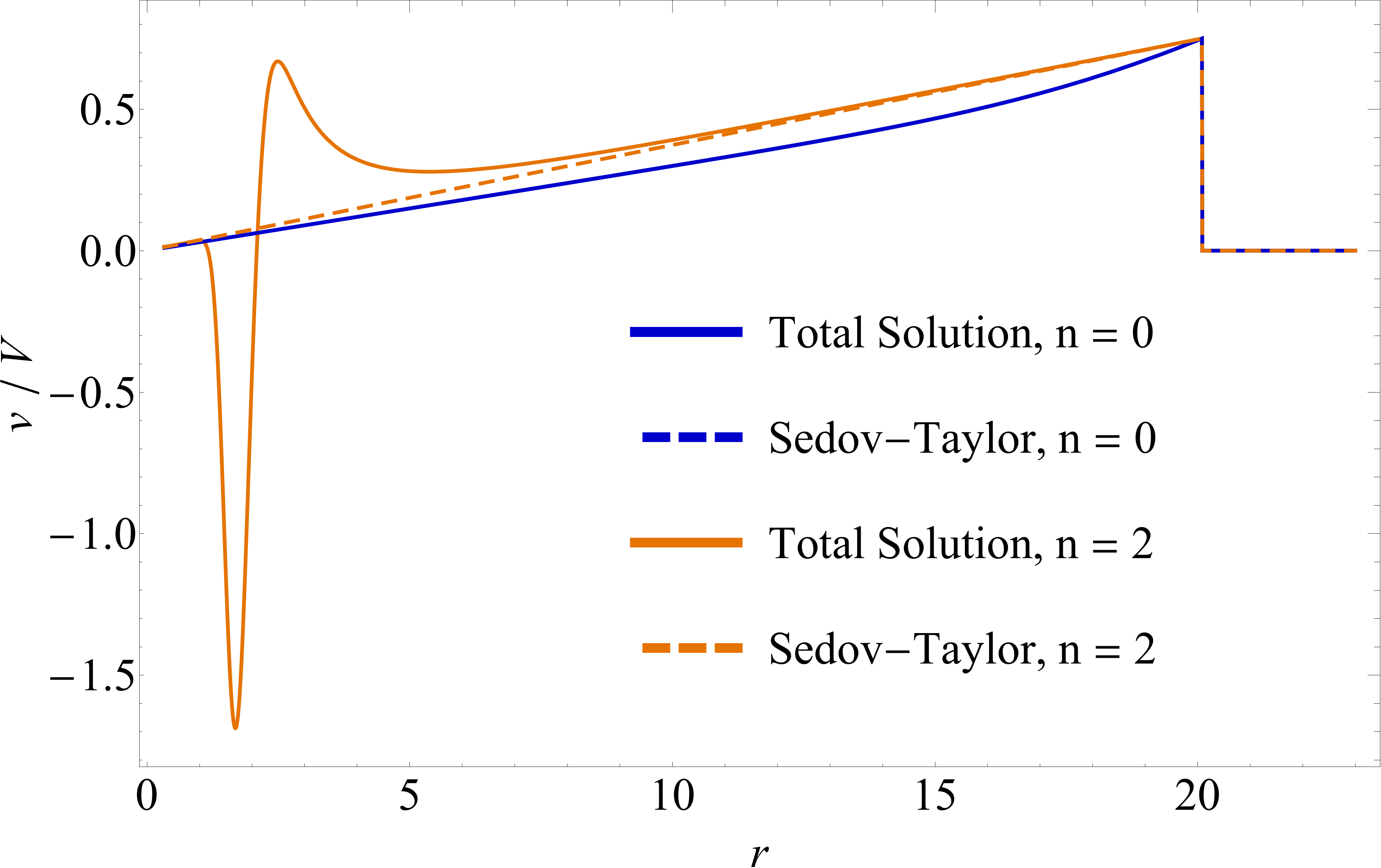}
     \includegraphics[width=0.495\textwidth]{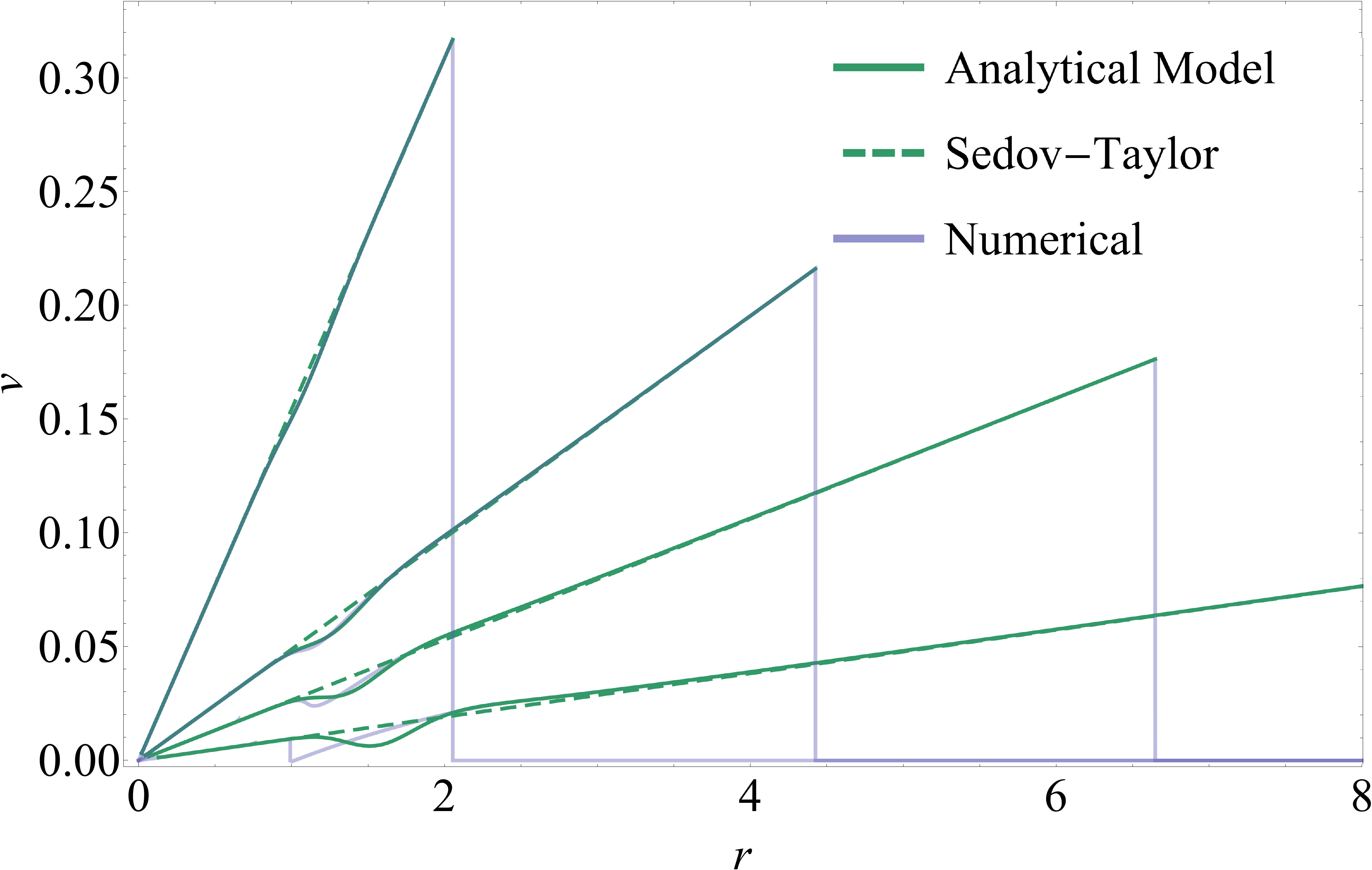}
    \caption{Left: The fluid velocity normalized by the shock velocity for the same, Gaussian perturbation as in Figure \ref{fig:vel_profile} for $n = 0$ and $n = 2$ (solid curves) alongside the Sedov-Taylor solutions (dashed curves) when the shock has expanded to $e^{3} \simeq 20$. While the perturbations to the post-shock flow for a constant-density medium ($n = 0$) have decayed, making the solution with the perturbation indistinguishable from the Sedov-Taylor (self-similar) solution, the amplitude of the velocity perturbation for the wind-fed medium has increased substantially. This increase arises from the declining sound speed in the interior of the blastwave, and would nonlinearly steepen into a reverse shock. The $n = 2$, Sedov-Taylor blastwave is unstable from the standpoint that any small perturbations arising from the ambient medium generate reverse shocks that travel into the interior of the blast. Right: The velocity as a function of radius for a Gaussian bump of magnitude $\delta \rho = 0.04$ for $n = 2$ and $\gamma = 5/3$ at four different times. The light-blue curves give the numerical solution from a one-dimensional hydrodynamics code, the solid, green curves are from the analytical, perturbative solution, and the dashed, green curves are the Sedov-Taylor solution. We see that the perturbative and numerical solutions agree extremely well with one another until the wave starts to steepen substantially, after which point the numerical solution self-consistently results in the formation of a reverse shock.}
    \label{fig:vel_n0_n2}
\end{figure}

On the other hand, it can be seen from the right panel of this figure (i.e., when $n = 2$) that the amplitude of the perturbation to the post-shock velocity increases as the shock continues to move out in a wind-fed medium. The left panel of Figure \ref{fig:vel_n0_n2} compares the post-shock velocity profile for $n=0$ and $n=2$ when the shock has advanced to $e^{3} \simeq 20$ for the same density perturbation (solid curves) and the Sedov-Taylor solutions (dashed curves). For a constant-density medium, the solution that includes the perturbation is indistinguishable from the Sedov-Taylor blastwave, indicating the stability of the post-shock flow. For a wind-fed medium, it is apparent that the amplitude of the perturbation has increased, while the width has simultaneously decreased. Over time, and with the inclusion of nonlinear terms, this reverse wave would steepen into a reverse shock, and arises from the declining sound speed in the interior of the blastwave for this combination of $n$ and $\gamma$. Thus, even though the shock front is stable to such perturbations, meaning that the shock position and velocity will regain their Sedov-Taylor scaling after long enough, the post-shock flow is \emph{unstable} from the standpoint that any small perturbations will grow into shockwaves that propagate back into the flow. 

The perturbation analysis does not incorporate the nonlinearities that will eventually cause the reverse wave to steepen into a shock. To validate this notion -- and to demonstrate that capturing the nonlinearities does indeed lead to the formation of a shock -- the right panel of Figure \ref{fig:vel_n0_n2} shows the result from linear perturbation theory (solid, green curves) alongside a numerical simulation (solid, light-blue curves) for a Gaussian bump of magnitude $\delta \rho = 0.04$ in an $n = 2$ medium with $\gamma = 5/3$; the dashed curves in this panel show the Sedov-Taylor solution.

The numerical solutions are obtained using a second-order finite volume solution scheme for the compressible, energy-conserving Euler equations in spherical symmetry. The numerical solutions are generated on a mesh with $8 \times 10^5$ zones, evenly spaced in the radial coordinate. The innermost zone extends to the origin at $r=0$ (the inner radial face of the innermost zone has zero area, so no flux needs to be computed there). The geometrical source term $2p/r$ on the radial momentum $\rho v$ is volume-integrated to improve robustness in the vicinity of the origin. The outer domain boundary is zero-gradient, and placed far enough away that the shock front does not reach it in any of the snapshots shown in Figure \ref{fig:vel_n0_n2}. The scheme is second-order accurate in space, using a piecewise-linear reconstruction applied to the primitive hydrodynamic variables $\rho$, $p$, and $v$. It uses a second-order total variation diminishing explicit Runge-Kutta procedure to advance the solution in time.

From Figure \ref{fig:vel_n0_n2}, we see that the perturbed solution and the numerical solution agree extremely well with one another -- and are effectively indistinguishable -- at early times and when the wave has not yet substantially steepened. At later times (when the shock is at $r \simeq 6.6$) the two solutions agree well, but also exhibit noticeable differences in the region where the wave has started to steepen. At a sufficiently late time (bottom-most curve; the position of the shock is off of the figure by this time to maintain the clarity of the figure and to focus on the region of interest, but is at a position of $r \simeq 13$) the numerical solution self-consistently steepens into a reverse shock, while the perturbed solution remains smooth. Note, however, that away from the immediate vicinity of the reverse shock the three solutions agree well with one another; this feature demonstrates that the unstable region of the post-shock flow is disconnected from and does not influence the stability and propagation of the forward shock and the fluid in the immediate vicinity thereof.

\subsubsection{Rectangular bump}
\label{sec:Rectangular}
Let the perturbation to the ambient density take the form of a localized and discontinuous, ``rectangular bump,'' such that

\begin{equation}
    \delta\rho(\tau) = \begin{cases}
    \delta\rho_0 \textrm{ for }-\Delta\tau \le \tau \le \Delta \tau \\
    0 \quad \textrm{ otherwise }
    \end{cases}
    \quad \Rightarrow \quad \delta\tilde{\rho} = \frac{2}{\sigma}\sin\left(\sigma\Delta\tau\right).
\end{equation}
The discontinuity in the density implies that the acceleration will be characterized by a $\delta$-function at $-\Delta \tau$ and $\Delta \tau$, where $2\Delta\tau$ is the total distance over which the density is increased by $\delta\rho_0$. However, the shock velocity will be piecewise-continuous, and we can numerically calculate the integral of $\alpha_1$ to determine the change in the energy parameter $\mathscr{E}_1$, which measures the instantaneous deviation of the shock properties from the Sedov-Taylor solution. This parameter is, for reference, given in Equation \eqref{escript}, and is just directly proportional to the integral of the shock acceleration parameter.

\begin{figure}[htbp] 
   \centering
   \includegraphics[width=0.325\textwidth]{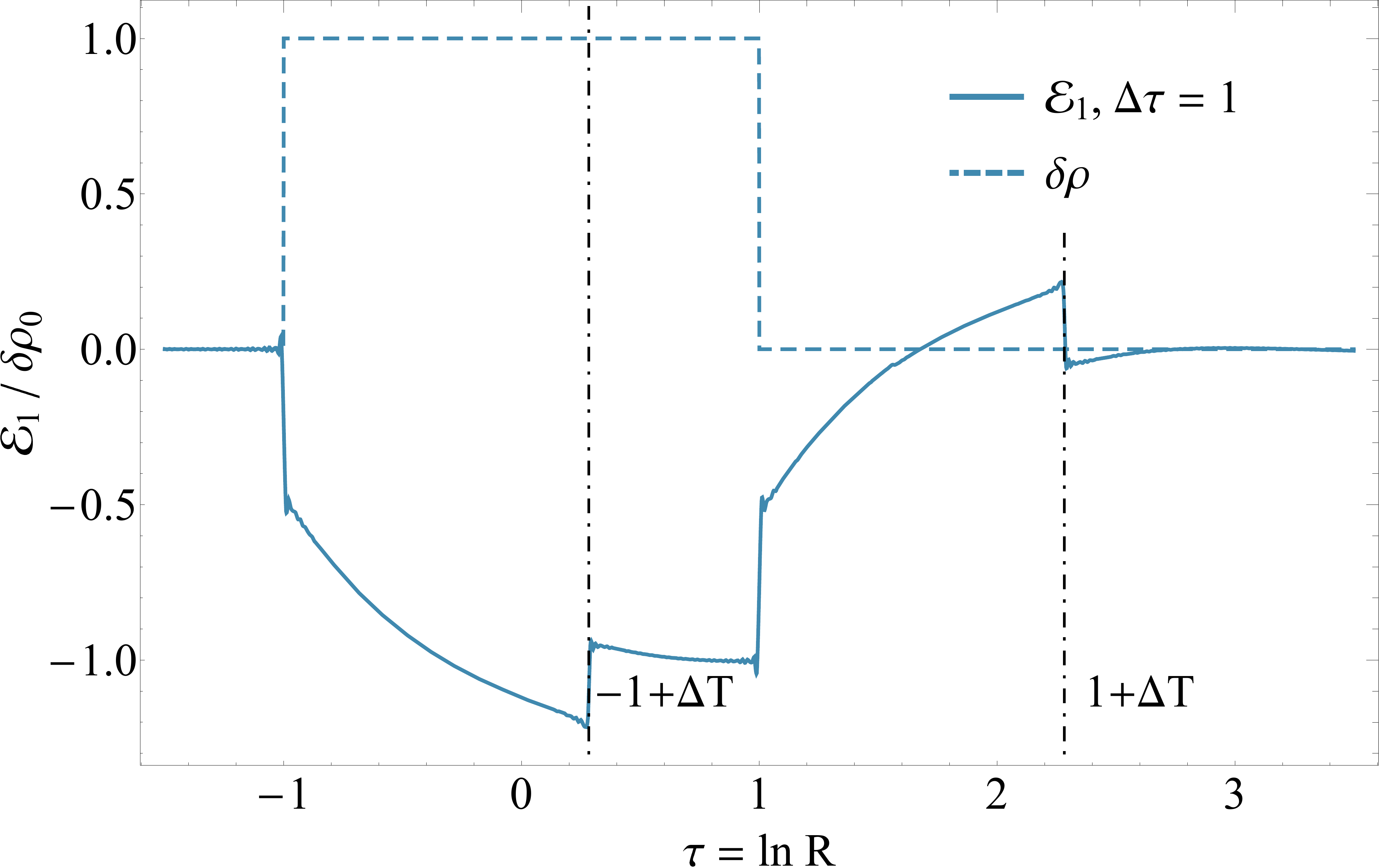} 
   \includegraphics[width=0.325\textwidth]{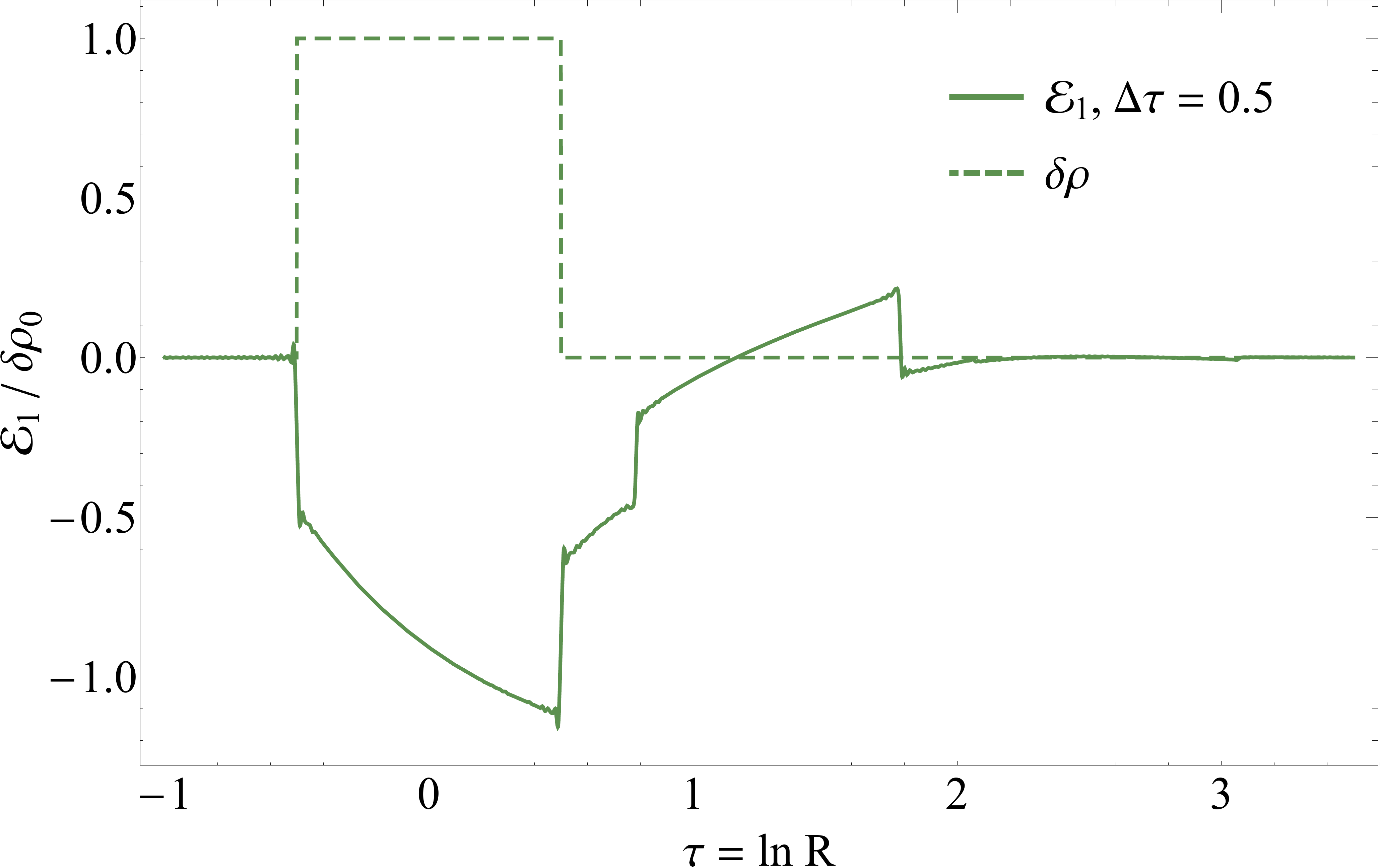} 
   \includegraphics[width=0.325\textwidth]{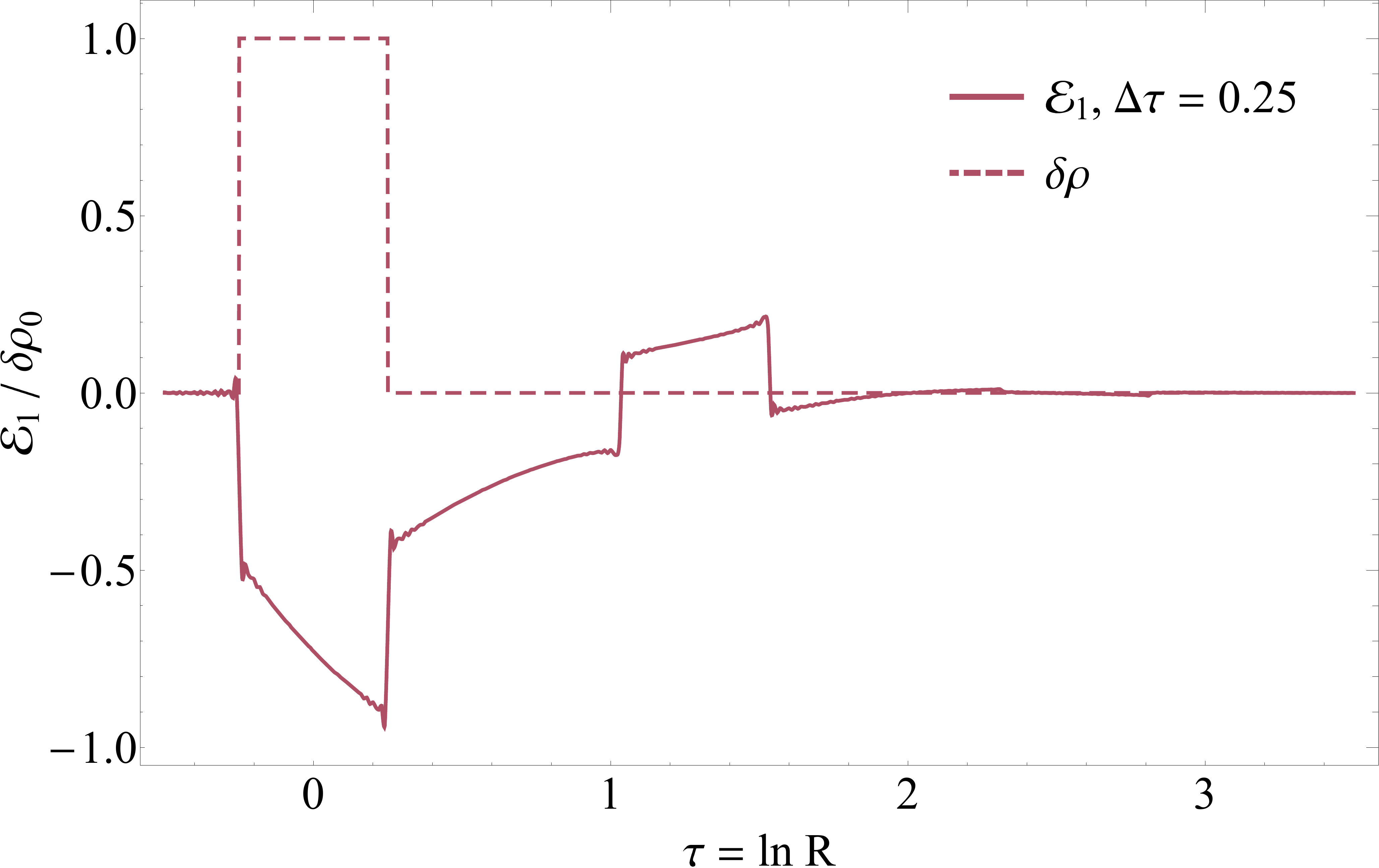} 
     \includegraphics[width=0.325\textwidth]{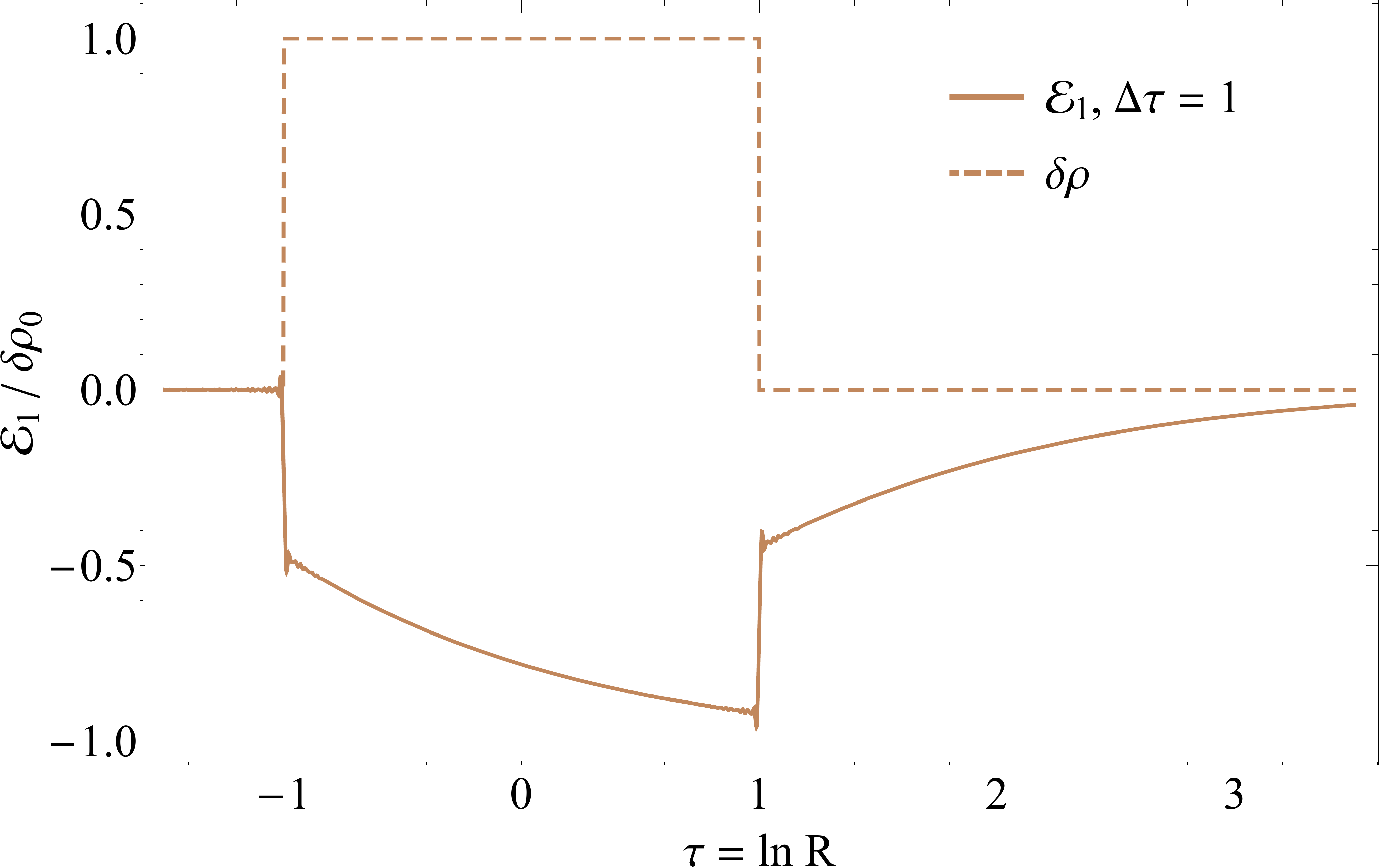} 
   \includegraphics[width=0.325\textwidth]{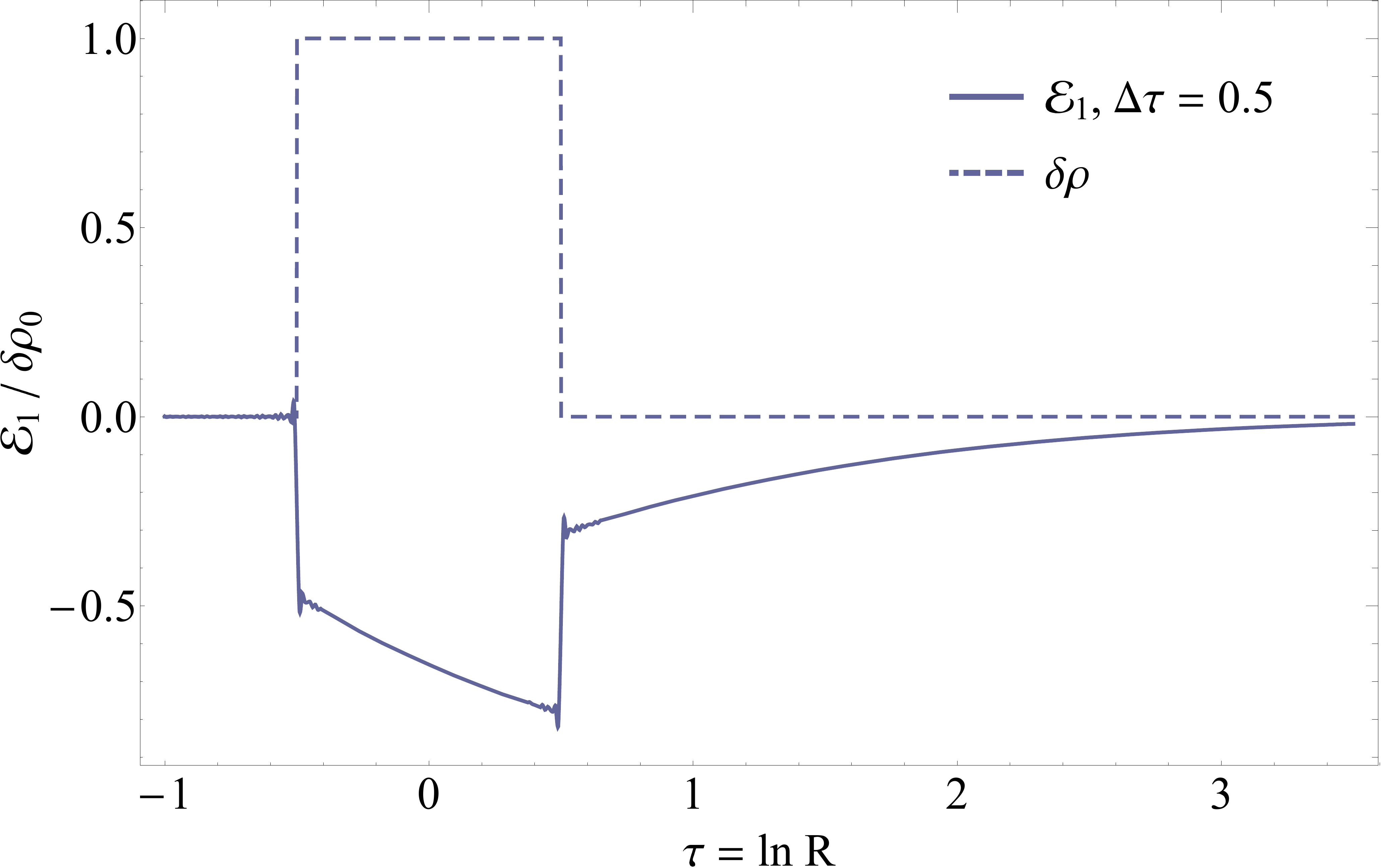} 
   \includegraphics[width=0.325\textwidth]{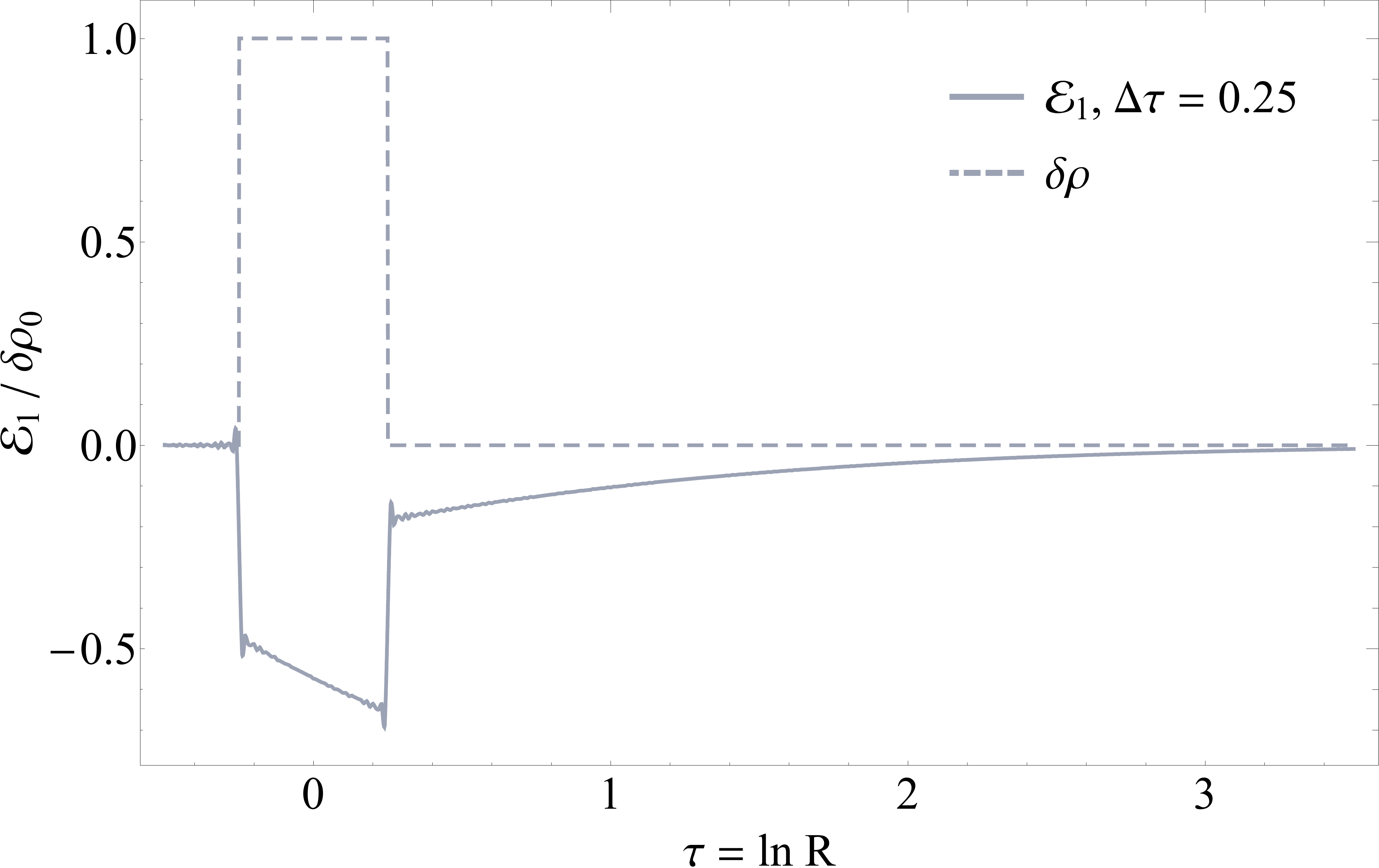}
   \caption{The correction to the energy variable, normalized by the perturbation to the density, as the shock encounters a ``rectangular bump'' of width $\Delta \tau$, with $\Delta \tau$ shown in the legend of each panel; the dashed curves give the scaling of the density perturbation for reference. The top row is for a constant-density ambient medium ($n = 0$), and the shock exhibits sudden and discontinuous changes at $\tau = \Delta T \simeq 1.286$ after encountering the discontinuity in the ambient medium. For the wind-like medium (bottom row), the shock does not exhibit any additional deviations, which arises from the causal disconnectedness of the post-shock flow.}
   \label{fig:eps_rectangle}
\end{figure}

Figure \ref{fig:eps_rectangle} shows the perturbation to the energy, $\mathscr{E}_1$, normalized by the magnitude of the density perturbation for three rectangular bumps, where the width of the bump $\Delta \tau$ is shown in the legend. The dashed curves in these figures show the shape of the density perturbation for reference. The top row illustrates the solution for a constant-density medium, from which it is apparent that the shock properties display discontinuities not only when encountering the discontinuous change in the density profile, but also at discrete times later. These additional, sudden changes in the shock properties arise from the fact that the information about the discontinuous change in the ambient properties propagates through the post-shock fluid, reflects off the origin, and travels back to the shock front in a finite time, inducing additional variation in the shock properties in the absence of any perturbation in the ambient properties. The timescale over which this happens is given by $\Delta T \simeq 1.286$ after encountering a density discontinuity for $n = 0$ and $\gamma = 5/3$; the times $\tau = -1+\Delta T$ and $1+\Delta T$ are shown by the vertical, dot-dashed lines in the left panel of this figure, and coincide precisely with when the shock shows a sudden change in the energy variable despite the absence of any variations in the ambient density. On the other hand, for wind-like media where the sound-crossing time behind the shock is infinite (bottom three panels), the shockwave only exhibits variations in its properties upon encountering the changes in the ambient medium.

\begin{figure}
    \centering
    \includegraphics[width=0.495\textwidth]{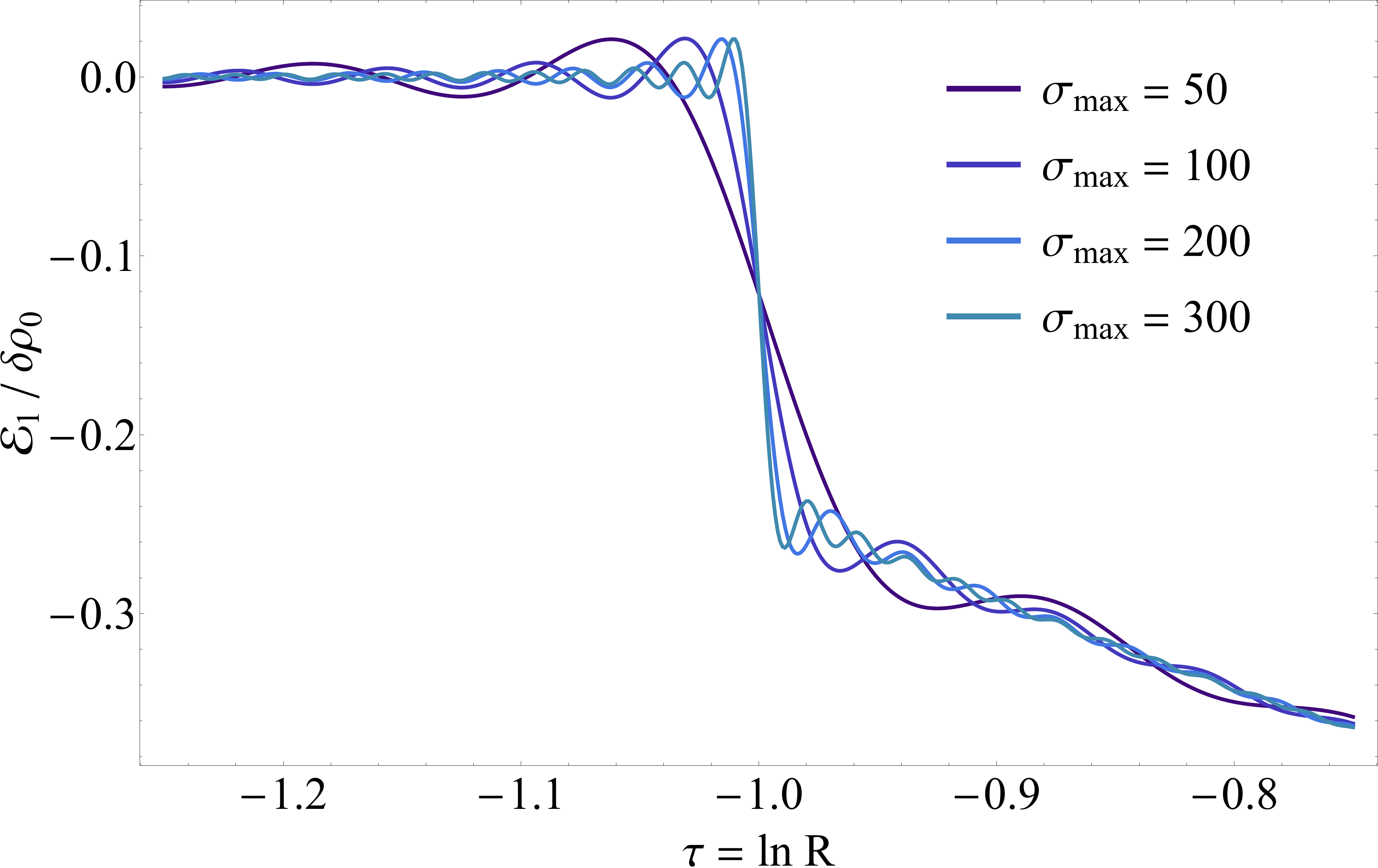}
    \includegraphics[width=0.495\textwidth]{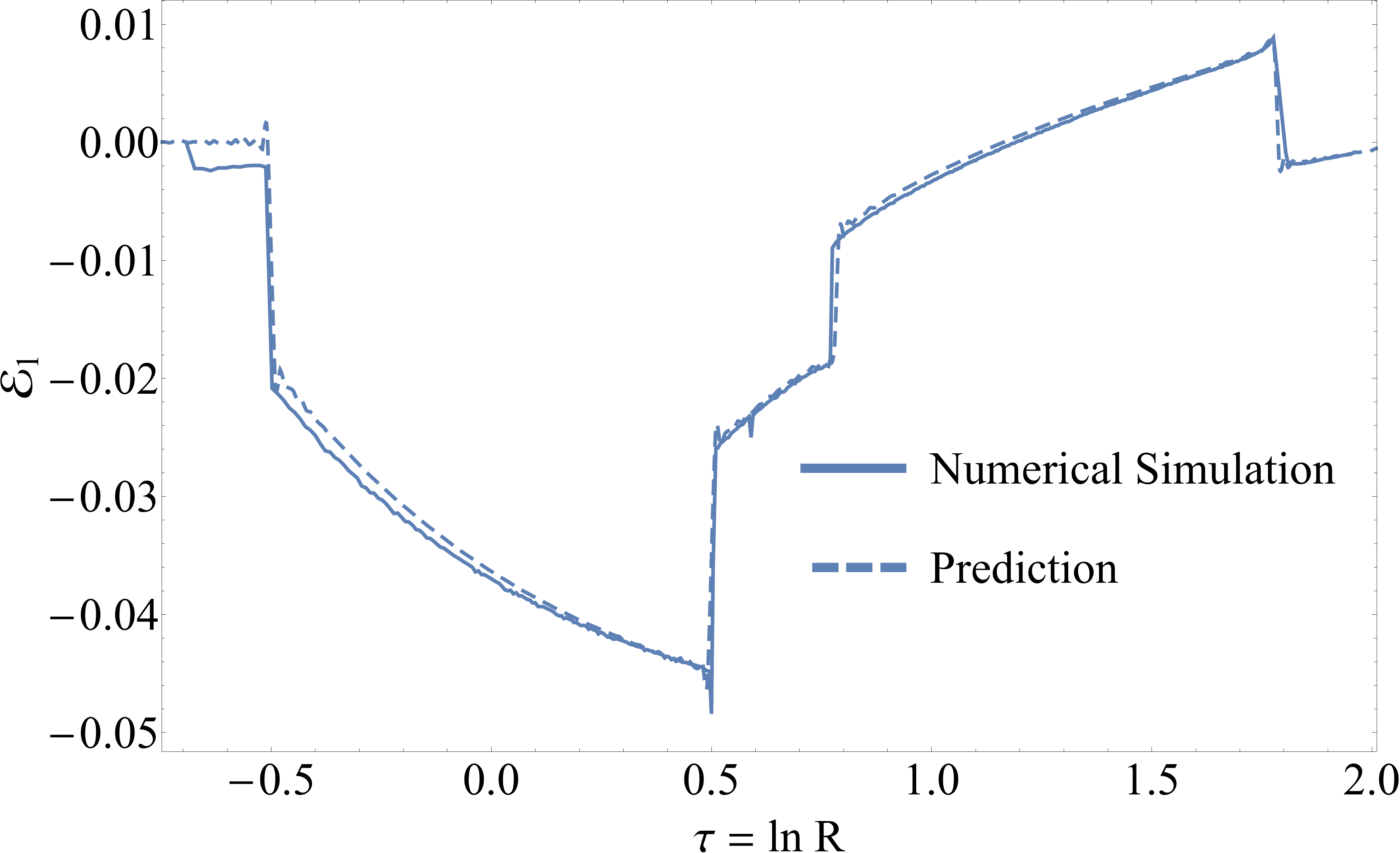}
    \caption{Left: A closeup of the response of the blastwave in the immediate vicinity of the first discontinuity in a rectangular wave of width $\Delta \tau = 1$. The different curves correspond to the upper limit of the integral of the inverse Fourier transform as shown in the legend. As the upper limit increases, the oscillations of the solution become more confined to the discontinuity itself, though the magnitude of the oscillations does not decrease (the Gibbs phenomenon). Right: The evolution of the perturbation to the energy (Equation \ref{escript}) when $\delta \rho = 0.04$, $n = 0$, and $\gamma = 5/3$ when the shock encounters a rectangular bump with $\Delta \tau = 0.5$. The dashed curve is the solution calculated from linear perturbation theory, and the solid curve is obtained from a one-dimensional, hydrodynamical simulation.}
    \label{fig:gibbs}
\end{figure}

It is also apparent from this figure that the discontinuities in the energy variable possess oscillations in the immediate vicinity of the discontinuities themselves. This is a consequence of the familiar Gibbs phenomenon, where the Fourier transform over and under-shoots the true value of the function at a discontinuity by an amount that does not decrease as the number of terms in the Fourier transform increases. However, the region around the discontinuity at which the disparity arises becomes smaller as the number of terms in the Fourier transform increases. The left panel of Figure \ref{fig:gibbs} shows the solution for the response of the shock in the vicinity of the first discontinuity in the density profile when $\Delta \tau = 1$, and the different curves correspond to the upper limits for the integral in the inverse Fourier transform shown in the legend. It is apparent that, as the number of terms increases, the region within which the oscillations occur becomes smaller, even though the magnitude of the oscillations does not change appreciably.

Also, there are two small, additional discontinuities in $\mathscr{E}_1$ at even larger $\tau$ separated by $2\Delta \tau$ (this is most noticeable in the top-right panel, around $\tau \simeq 2.4$ and 2.9; see also Figure \ref{fig:alpha_tri} below). These arise when the discontinuity propagates \emph{twice} through the interior of the blastwave, and hits the back of the shockwave a second time. As we argued heuristically above, we can think of the imprint of the discontinuity in the density profile as being lessened over time as a consequence of the shear in the velocity profile behind the blast. As such, these additional changes in the shock properties are much smaller, but are nonetheless apparent in the solution. 

The right panel of Figure \ref{fig:gibbs} shows a comparison between the analytical, perturbative method outlined here (dashed curve) and a one-dimensional, hydrodynamical simulation (solid curve) when the shock encounters a rectangular overdensity with magnitude $\delta \rho = 0.04$ and $\Delta \tau = 0.5$; the hydrodynamical simulation used the same methodology as applied to Figure \ref{fig:vel_n0_n2} (see the discussion pertaining to that figure). The agreement between the two approaches illustrates the accuracy of the perturbative approach in capturing the evolution of the blastwave in response to changes in the ambient medium. 

\subsubsection{Triangular bump}
\label{sec:triangular}
The Gaussian perturbation considered in Section \ref{sec:Gaussian} is infinitely differentiable but non-compact, while the Rectangular bump in Section \ref{sec:Rectangular} is compact but is discontinuous. Consequently, the acceleration (and all higher-order derivatives) of the shock as it encounters a Gaussian bump is continuous, while the acceleration undergoes a $\delta$-function-like response as it hits the discontinuity in the density provided by the rectangular bump. We anticipate in general that as the density perturbation becomes increasingly smooth with respect to its connectedness to the background density, the corresponding response of the blastwave should likewise be more continuous. To demonstrate this behavior, we consider a ``triangular bump'' in the density:

\begin{equation}
\delta\rho = \frac{\delta\rho_0}{\Delta \tau}\times
\begin{cases}
\tau+\Delta\tau \textrm{ for } -\Delta\tau \le \tau \le 0 \\
-\tau+\Delta\tau \textrm{ for } 0 \le \tau \le \Delta\tau \\
0 \textrm{ otherwise }
\end{cases} \quad \Rightarrow \quad \delta\tilde{\rho} = \frac{4\delta\rho_0}{\Delta\tau}\frac{\sin^2\left(\frac{\sigma\Delta\tau}{2}\right)}{\sigma^2}.
\end{equation}
This density perturbation smoothly joins onto the background density, but its derivative changes discontinuously at $\pm \Delta \tau$. Since $\delta\tilde{\rho}$ scales as $\propto 1/\sigma^2$ for $|\sigma| \gg 1$, we expect the perturbation to the acceleration to be finite but to jump discontinuously upon encountering the density perturbation. 

\begin{figure}[htbp] 
   \centering
   \includegraphics[width=0.325\textwidth]{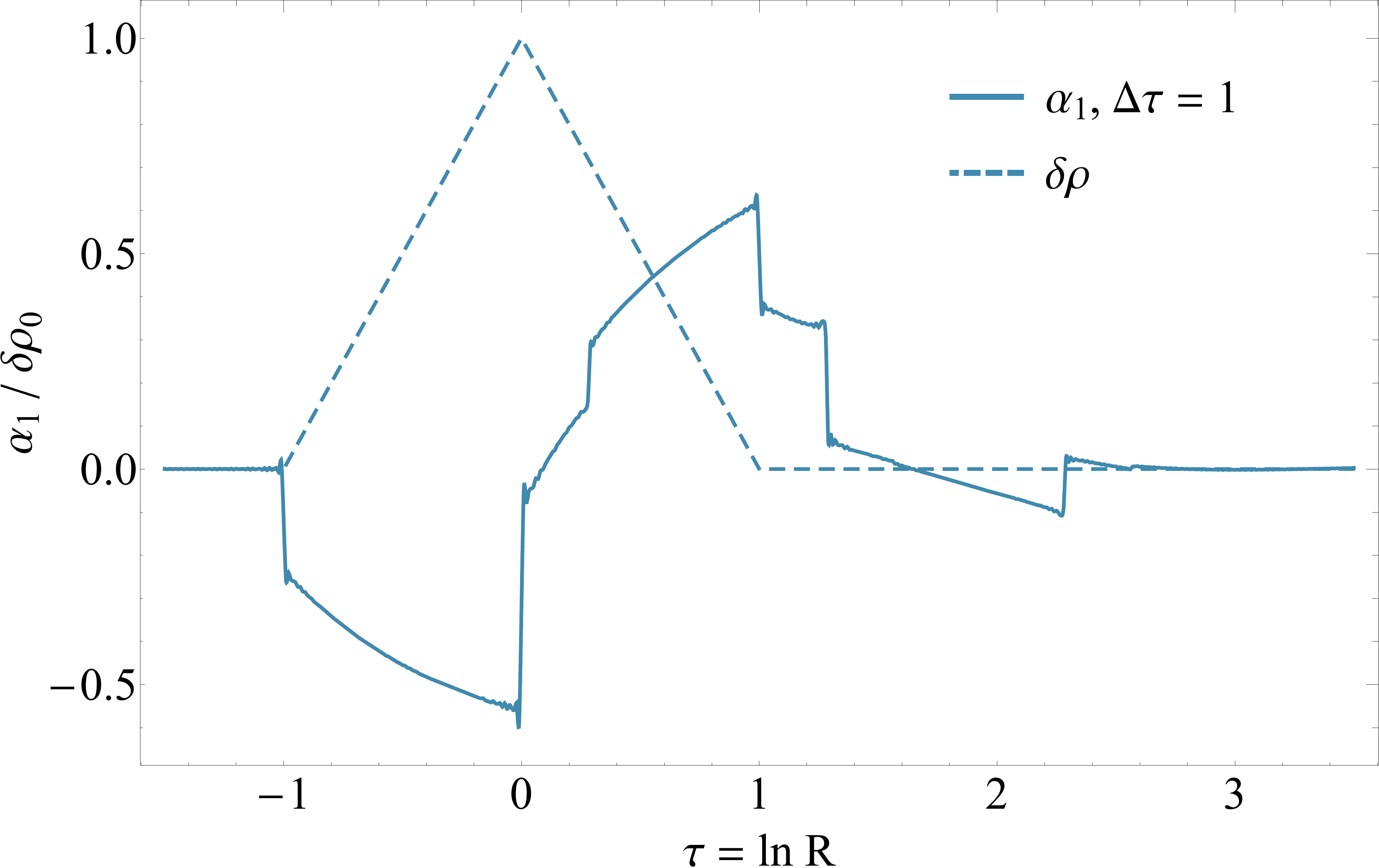} 
   \includegraphics[width=0.325\textwidth]{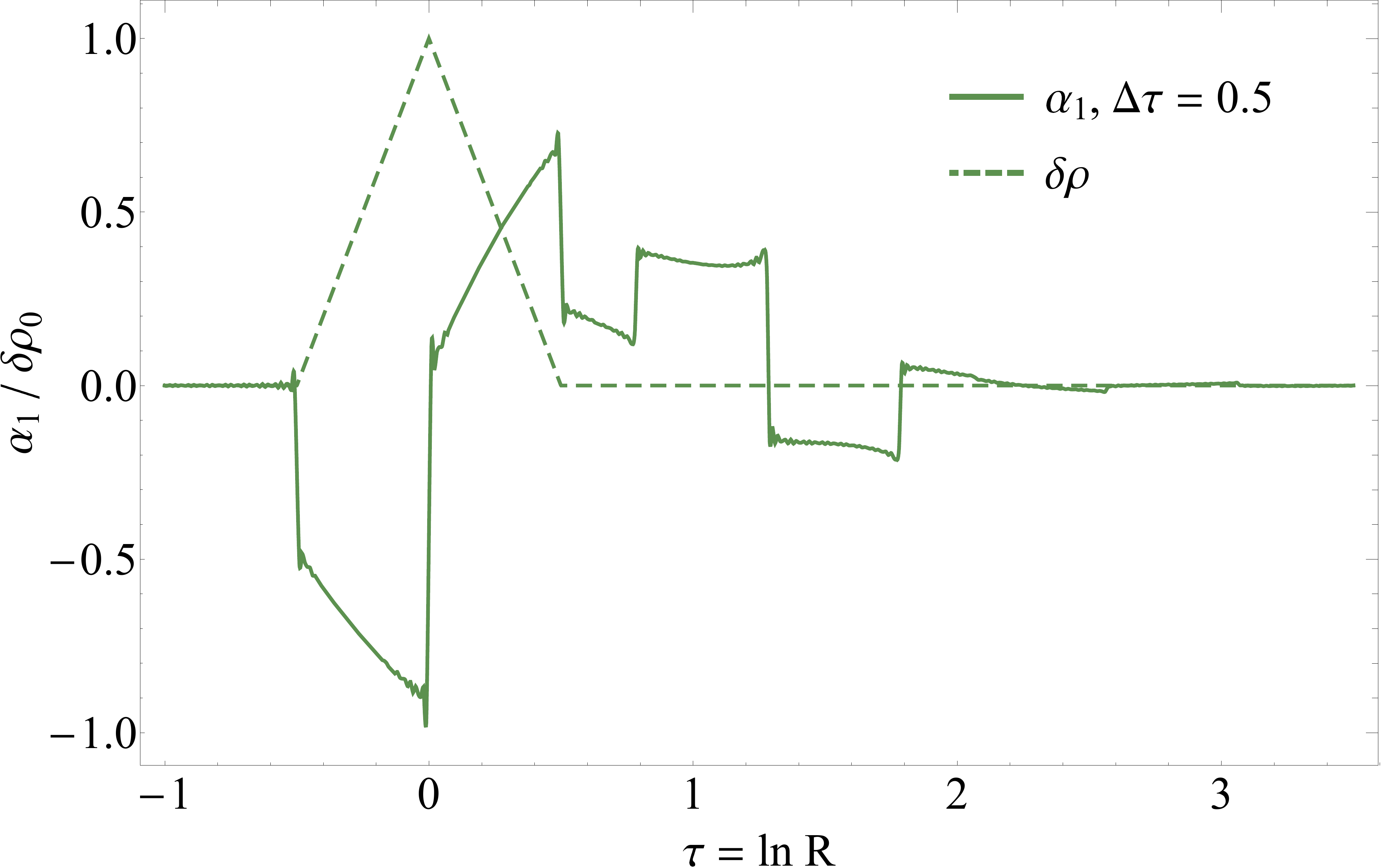} 
   \includegraphics[width=0.325\textwidth]{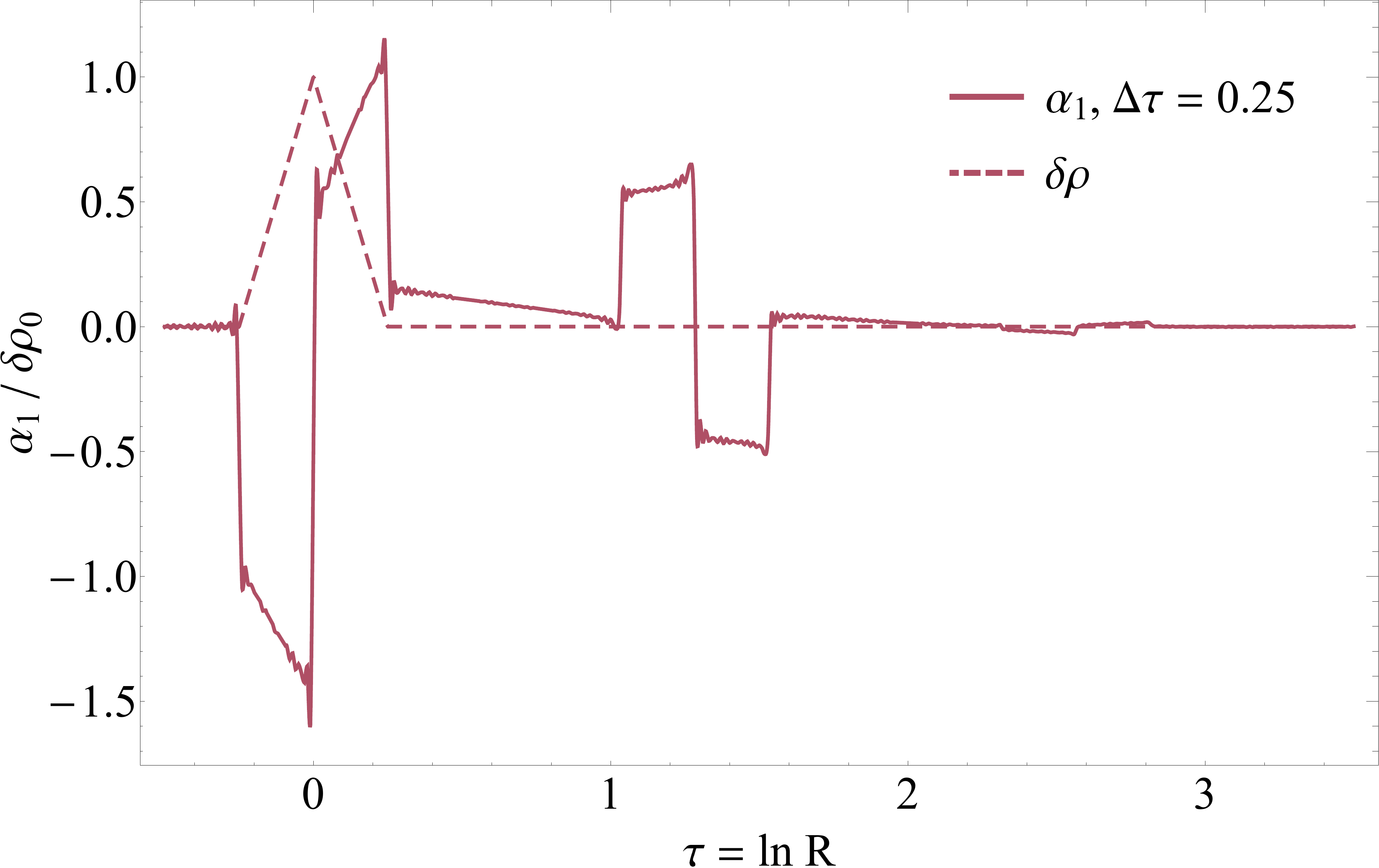} 
     \includegraphics[width=0.325\textwidth]{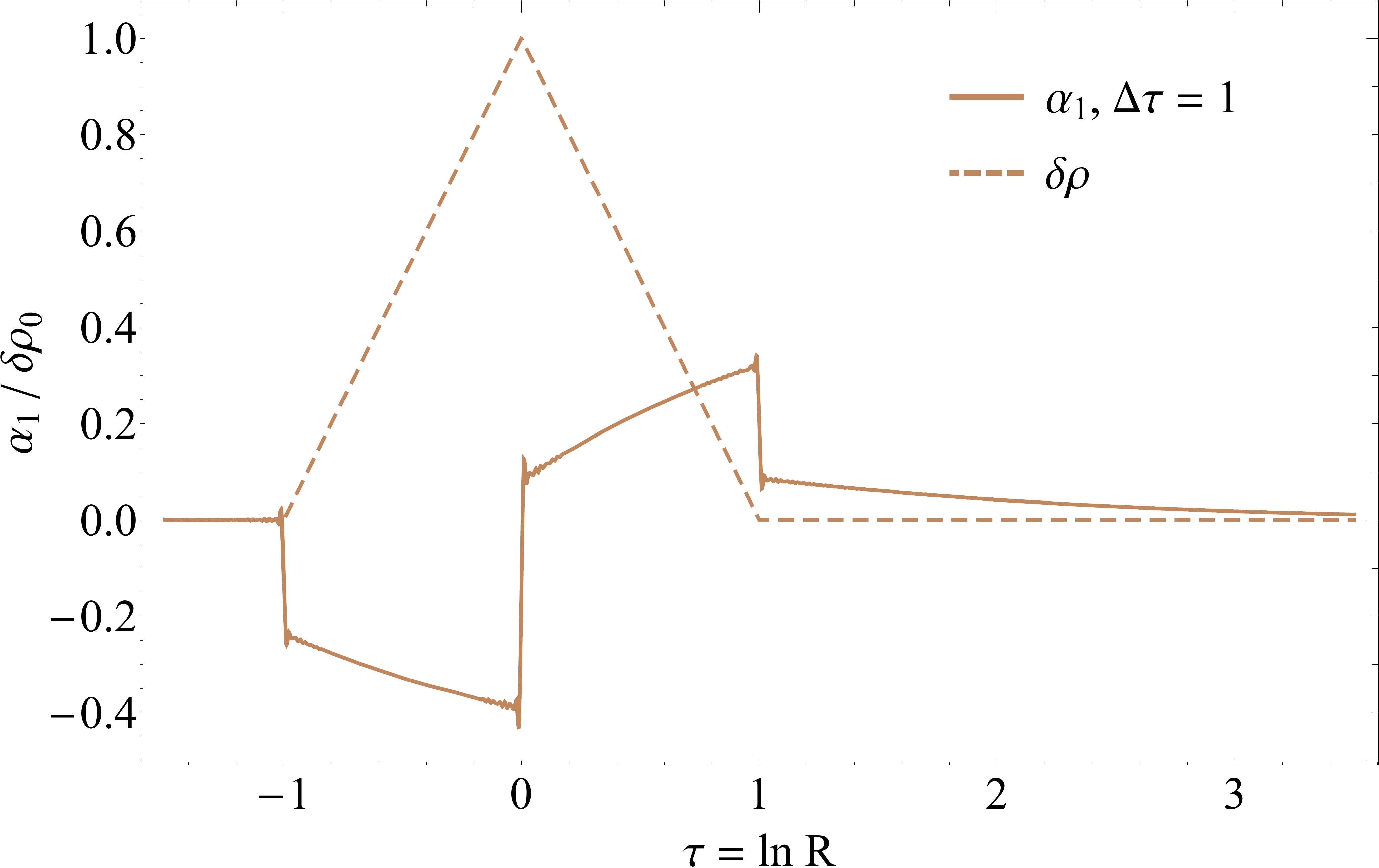} 
   \includegraphics[width=0.325\textwidth]{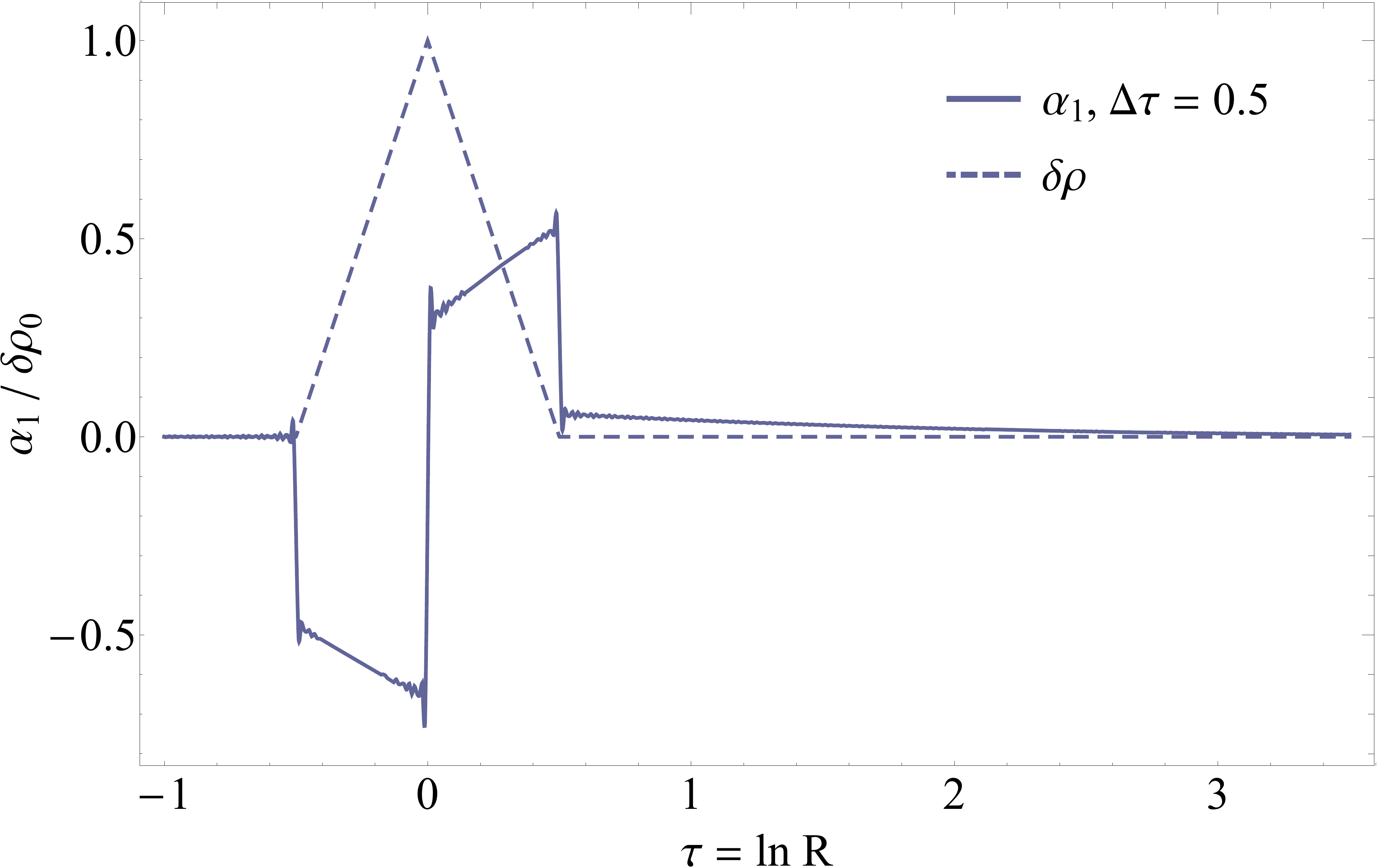} 
   \includegraphics[width=0.325\textwidth]{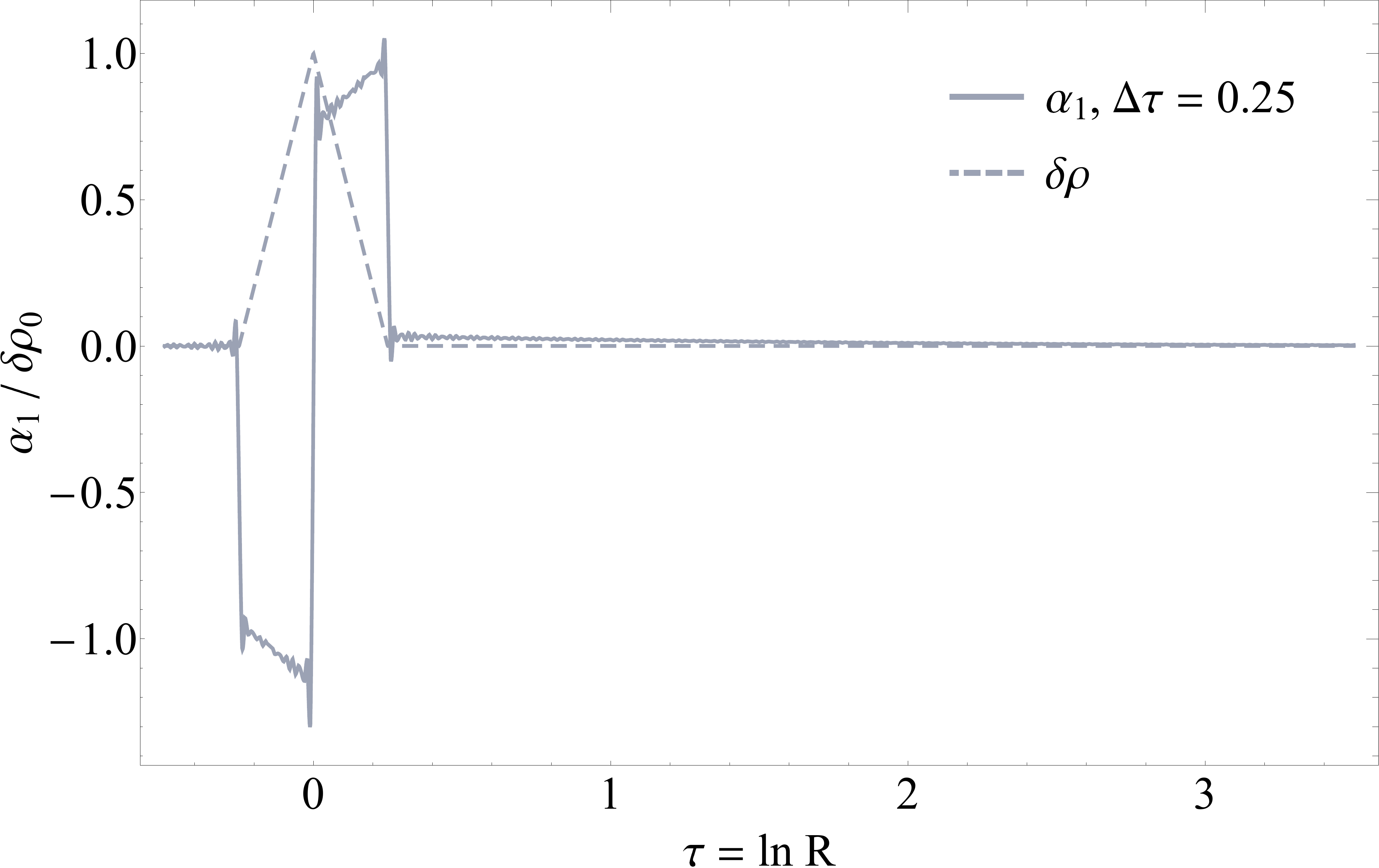}
   \caption{The same as Figure \ref{fig:alpha_Gaussian_n0}, but with a ``triangular bump'' of width $\Delta \tau$; the top row is for $n = 0$ (constant-density ambient medium), bottom for $n = 2$ (wind-fed ambient medium), the value of $\Delta\tau$ is shown in the top-right of each panel, and the dashed curves show the perturbation to the density for reference.}
   \label{fig:alpha_tri}
\end{figure}

Figure \ref{fig:alpha_tri} shows the shock acceleration parameter for $n = 0$ and $\gamma = 5/3$ (top row) and $n = 2$ (bottom row) for three different triangular bumps, the arbitrarily scaled density profiles for which are shown by the dashed lines in these figures. As we anticipated, the shock acceleration changes discontinuously as it first encounters the linear increase in the density. We also see a number of additional, discontinuous jumps that occur both as the shock encounters changes in the density and as discontinuities propagate through the post-shock fluid and back to the shock itself.

\subsubsection{Sinusoidal bump}
There is clearly an infinite number of density perturbations to which we can subject an energy-conserving blastwave, and the three previous examples illustrate the effects on the shockwave as it encounters perturbations that are more or less continuous in their deviation from the background properties. As a last, illustrative example, consider a ``sinusoidal bump,'' given by

\begin{equation}
    \delta\rho = \delta\rho_0\times
    \begin{cases}
    \cos\left[\frac{1}{2}\left(2m+1\right)\frac{\pi\tau}{\Delta\tau}\right] \textrm{ for } -\Delta\tau \le \tau \le \Delta \tau \\
    0 \textrm{ otherwise }
    \end{cases}
    \quad \Rightarrow \quad \delta\tilde{\rho} = \delta\rho_0\pi\Delta\tau\left(2m+1\right)\left(-1\right)^{m}\frac{\cos\left(\sigma\Delta\tau\right)}{\pi^2\left(m+1/2\right)^2-\sigma^2\Delta\tau^2},
\end{equation}
where $m \ge 0$ is an integer. Such a perturbation corresponds to a sinusoidally varying density oscillation with $2m+1$ maxima and minima within the range $\{-\Delta\tau,\Delta\tau\}$. For example, $m=0$ has one density maximum at $\tau = 0$, $m = 1$ has one density maximum and two density minima, $m = 2$ has three density maxima and two minima, etc. 

\begin{figure}[htbp] 
   \centering
   \includegraphics[width=0.325\textwidth]{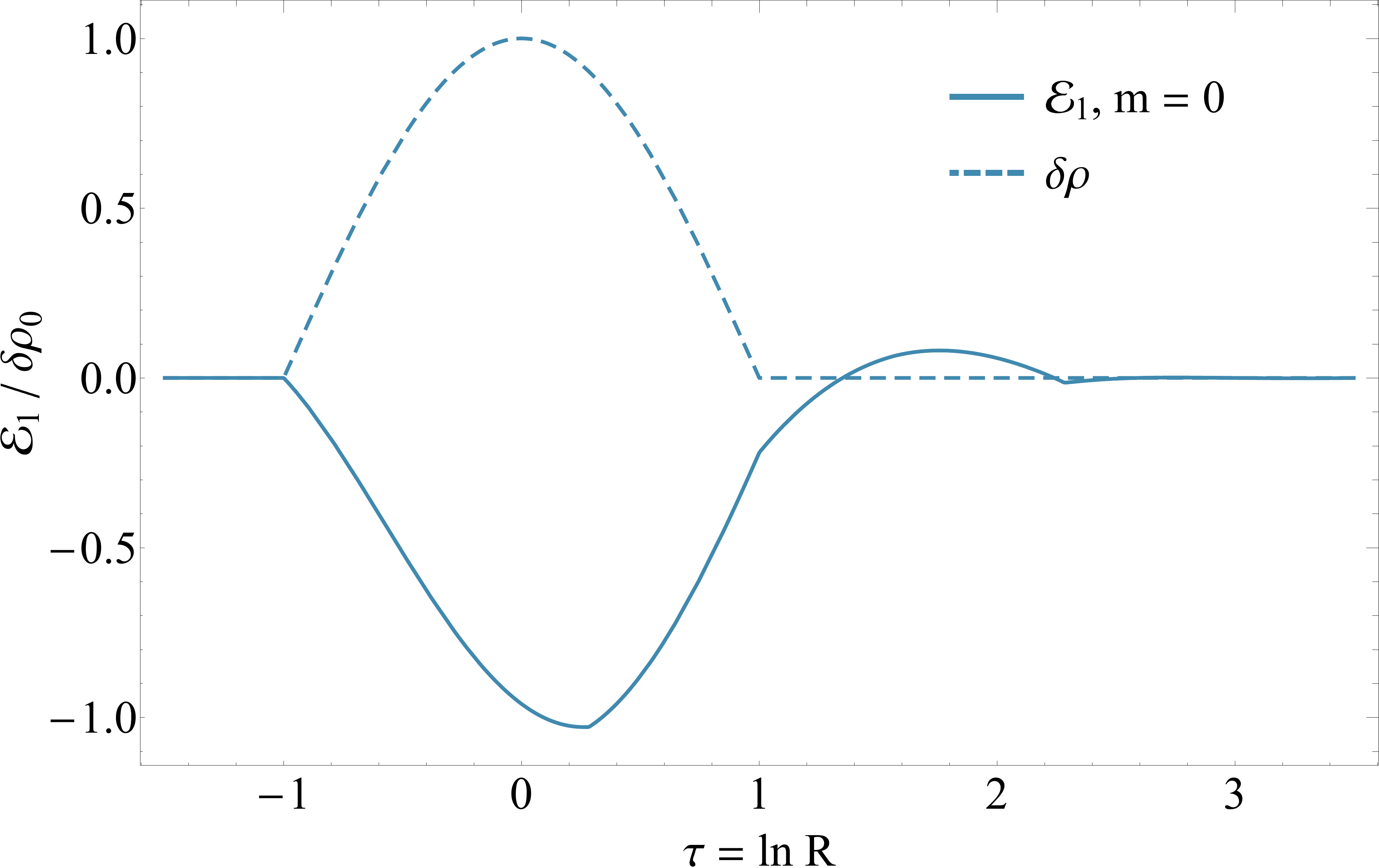} 
   \includegraphics[width=0.325\textwidth]{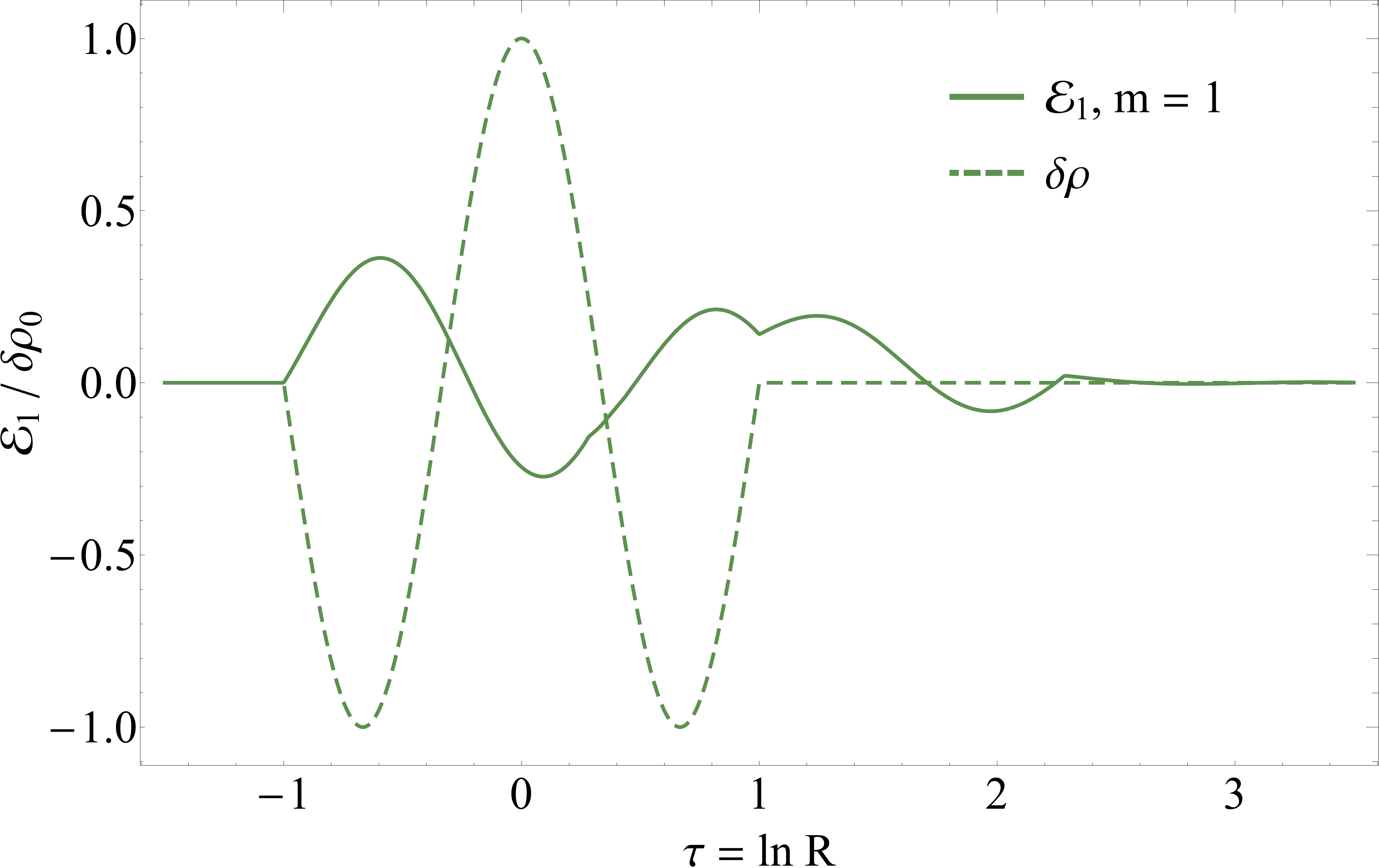} 
   \includegraphics[width=0.325\textwidth]{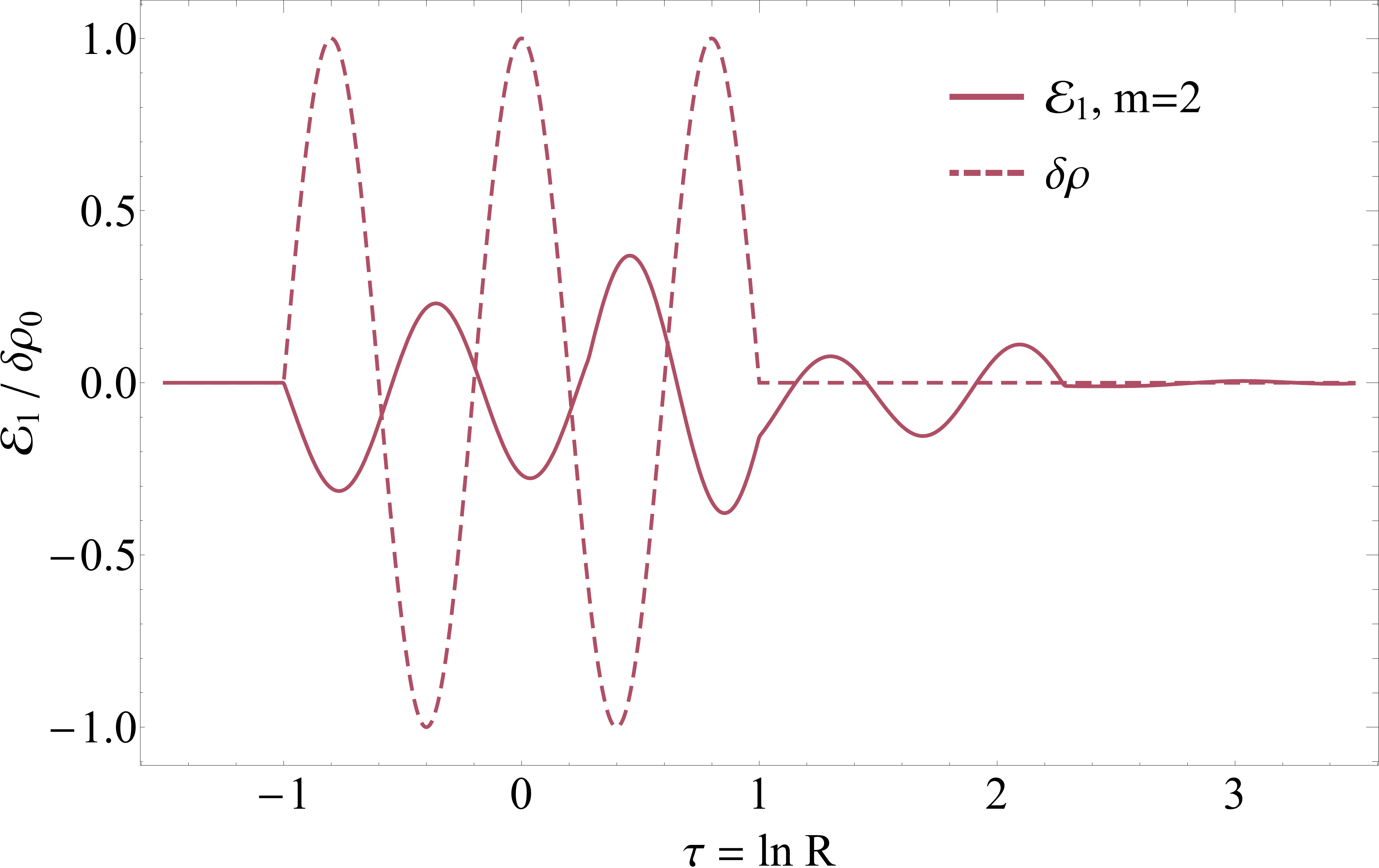} 
     \includegraphics[width=0.325\textwidth]{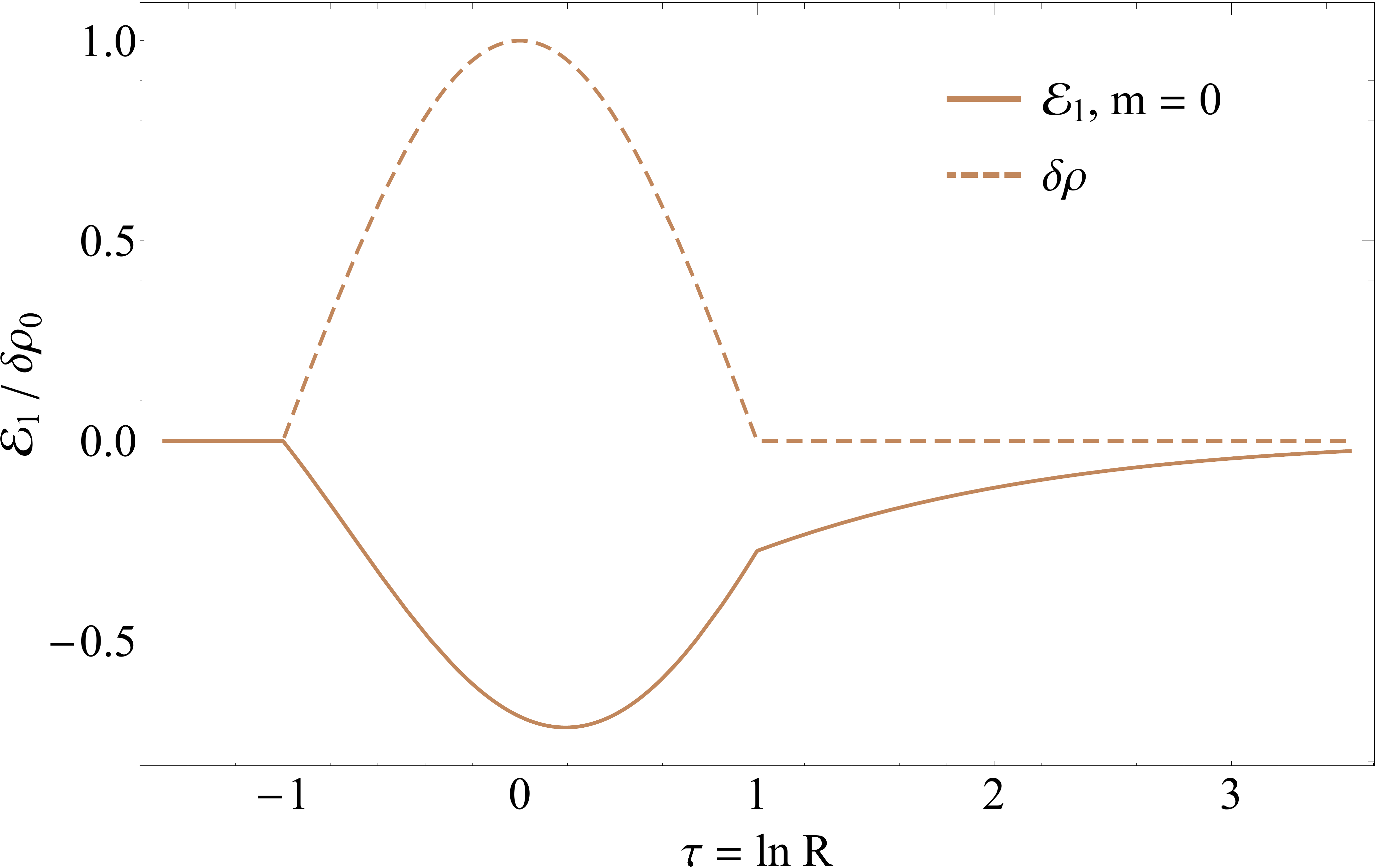} 
   \includegraphics[width=0.325\textwidth]{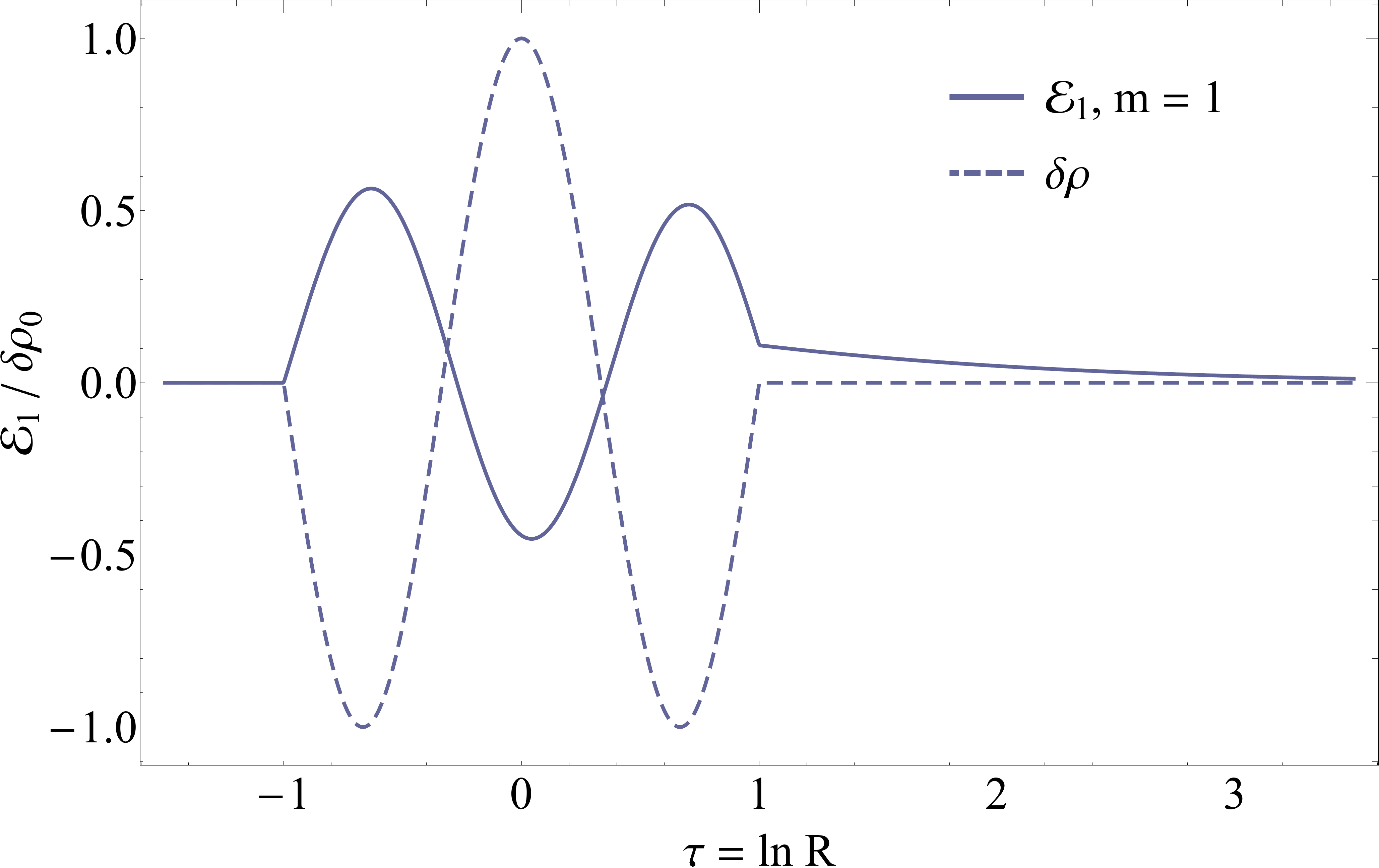} 
   \includegraphics[width=0.325\textwidth]{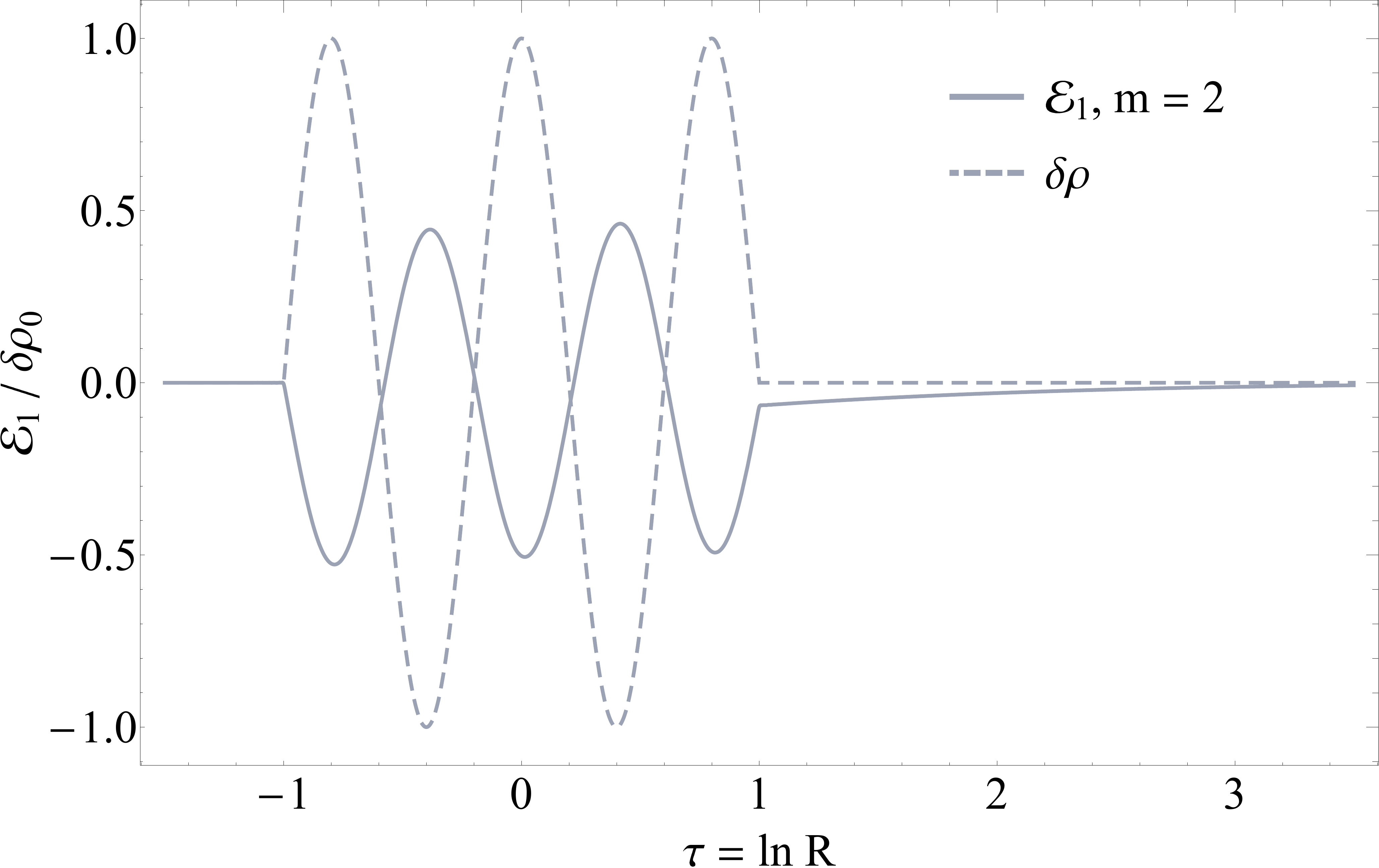}
   \caption{The same as Figure \ref{fig:eps_rectangle}, being the correction to the shock energy as a function of the log of the shock position, but with a ``sinusoidal bump'' perturbation on top of the ambient density profile. The dashed curves in this figure scale with the density perturbation.}
   \label{fig:eps_sine}
\end{figure}

Figure \ref{fig:eps_sine} shows the correction to the energy variable, $\mathscr{E}_1$, as a function of $\tau$ for three different sinusoidal ``bumps,'' with 1 maximum and no minima (left), one maximum and two minima (middle), and three maxima and two minima (right). The top panel is for a constant-density ambient medium, while the bottom is for a wind-like medium. We see that the response of the shock is approximately to decelerate in regions of increasing density, and accelerate in areas of decreasing density. However, there is a clear phase lag between the two; for example, in the left panel the minimum value of the perturbation to the energy is reached at a location that is slightly offset from the location at which the perturbation to the density reaches a maximum. It is also apparent that the solutions with $\gamma = 5/3$, for which there is a finite post-shock sound-crossing time, exhibit additional deviations (i.e., not just decelerating/accelerating within regions of increasing/decreasing density) both within the region occupied by the density perturbation (for $-\Delta \tau \le \tau \le \Delta \tau$) and outside of it. For example, the top-left panel of this figure shows a ``cuspy'' response of the shock energy just after $\tau = 0$ and at a time of $-1+\Delta T$, where $\Delta T$ is the dimensionless sound-crossing time behind the shock, and the top-right panel shows that the shock continues to oscillate when the perturbation to the density is exactly zero. This behavior arises from the finite sound-crossing time behind the shock, and that these perturbations continue to reverberate in the interior of the blast and affect the propagation of the shock.

\section{Astrophysical implications}
\label{sec:implications}
As a strong shock encounters a density enhancement, our analysis above indicates that the post-shock variables (the density, pressure, and velocity) show time-dependent and secular evolution that is not captured by the purely self-similar nature of the Sedov-Taylor solution. Similarly, the shock itself displays variation in its velocity and position that deviate from the Sedov prediction. As such, the observational appearance of the blastwave will change in ways that reflect the underlying nature of the density enhancement that the shock has encountered and -- because of the sound-crossing time over the post-shock flow -- will do so not only when it initially encounters the enhancement, but also at a delayed time and (conceivably) when there is no enhancement in the immediate vicinity of the shock front.

\begin{figure}
    \centering
    \includegraphics[width=0.495\textwidth]{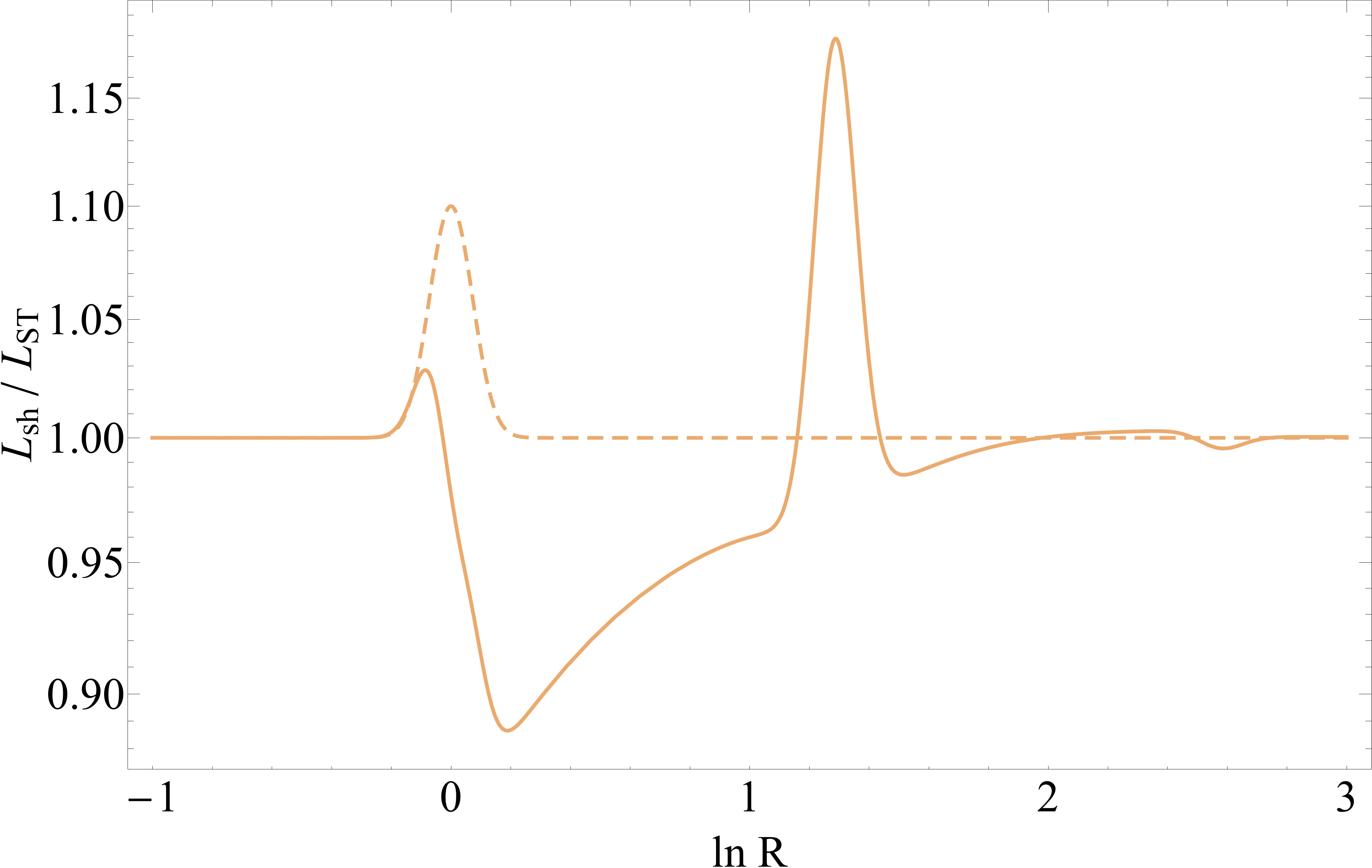}
      \includegraphics[width=0.495\textwidth]{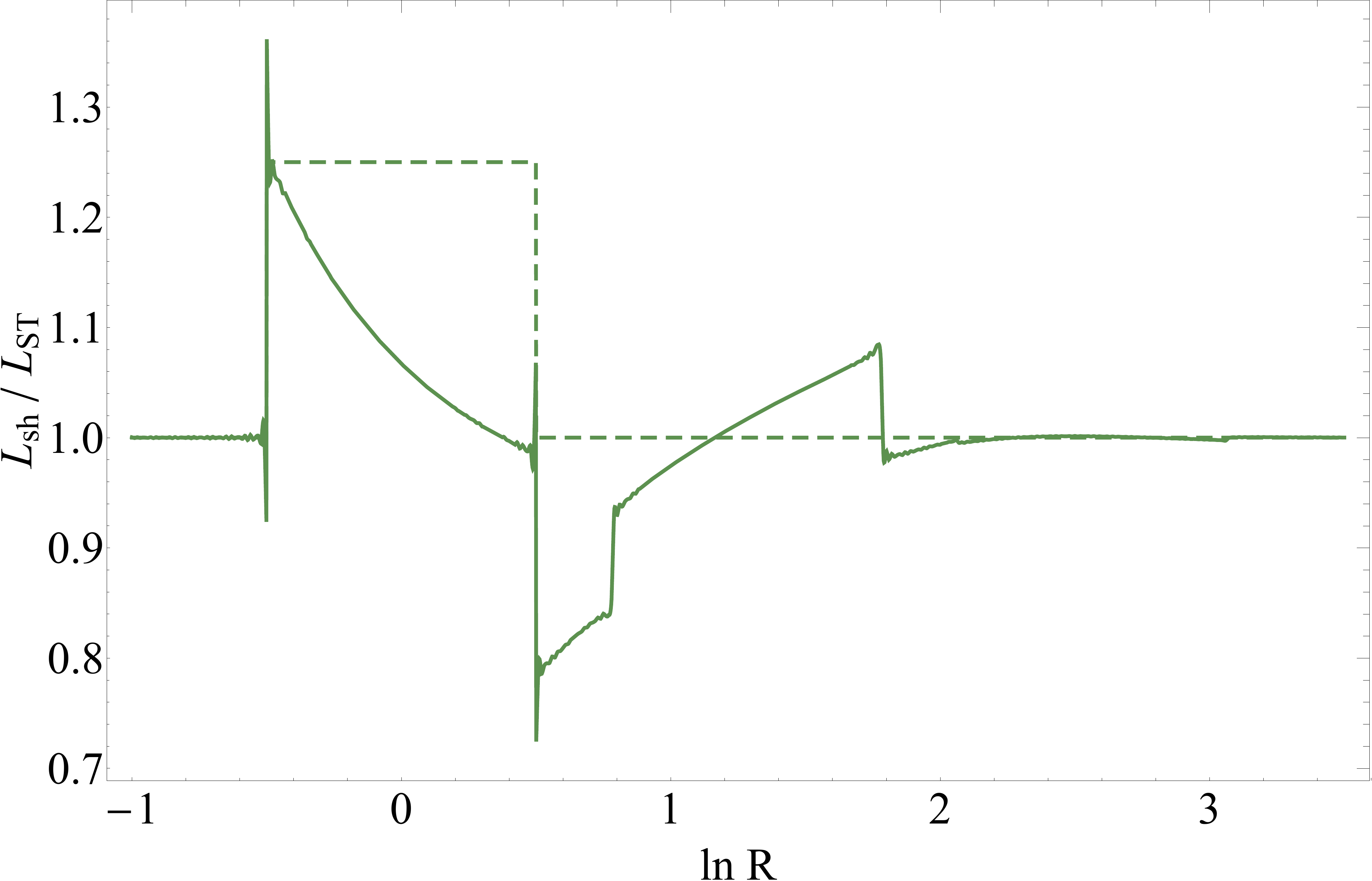}
    \caption{The shock luminosity, as defined via Equation \eqref{Lsh}, relative to the shock luminosity in the absence of a density perturbation (i.e., of the Sedov-Taylor blastwave) as a function of the natural log of the shock position for a Gaussian overdensity of width $\Delta \tau = 0.1$ (left) and a rectangular overdensity of width $\Delta \tau = 0.5$ (right). Here the unperturbed, ambient density is assumed to be constant, so $n = 0$. The dashed curves give the scaled density profiles. In each case the luminosity is characterized by a sharp change upon encountering the density enhancement, where the change is mediated by the simultaneous change in the shock velocity and the ambient density, and a later rebrightening as the pulse launched from the initial interaction returns to the shock.}
    \label{fig:Lsh}
\end{figure}

\begin{figure}
    \centering
    \includegraphics[width=0.49\textwidth]{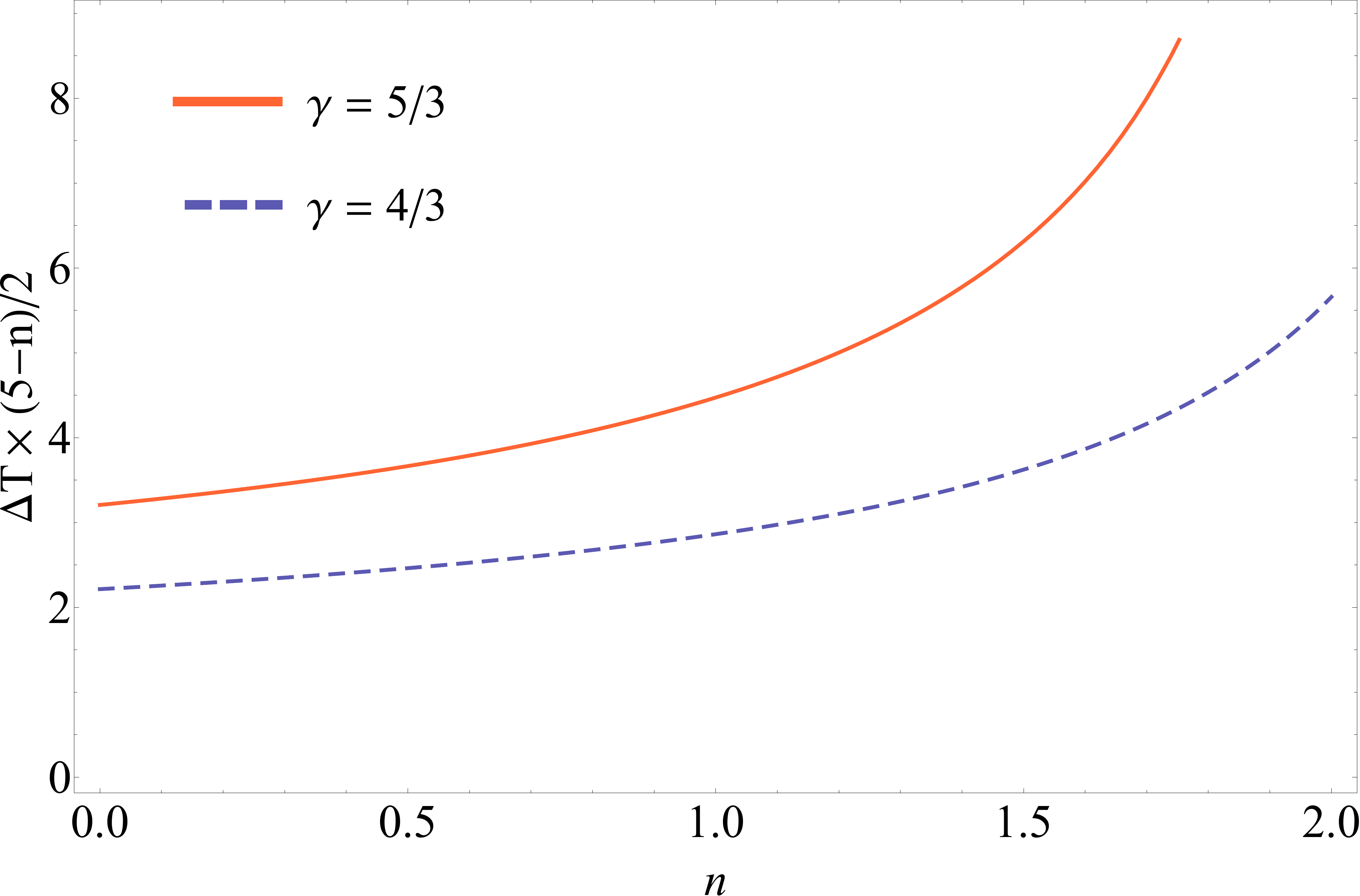}
    \includegraphics[width=0.5\textwidth]{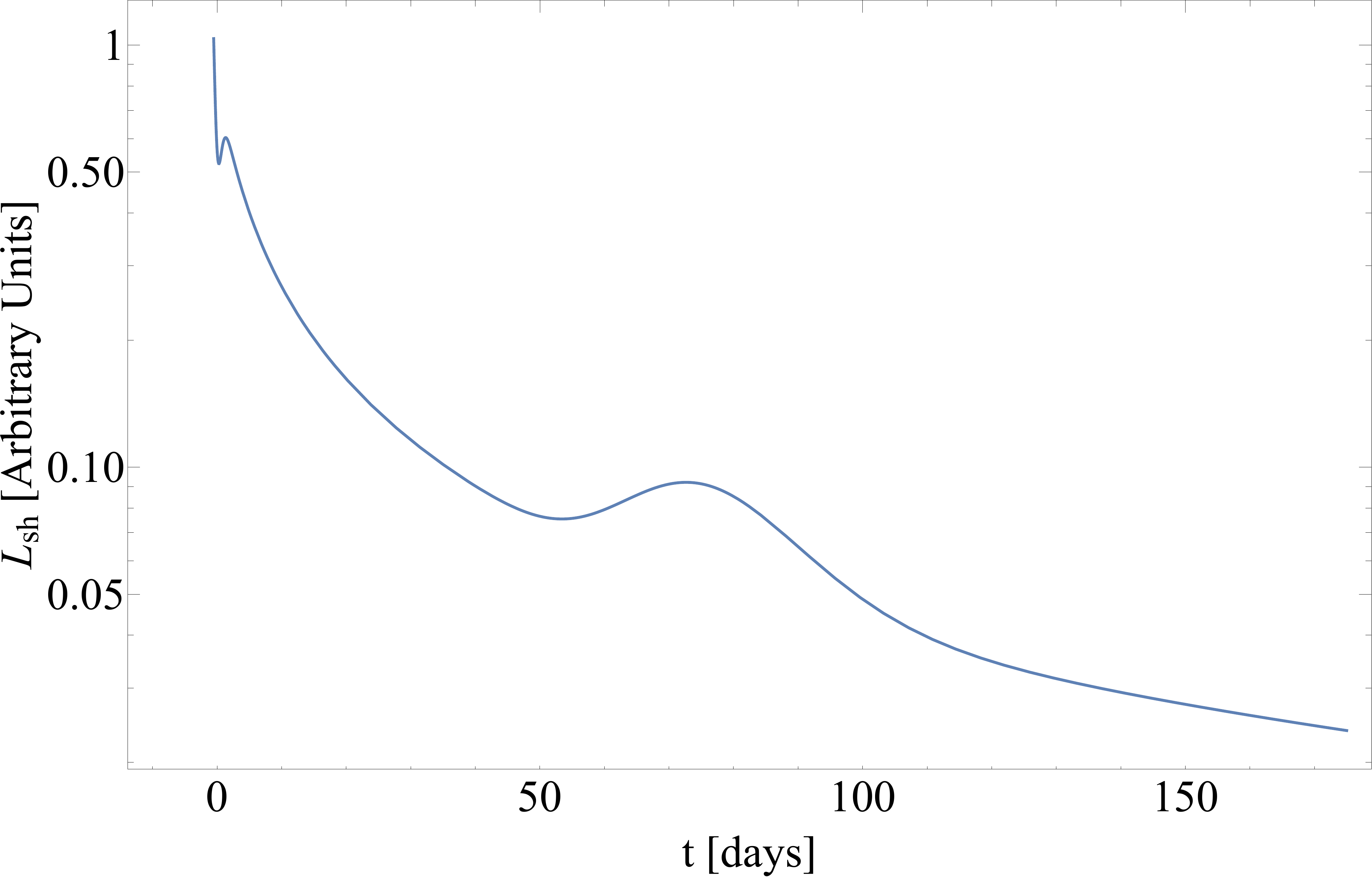}
    \caption{Left: The product $\Delta T\times (5-n)/2$, which is the log of the factor by which the shock expands in the sound crossing time behind the shock, as a function of $n$. The fact that this increases as $n$ shows that the time between when the shock first encounters a density enhancement and exhibits the delayed rebrightening increases as the density profile of the ambient medium steepens. The solid, orange curve is for $\gamma = 5/3$, while the dashed, purple curve is for $\gamma = 4/3$. Right: The shock luminosity, in arbitrary units, as a function of time in days for the same density perturbation as in the left panel of Figure \ref{fig:Lsh}; here the magnitude of the overdensity was set to $\delta\rho/\rho = 1.5$. The extended luminosity enhancement at $\sim 75$ days is in good, qualitative agreement with the evolution of the event SN 2019tsf. This figure also shows that the initial brightening of the source upon encountering the density enhancement, which occurs near $t = 0$ here and lasts for $\lesssim 1$ day, could have plausibly been missed in the early monitoring of the event.}
    \label{fig:DeltaT_of_n}
\end{figure}

We can more quantitatively assess the temporal evolution of the observed, shock properties by investigating the shock power, which is defined as\footnote{One could also define the shock power as the kinetic energy flux in the rest-frame of the shock; by virtue of the jump conditions, the resulting expression is identical to the one given in Equation \eqref{Lsh} to a factor of the order unity, and all of the plots that use this expression (Figures \ref{fig:Lsh} and the right panel of Figure \ref{fig:DeltaT_of_n}) would be unchanged.}

\begin{equation}
    {L}_{\rm sh} = u_{\rm int}\dot{M}_{\rm sh} = \frac{8\pi}{\left(\gamma+1\right)^2}\rho_{\rm a}R^2V^3. \label{Lsh}
\end{equation}
Here $u_{\rm int}$ is the specific internal energy of the post-shock gas, and is $u_{\rm int} = (p/\rho)/(\gamma-1)$, where here $p$ and $\rho$ are the post-shock pressure and density, and $\dot{M}_{\rm sh}$ is the mass flux across the shock front. Equation \eqref{Lsh} therefore represents the total energy able to be radiated at the shock front as a consequence of the dissipation of kinetic energy at the expense of heating the gas; the last equality in this expression results from using the jump conditions at the shock. Note that the $\rho_{\rm a}$ that appears here is the density of the ambient gas that includes the perturbations; therefore, for a density enhancement where the velocity declines owing to the increased momentum flux across the shock, the product of $\rho_{\rm a}$ and $V^{3}$ will behave non-trivially (i.e., the density enhancement offsets the reduction in the velocity and the two effects compete). 

Figure \ref{fig:Lsh} shows the shock luminosity as a function of the shock position for a Gaussian bump of with $\Delta \tau = 0.1$ (left) and a square bump of width $\Delta \tau = 0.5$ (right) in an otherwise-constant-density ambient medium (i.e., $n = 0$). Here we normalized the shock luminosity by the value that would result from the Sedov-Taylor explosion in the absence of any density perturbation, and we set the magnitude of the overdensity to $\delta \rho = 0.5$. The solid curves show the normalized luminosity, while the dashed curves show the density scaled arbitrarily. In each case, the response can be understood as an initial brightening as the shock encounters the overdensity; the brightening occurs because the increase in the density outweighs the reduction in the velocity that follows from the augmented momentum flux at the shock front. The luminosity then decays as a byproduct of the reduction in the shock velocity. For the Gaussian overdensity, the luminosity then sharply increases again after the shock has expanded by a factor of $\sim e^{\Delta T} \simeq 3.6$, where $\Delta T \simeq 1.283$ is the time taken for a sound wave to traverse the interior of the blast (see Table \ref{tab:DeltaT}), and at this time the shock is well past the maximum in the overdensity. The luminosity associated with the square bump abruptly declines at the outer-most edge of the bump owing to the reduction in the density, and is then quickly followed by a sudden rebrightening (though the luminosity relative to the Sedov-Taylor value is still less than one) as the wave from the first encounter with the bump hits the shock from behind. The luminosity slowly increases in amplitude again until displaying a sudden reduction, the latter feature imparted by the wave induced by the outer edge of the shell intersecting the shock. 

If the dense shell that the shock encounters has a width that is less than the shock radius, the behavior exhibited by the shock luminosity in this figure is the general response that we expect -- the shock should respond immediately to the overdensity and then exhibit a sudden change in brightness at a time delayed by a factor of $\sim 3.6$ in terms of the shock expansion, where the factor of 3.6 comes from Table \ref{tab:DeltaT} and is for the specific case of a constant-density ambient medium. To turn this dimensionless factor into a physical time, we need to set both a length scale for the shock ($R_0$), an energy scale for the explosion, and the total mass swept up by the ejecta by the time the shock has reached $R_0$. Adopting fiducial values for these parameters of $R_0 = 10^{14}$ cm, $E_0 = 10^{50}$ erg, and $M_0 = 1M_{\odot}$ and letting $\gamma = 5/3$ and $n = 0$, the velocity scale of the shock is

\begin{equation}
    V_0 \simeq \sqrt{4.07 E_0/(3M_0)} \simeq 2.6\times 10^{3}\textrm{ km s}^{-1}.
\end{equation}
Here the factor of 4.07 comes from the dimensionless, integrated energy behind the blast, i.e., from the fact that

\begin{equation}
    \int_0^{1}\left(\frac{1}{2}g_0f_0^2+\frac{1}{\gamma-1}h_0\right)\xi^2d\xi \simeq 0.245
\end{equation}
for $n = 0$ and $\gamma = 5/3$. The time in between the initial response -- when the shell is in the immediate vicinity of the shock -- and the delayed response -- well after the shock has passed the overdensity -- is approximately given by the time taken by the Sedov solution to expand by a factor $e^{\Delta T}$, which is

\begin{equation}
    \Delta t = \frac{2}{5}\frac{R_0}{V_0}\left(\left(\frac{R}{R_0}\right)^{5/2}-1\right) \simeq 42 \left(\frac{R_0}{10^{14}\textrm{ cm}}\right)\left(\frac{V_0}{2.6\times 10^{3}\textrm{ km s}^{-1}}\right)^{-1} \textrm{ d}. \label{Deltat}
\end{equation}
By this time the shock has expanded to a position of $R \simeq 3.6\times 10^{14}$ cm. Note that this timescale depends only on the radius at which the shock encounters the shell, $R_0$, and the shock velocity at that time, $V_0$, and not on the specific shape or other details of the ambient overdensity (provided that the width of the overdensity or, more specifically, the distance over which the shell increses in density, is $\lesssim R_0$). However, as evidenced by Figure \ref{fig:Lsh}, the behavior of the lightcurve in between the initial and delayed rebrightening does depend on the specific nature of the overdensity.

Equation \eqref{Deltat} was derived for the specific case of a constant-density ambient medium. When the ambient density profile scales as $\rho \propto r^{-n}$, then the more general expression for the time between when the shock first encounters the overdensity (at shock radius $R_0$) and the secondary rebrightening occurs that arises from the Sedov solution for the shock radius as a function of time is 

\begin{equation}
    \Delta t = \frac{2}{5-n}\frac{R_0}{V_0}\left(\left(\frac{R}{R_0}\right)^{\frac{5-n}{2}}-1\right), \label{DeltatSedov}
\end{equation}
and at this time the shock has expanded by a factor of $e^{\Delta T}$ with $\Delta T$ a function of $n$ and provided in Table \ref{tab:DeltaT}. Interestingly, there are two competing effects that modify this timescale as a function of $n$, the first of which is that for small $n$ the value of $\Delta T$ is smaller (see Table \ref{tab:DeltaT}). This behavior arises from the fact that the dimensionless sound-crossing time is shorter for smaller $n$, and more of the interior of the blastwave is a high-pressure, low-density region (see Figure \ref{fig:sedov}). The second effect, which offsets the larger-$\Delta T$ trend, is that the factor of $(5-n)/2$ decreases as $n$ increases, and occurs because the shock decelerates less rapidly as $n$ increases (e.g., for $n = 0$ $R\propto t^{2/5}$, while for $n = 2$, $R\propto t^{2/3}$). The left panel of Figure\ref{fig:DeltaT_of_n} shows the product $\Delta T\times\left(5-n\right)/2$ as a function of $n$ for $\gamma = 5/3$ (orange, solid curve) and $\gamma = 4/3$ (purple, dashed curve), and demonstrates that for small $n$, these two effects nearly cancel one another, resulting in a shallow rise of $\Delta T\times(5-n)/2$ with $n$. For $\gamma = 5/3$, the fact that the solution is causally disconnected as $n$ approaches 2 implies that the sound-crossing time increases in this limit, and correspondingly the amount by which the shock expands increases enormously by the time the sound wave traverses the post-shock flow (recall that the shock expands exponentially with the quantity plotted in the left panel of this figure).

As we noted above, a specific supernova that showed anomalous rebrightening -- and did not exhibit the formation of narrow spectral lines -- is SN 2019tsf \citep{sollerman20}. SN 2019tsf was found on the decline and was sparsely covered initially, and hence the time of the maximum maximum optical luminosity is unknown, but the peak in the rebrightening occurred at roughly 75 days post-discovery (see Figure 1 of \citealt{sollerman20}). The right panel of Figure \ref{fig:DeltaT_of_n} shows the shock luminosity (arbitrarily scaled) as a function of time in days for the same, Gaussian bump as in Figure \ref{fig:Lsh} (which is for a constant-density background, i.e., $n = 0$, and a gas-pressure dominated equation of state with $\gamma = 5/3$), but here we set the magnitude of the overdensity to $\delta\rho/\rho = 1.5$ (i.e., a shell with a density enhancement of order unity). Here we also set $2/5\times R_0 / V_0 = 8$ days as the fundamental timescale entering the Sedov blastwave evolution (see Equation \ref{DeltatSedov}); for a shock scale radius of $R_0 = 5\times10^{14}$ cm, this corresponds to a characteristic shock velocity of $V_0 \simeq 3\times 10^{3}$ km s$^{-1}$. The time $t = 0$ is defined to coincide with when the shock is at a radius $R_0$.

In addition to showing good, qualitative agreement between the model proposed here and the optical lightcurve evolution of SN 2019tsf, this figure also demonstrates that the first peak in the lightcurve -- when the shock first encounters the density enhancements -- rises and fades over a short timescale (in the figure it occurs over a timescale of $\sim 1$ day around $t = 0$). Thus, this first brightening of the source could have easily been missed during the monitoring of SN 2019tsf, as the event was much more sparsely observed at early times when it was not doing anything interesting. In general, however, this is a prediction of the model -- the blastwave should show variation upon first encountering the density enhancement, though it may rise and fade on a shorter timescale, and in general be harder to detect, than the secondary rebrightening that arises from the internal reflection of the sound wave generated from the first interaction.

\subsection{Qualitative differences between this model and observations and more realistic scenarios}
The approach we have taken in this paper is to generate a strong explosion via a $\sim$ impulsive injection of energy that launches a strong blastwave into an assumed power-law ambient medium (without the perturbation). As such, the shock is constantly interacting with the ambient medium and converting the kinetic energy of the ejecta into thermal and radiated energy (though the latter is not explicitly accounted for in the model in that the back reaction of radiative losses is not included on the hydrodynamic evolution of the fluid; this is a valid assumption until much later times when the total radiated energy is a significant fraction of the initial energy). The interaction with an inhomogeneity in the ambient fluid then generates an instantaneous change in the shock properties, i.e., immediately upon encountering the density enhancement, and also at a delayed time. The delayed reaction occurs because, in the limit that the change in the density profile is small relative to the power-law background, the interaction creates a wave-like disturbance that propagates back through the shocked fluid, reflects off of the origin, and returns to the shock front in a finite time. The interaction between the wave and the shock then generates a distinct change in the shock properties that can occur in the absence of any density enhancement in the ambient gas at that time.

This picture is qualitatively distinct from the usual interaction picture of supernovae, where the outer envelope of the star -- ejected by the passage of a shock and expanding $\sim$ homologously -- initially propagates freely and suddenly encounters a circumstellar shell of material \citep{chevalier82}. The interaction between the ejecta and the shell then provides a distinct mechanism for tapping into the kinetic energy of the outflowing material and the source brightens as the radioactive decay energy is supplemented by the shock-heated emission. How the photons ultimately escape from the dense shell and reach the observer depends on the detailed geometry of the circumstellar material and the orientation of the observer viewing angle with respect to the source (e.g., \citealt{andrews18}). 

We emphasize, however, that while the phenomenological picture we have developed and employed here is distinct from the standard, interacting-supernova paradigm, the physical response of the supernova ejecta should generate a similar, delayed reaction to what we described here. In the limit that the ejecta is cold and described by a power-law with distance from the explosion center, the reverse shock (and contact discontinuity) propagates into the supernova ejecta at a well-defined rate that yields the self-similar expansion of the reverse shock-contact discontinuity-forward shock structure \citep{chevalier82}. However, eventually the reverse shock will decouple from the forward shock as it encounters the material that initially comprised the stellar interior, and will propagate at a faster rate through the more highly pressurized material and will likely decrease in Mach number. As the reverse shock reflects off the origin, it will propagate back to the reverse shock front and induce a similar effect to the one outlined here. The details of the length of time over which this occurs clearly depends on the structure of the interior of the blast and the nature of the spent ejecta -- the ``dross'' left behind from the supernova that inhabits the region near the explosion site -- but we expect the same qualitative behavior.

Furthermore, it should be straightforward to consider and analyze the same problem as outlined here, but with the background solution given by the self-similar, expanding shell described in \citet{chevalier82}. As we noted above, doing so merely consists of changing the value of $\alpha_0$ that appears in Equation \eqref{unperts}. In this case one could consider a bump within the shell, or a ``shell within a shell,'' that would launch a wave that would partially reflect off of the contact discontinuity and interact with the forward shock. It is likely the case that the timescales would all be significantly reduced as compared to the scenario in which the wave has to traverse the entire, post-shock blast.

In general, we expect the Sedov-Taylor regime of shock propagation to hold once the ambient mass swept up by the blastwave is comparable to the initial mass involved in the explosion; prior to this time the speed of the forward shock should be comparable to $V_{\rm ej} \simeq \sqrt{E/M_{\rm ej}}$, where $M_{\rm ej}$ is the ejecta mass and $E$ is the energy involved in the explosion. Before this time the post-shock variation of the fluid variables could display much more deviation from the Sedov-Taylor solution. Setting $E = 10^{50}$ erg and $M_{\rm ej} = 0.1 M_{\odot}$, the (assumed-constant) ambient density would be $\rho_{\rm a} \simeq 4.7\times10^{-10}$ g cm$^{-3}$ if the radius at which the swept-up mass equals the initial ejecta mass is $R = 10^{14}$ cm, while the ambient density would need to be $\sim 4.7\times 10^{-13}$ g cm$^{-3}$ if $R = 10^{15}$ cm. Such high densities could conceivably be attained during periods of extreme mass loss toward the end of the life of the star. We reiterate, however, that the qualitative response of the blastwave should be similar to what we described here, independent of the background state (i.e., Sedov-Taylor or ejecta-driven), in that the reverse wave generated by the interaction between the shock and the ambient density perturbation will intersect the shock at a later time and generate a rebrightening of the source in the absence of ambient density enhancements.  

\section{Summary and Conclusions}
\label{sec:summary}
We analyzed the response of a strong, energy-conserving shockwave (i.e., the Sedov-Taylor blastwave) to spherically symmetric changes in the density profile of the ambient medium into which the shock advances; these changes induce non-self-similar behavior in the propagation of the shock itself and the post-shock fluid variables (the density, velocity, and pressure). Our methodology treated the changes in the density as perturbations on top of an otherwise power-law decline of the ambient density, which allowed us to derive a set of linearized equations from the fluid equations (Section \ref{sec:equations}). Upon taking the Laplace transform of that set of equations, we analyzed the eigenvalues (Section \ref{sec:eigenmodes}) and found that (in agreement with the recent results of \citealt{sanz16}) there is a continuum of modes in addition to a discrete set. For a constant-density ambient medium, the discrete eigenvalue with the largest real part for $\gamma = 5/3$ ($\gamma = 4/3$) was found to be $\sigma = -2.77+7.05i$ ($\sigma = -2.36+9.58i$). On the other hand, for a wind-like medium, there was only a continuum of poles with a real part around $\sigma \simeq -0.8$ when $\gamma = 5/3$, while for $\gamma = 4/3$ the largest, discrete eigenvalue was $\sigma \simeq -1.01+4.48i$; see Table \ref{tab:eigenvalues} and Figure \ref{fig:poles}. 

Because the eigenmodes are characterized by a continuum, they are not particularly useful for numerically reconstructing the response of a blastwave to density inhomogeneities. We showed, however, that it is also possible to reconstruct the response of a blastwave to an arbitrary density perturbation by numerically calculating the Laplace-transformed solutions for the fluid variables and the shock acceleration along the imaginary axis, i.e., with the Laplace-transformed variable $\sigma$ equal to an imaginary number. Because the Laplace-transformed fluid equations are able to be normalized by the magnitude of the Laplace-transformed density perturbation, these solutions (as shown in Figure \ref{fig:Realpha} and \ref{fig:funsofn}) are valid for any given density perturbation; the response to a particular density perturbation can then be calculated by convolving the solution with the specific perturbation to the density. We considered a number of such specific perturbations in Section \ref{sec:examples}. We delinated the astrophysical implications of our findings in Section \ref{sec:implications}, and in particular calculated the shock luminosity and compared a specific example to, and found good agreement with, the supernova SN 2019tsf (see Figure \ref{fig:DeltaT_of_n} and Figure 1 of \citealt{sollerman20}).

We summarize our main results here:

\begin{enumerate}
    \item The time-like variable $\tau$ that results in the separability of the fluid equations (and hence their ability to be Laplace transformed and analyzed through standard linear perturbation theory) is $\tau = \ln R/R_0$, where $R$ is the shock position and $R_0$ is the shock position at some initial time. As a consequence, the perturbations to the shockwave that arise from interactions with density inhomogeneities vary logarithmically with time and can therefore be long lived. 
    \item Causally connected blastwaves possess a finite sound-crossing time $\Delta T$ (where by time we mean the time-like variable $\tau = \ln R/R_0$), that for $n = 0$ and $\gamma = 5/3$ ($\gamma = 4/3$) is $\Delta T = 1.283$ ($\Delta T = 0.886$). The existence of a finite sound-crossing time and the solution to the Riemann problem implies that any sudden (i.e., nearly or effectively discontinuous) changes in the density of the ambient medium propagate through the post-shock fluid on a timescale of $\Delta T$, over which the shockwave expands by a factor of $e^{\Delta T}$. Therefore, the shock properties not only change abruptly upon encountering any sudden change in density, but also after the shock expands by a factor of $e^{\Delta T}$ as the shock is hit \emph{from behind} by a discontinuity. This feature is illustrated most clearly in the left panel of Figure \ref{fig:abs_alpha}, the top row of Figure \ref{fig:eps_rectangle}, the top row of Figure \ref{fig:alpha_tri}, and Figure \ref{fig:Lsh}. This can also be seen as a function of time in the right panel of Figure \ref{fig:DeltaT_of_n}, which can be compared to the right panel of Figure 1 in \citet{sollerman20}, which illustrates the optical lightcurve evolution of the narrow line-less type Ib supernova SN 2019tsf.
    \item The solution for a wind-like medium when the post-shock flow is gas-pressure dominated ($n = 2$, $\gamma = 5/3$, also known as the ``Primakoff blastwave'') is causally disconnected, such that $\Delta T = \infty$. The declining sound speed of the post-shock flow implies that any sudden changes to the ambient medium are not communicated back to the shock in a finite time, and hence, according to the linear analysis, the blastwave only displays changes in its properties upon encountering the density perturbation. This property is shown most clearly in the bottom three panels of Figure \ref{fig:eps_rectangle} and Figure \ref{fig:alpha_tri}.
    \item The declining sound speed of the post-shock flow of the Sedov-Taylor blastwave dominated by gas pressure and expanding into a wind-like medium (specifically, the sound speed declines as $\propto r$ with $r$ the spherical radius from the center of the blast) implies that the magnitude of any perturbation to the fluid variables (seeded by a perturbation to the ambient density) is amplified as it propagates inward, which would result in the formation of a shock in a nonlinear analysis. This feature is shown most clearly in the right panel of Figure \ref{fig:vel_profile} and in Figure \ref{fig:vel_n0_n2}. The results of the one-dimensional, hydrodynamical simulation shown in Figure \ref{fig:vel_n0_n2} also demonstrate directly that any inward-propagating wave, generated by a small density perturbation, steepens into a shock (note the discontinuity that has formed around $r = 1$ at the latest time in this figure).
    \item The existence of discrete eigenvalues, one of which has a largest real part, implies that the late-time behavior of the blastwave will always be characterized by that eigenvalue. However, the coefficient multiplying that eigenvalue in the eigenmode expansion depends on the nature of the density perturbation, meaning that the largest eigenvalue may not dominate the evolution until very late times (in $\ln R$). This behavior is shown most clearly in the left panel of Figure \ref{fig:abs_alpha}. 
\end{enumerate}

A consequence of the combination of points 1 and 5 is that, when the single-eigenvalue behavior dominates the evolution of the shock (which can arise, as is apparent from the top-left panel of Figure \ref{fig:abs_alpha}), any perturbation to the shock properties varies as $\propto e^{\sigma\tau} \propto R^{\sigma}$. Since the solution is approximately given by the Sedov-Taylor blastwave, this implies that perturbations to the blast vary in time as $\propto t^{2\sigma/5}$. For the values of $\sigma$ recovered for $\gamma = 5/3$ and $\gamma = 4/3$ (see Table \ref{tab:eigenvalues}), this implies that the perturbations decay in time as $\propto t^{-1.1+2.8 i}$ and $\propto t^{-0.94+3.8i}$, respectively. The fact that these only decay as fairly weak (but nonetheless oscillatory) power-laws implies that they can impact the behavior of the blastwave over long timescales.

Point 2 suggests that even when there is no enhancement in the density of the ambient medium at the location of the shock and consequently no narrow line emission, the shock may exhibit variations in its observable properties, and Figure \ref{fig:DeltaT_of_n} shows good qualitative agreement with the observations of supernova SN 2019tsf (Figure 1 of \citealt{sollerman20}). We note, however, that the optical depth of the gas may prevent the radiation from escaping the shock if it is still optically thick. In particular, since the speed of a photon in an optically thick gas is $c_{\rm eff} \simeq c/\tau$, where $\tau$ is the optical depth as measured from infinity, it follows that any photons radiated from the immediate vicinity of the shock can only successfully escape if $V \lesssim c/\tau$, i.e., if the optical depth satisfies $\tau \lesssim c/V$ (e.g., \citealt{sapir11}). If the gas in front of the blastwave is still highly optically thick, then of course there is very little observational consequence of the interaction at all (or, indeed, the presence of the blastwave itself). In general, however, the presence of a finite optical depth implies that the photons will ``leak out'' of the shock over some finite time $\simeq R\tau/c$. Additionally, even if all of the radiation could be emitted instantaneously from the shock surface at the moment it encounters a discontinuous density jump, the \emph{received} emission would still be smeared out over the light-crossing time of the shock radius $\sim R/c$ \citep{ensman92, fernandez18}. 

Nevertheless, it seems probable that prior interactions between a shock and ambient density variations can impact the observational characteristics of a blastwave at a later time, when there may be no modifications to the ambient medium. Such a scenario would imply a re-brightening of the blastwave, as the post-shock pressure increases as a consequence of being hit by the discontinuity from behind, but -- since there are no enhancements to the density of the ambient medium -- there would be no associated increase in narrow line emission. As we have already noted, precisely this behavior was seen in the stripped-envelope (type Ib) supernova SN 2019tsf, as described in \citet{sollerman20}. Delayed re-brightenings also could partially explain the highly erratic and extreme behavior of iPTF14hls, which exhibited at least 5 surges in luminosity over a period of $\sim 500$ days in which the transient was bright \citep{arcavi17}; similar arguments can be made for the more recent event SN2020faa, which also displayed numerous re-brightenings \citep{yang20}.

Point 3 implies that a gas-pressure-dominated shockwave propagating through a wind-like medium should not display any secular (i.e., not immediately related to the presence of highly localized density changes) deviations as a consequence of the infinite sound-crossing time of the post-shock fluid. However, Point 4 shows that the \emph{nonlinear} behavior of the perturbations becomes more important as they propagate into the post-shock flow. Once the disturbances amplify to the point where the velocity perturbation is comparable to the local sound speed, the speed at which the disturbances propagate will approach a finite value. As such, the sound crossing time of the blastwave is finite when nonlinear effects are included, but will still be asymptotically large as the magnitude of the perturbation goes to zero (as in this case it will take infinitely long for the amplitude of the velocity perturbation to grow to the sound speed and, hence, steepen into a shock).

The steepening of arbitrarily small perturbations into shockwaves, as noted in Point 4, will be hindered by any finite amount of dissipation. In particular, the presence of viscosity will cause the wave to steepen less rapidly, and could result in its dissolution if the rate of dissipation is larger than the rate at which the wave steepens. Nonetheless, the fact that waves steepen in the interior of the Primakoff blastwave suggests that the interiors of these flows should be highly chaotic and disordered. Indeed, this is another interpretation of the erratic behavior of the velocity and density in Figure \ref{fig:funsofn} for $n = 2$ (shown in the bottom row): these solutions correspond to a single value of $\sigma$, which as described in Section \ref{sec:eigenmodes}, are appropriate to the scenario in which the ambient density perturbation is sinusoidally varying for all $\tau \in \{-\infty,\infty\}$. Thus, the velocity profile in the interior consists of a series of waves launched into the interior as the blastwave encounters the ripples in the density of the ambient medium, each of which steepens as it propagates into the interior. If nonlinear effects were included, all of these pulses would steepen into a train of shocks, which would propagate through the interior and interact with one another. The right panel of Figure \ref{fig:vel_n0_n2} shows that, when nonlinear effects are included (as in the simulation results in this figure), small-amplitude waves do steepen into shocks when there is no (explicit, i.e., non-numerical) dissipation.

In all of our analysis, we considered perturbations to the post-shock flow that were induced by changes in the ambient density profile, and in which the initial (i.e., pre-density perturbation) blastwave was exactly characterized by the Sedov-Taylor solution. We could use precisely the same formalism to understand the evolution of the blastwave in response to any \emph{initial} deviations in the post-shock fluid variables from the Sedov-Taylor blastwave. This can be accomplished simply by letting the right-hand side of any of Equations \eqref{L1} -- \eqref{L3} be functions of $\xi$, which would represent the initial deviation of the density (Equation \ref{L1}), velocity (Equation \ref{L2}), or entropy (Equation \ref{L3}) from the Sedov-Taylor solution. Unlike the case of an ambient density perturbation, however, the dependence of the post-shock fluid variables on $\xi$ implies that there is a distinct solution for $\tilde{\alpha}$, $\tilde{f}_1$, etc., for any initial perturbation.

We only considered spherically symmetric perturbations to the ambient density profile and the corresponding response of the blastwave. Extending the analysis to the scenario in which the ambient density perturbations have an angular dependence has obvious relevance for more realistic scenarios, such as a ``clumpy'' stellar wind or convective envelope of a supergiant. The methodology for extending the analysis presented here to anisotropic perturbations is straightforward, with one of the main differences being that the perturbations to the shock no longer respect the invariance of the shock position to arbitrary renormalizations; this arises from the fact that the Sedov-Taylor blastwave has a well-defined spherical harmonic of $\ell = 0$. As such, the shock must be perturbed relative to a solution that respects the spherical symmetry and acceleration of the unperturbed blastwave, and there is no freedom to renormalize the $\ell \neq 0$ component of the shock position and velocity.

When the ambient density perturbation is not spherically symmetric, the shock can be unstable when the adiabatic index is sufficiently small and the spherical harmonic number is sufficiently high (i.e., \citealt{ryu87} show that the shock is unstable once $\gamma \lesssim 1.2$ and $\ell \gtrsim 10$). When the shock is not radiative, i.e., during the early stages of the supernova explosion, we expect angular perturbations to be stable from the standpoint that the shock will eventually return to a spherically symmetric state (assuming that the angular perturbations are spatially compact, i.e., that the ambient density profile is spherically symmetric at sufficiently large radii). However, it may be the case that the non-radial response of the blastwave to angularly confined perturbations is temporally ``smeared,'' as sound waves excited at different spherical harmonic $\ell$'s traverse the post-shock flow in varying amounts of time. The bulk response of the blast to a non-spherically-symmetric, ambient density perturbation could therefore be reduced relative to a spherically symmetric perturbation, as considered here. We plan to return to the asymmetric limit in future work.

Finally, even though the perturbation equations we derived (Equations \ref{contpert} -- \ref{entpert}) can be applied to any shock solution that behaves as a power-law in time (through Equation \ref{Vss}), the analysis of the perturbation equations that we performed in this paper was specific to the case of energy-conserving explosions, the self-similar solution for which is the Sedov-Taylor blastwave. However, the early stages of the propagation of the forward shock into its ambient environment (as generated by a core-collapse supernova) can also be described by an interaction regime, in which the forward shock is advanced at a faster rate than the energy-conserving prediction by a contact discontinuity that separates the shocked ambient gas from the shocked supernova ejecta \citep{chevalier82}. While this case should be studied in detail in a future investigation, many of the same general conclusions can be reached. In particular, when a shock that is in the interaction regime encounters a density perturbation, deviations to the post-shock flow will be communicated back through the fluid on the sound-crossing time appropriate to that self-similar solution. There is then a characteristic timescale $\Delta T$ over which sharp changes in the ambient density will propagate through the post-shock flow, reflect off of the contact discontinuity, and impact the forward shock from behind. Since the post-shock gas is confined to a thin shell during this regime, the dimensionless sound crossing time will be shorter, and we expect the shock to show corresponding deviations over shorter timescales.

\acknowledgements
E.R.C.~acknowledges support from the National Science Foundation through grant AST-2006684. We thank the anonymous referee for useful comments and suggestions.


\begin{thebibliography}{}
\expandafter\ifx\csname natexlab\endcsname\relax\def\natexlab#1{#1}\fi
\providecommand{\url}[1]{\href{#1}{#1}}
\providecommand{\dodoi}[1]{doi:~\href{http://doi.org/#1}{\nolinkurl{#1}}}
\providecommand{\doeprint}[1]{\href{http://ascl.net/#1}{\nolinkurl{http://ascl.net/#1}}}
\providecommand{\doarXiv}[1]{\href{https://arxiv.org/abs/#1}{\nolinkurl{https://arxiv.org/abs/#1}}}

\bibitem[{{Andrews} \& {Smith}(2018)}]{andrews18}
{Andrews}, J.~E., \& {Smith}, N. 2018, \mnras, 477, 74,
  \dodoi{10.1093/mnras/sty584}

\bibitem[{{Arcavi} {et~al.}(2017){Arcavi}, {Howell}, {Kasen}, {Bildsten},
  {Hosseinzadeh}, {McCully}, {Wong}, {Katz}, {Gal-Yam}, {Sollerman}, {Taddia},
  {Leloudas}, {Fremling}, {Nugent}, {Horesh}, {Mooley}, {Rumsey}, {Cenko},
  {Graham}, {Perley}, {Nakar}, {Shaviv}, {Bromberg}, {Shen}, {Ofek}, {Cao},
  {Wang}, {Huang}, {Rui}, {Zhang}, {Li}, {Li}, {Zhang}, {Valenti}, {Guevel},
  {Shappee}, {Kochanek}, {Holoien}, {Filippenko}, {Fender}, {Nyholm}, {Yaron},
  {Kasliwal}, {Sullivan}, {Blagorodnova}, {Walters}, {Lunnan}, {Khazov},
  {Andreoni}, {Laher}, {Konidaris}, {Wozniak}, \& {Bue}}]{arcavi17}
{Arcavi}, I., {Howell}, D.~A., {Kasen}, D., {et~al.} 2017, \nat, 551, 210,
  \dodoi{10.1038/nature24030}

\bibitem[{{Bellm}(2014)}]{bellm14}
{Bellm}, E. 2014, in The Third Hot-wiring the Transient Universe Workshop, ed.
  P.~R. {Wozniak}, M.~J. {Graham}, A.~A. {Mahabal}, \& R.~{Seaman}, 27--33

\bibitem[{{Bernstein} \& {Book}(1980)}]{bernstein80}
{Bernstein}, I.~B., \& {Book}, D.~L. 1980, \apj, 240, 223,
  \dodoi{10.1086/158226}

\bibitem[{{Blondin} {et~al.}(1996){Blondin}, {Lundqvist}, \&
  {Chevalier}}]{blondin96}
{Blondin}, J.~M., {Lundqvist}, P., \& {Chevalier}, R.~A. 1996, \apj, 472, 257,
  \dodoi{10.1086/178060}

\bibitem[{{Burrows}(1988)}]{burrows88}
{Burrows}, A. 1988, \apj, 334, 891, \dodoi{10.1086/166885}

\bibitem[{{Burrows} \& {Lattimer}(1986)}]{burrows86}
{Burrows}, A., \& {Lattimer}, J.~M. 1986, \apj, 307, 178,
  \dodoi{10.1086/164405}

\bibitem[{{Chambers} {et~al.}(2016){Chambers}, {Magnier}, {Metcalfe},
  {Flewelling}, {Huber}, {Waters}, {Denneau}, {Draper}, {Farrow}, {Finkbeiner},
  {Holmberg}, {Koppenhoefer}, {Price}, {Rest}, {Saglia}, {Schlafly}, {Smartt},
  {Sweeney}, {Wainscoat}, {Burgett}, {Chastel}, {Grav}, {Heasley}, {Hodapp},
  {Jedicke}, {Kaiser}, {Kudritzki}, {Luppino}, {Lupton}, {Monet}, {Morgan},
  {Onaka}, {Shiao}, {Stubbs}, {Tonry}, {White}, {Ba{\~n}ados}, {Bell},
  {Bender}, {Bernard}, {Boegner}, {Boffi}, {Botticella}, {Calamida},
  {Casertano}, {Chen}, {Chen}, {Cole}, {Deacon}, {Frenk}, {Fitzsimmons},
  {Gezari}, {Gibbs}, {Goessl}, {Goggia}, {Gourgue}, {Goldman}, {Grant},
  {Grebel}, {Hambly}, {Hasinger}, {Heavens}, {Heckman}, {Henderson}, {Henning},
  {Holman}, {Hopp}, {Ip}, {Isani}, {Jackson}, {Keyes}, {Koekemoer}, {Kotak},
  {Le}, {Liska}, {Long}, {Lucey}, {Liu}, {Martin}, {Masci}, {McLean}, {Mindel},
  {Misra}, {Morganson}, {Murphy}, {Obaika}, {Narayan}, {Nieto-Santisteban},
  {Norberg}, {Peacock}, {Pier}, {Postman}, {Primak}, {Rae}, {Rai}, {Riess},
  {Riffeser}, {Rix}, {R{\"o}ser}, {Russel}, {Rutz}, {Schilbach}, {Schultz},
  {Scolnic}, {Strolger}, {Szalay}, {Seitz}, {Small}, {Smith}, {Soderblom},
  {Taylor}, {Thomson}, {Taylor}, {Thakar}, {Thiel}, {Thilker}, {Unger},
  {Urata}, {Valenti}, {Wagner}, {Walder}, {Walter}, {Watters}, {Werner},
  {Wood-Vasey}, \& {Wyse}}]{chambers16}
{Chambers}, K.~C., {Magnier}, E.~A., {Metcalfe}, N., {et~al.} 2016, arXiv
  e-prints, arXiv:1612.05560.
\newblock \doarXiv{1612.05560}

\bibitem[{{Chatzopoulos} {et~al.}(2016){Chatzopoulos}, {Wheeler}, {Vinko},
  {Nagy}, {Wiggins}, \& {Even}}]{chatzoploulos16}
{Chatzopoulos}, E., {Wheeler}, J.~C., {Vinko}, J., {et~al.} 2016, \apj, 828,
  94, \dodoi{10.3847/0004-637X/828/2/94}

\bibitem[{{Chen} {et~al.}(2018){Chen}, {Inserra}, {Fraser}, {Moriya}, {Schady},
  {Schweyer}, {Filippenko}, {Perley}, {Ruiter}, {Seitenzahl}, {Sollerman},
  {Taddia}, {Anderson}, {Foley}, {Jerkstrand}, {Ngeow}, {Pan}, {Pastorello},
  {Points}, {Smartt}, {Smith}, {Taubenberger}, {Wiseman}, {Young}, {Benetti},
  {Berton}, {Bufano}, {Clark}, {Della Valle}, {Galbany}, {Gal-Yam},
  {Gromadzki}, {Guti{\'e}rrez}, {Heinze}, {Kankare}, {Kilpatrick},
  {Kuncarayakti}, {Leloudas}, {Lin}, {Maguire}, {Mazzali}, {McBrien},
  {Prentice}, {Rau}, {Rest}, {Siebert}, {Stalder}, {Tonry}, \& {Yu}}]{chen18}
{Chen}, T.~W., {Inserra}, C., {Fraser}, M., {et~al.} 2018, \apjl, 867, L31,
  \dodoi{10.3847/2041-8213/aaeb2e}

\bibitem[{{Chevalier}(1982)}]{chevalier82}
{Chevalier}, R.~A. 1982, \apj, 258, 790, \dodoi{10.1086/160126}

\bibitem[{{Chevalier}(1989)}]{chevalier89}
---. 1989, \apj, 346, 847, \dodoi{10.1086/168066}

\bibitem[{{Chevalier}(1990)}]{chevalier90}
---. 1990, \apj, 359, 463, \dodoi{10.1086/169078}

\bibitem[{{Chevalier} \& {Fransson}(1994)}]{chevalier94}
{Chevalier}, R.~A., \& {Fransson}, C. 1994, \apj, 420, 268,
  \dodoi{10.1086/173557}

\bibitem[{{Chugai} \& {Danziger}(1994)}]{chugai94}
{Chugai}, N.~N., \& {Danziger}, I.~J. 1994, \mnras, 268, 173,
  \dodoi{10.1093/mnras/268.1.173}

\bibitem[{{Coughlin}(2020)}]{coughlin20}
{Coughlin}, E.~R. 2020, \mnras, 496, L43, \dodoi{10.1093/mnrasl/slaa087}

\bibitem[{{Coughlin} {et~al.}(2018{\natexlab{a}}){Coughlin}, {Quataert},
  {Fern{\'a}ndez}, \& {Kasen}}]{coughlin18}
{Coughlin}, E.~R., {Quataert}, E., {Fern{\'a}ndez}, R., \& {Kasen}, D.
  2018{\natexlab{a}}, \mnras, 477, 1225, \dodoi{10.1093/mnras/sty667}

\bibitem[{{Coughlin} {et~al.}(2018{\natexlab{b}}){Coughlin}, {Quataert}, \&
  {Ro}}]{coughlin18b}
{Coughlin}, E.~R., {Quataert}, E., \& {Ro}, S. 2018{\natexlab{b}}, \apj, 863,
  158, \dodoi{10.3847/1538-4357/aad198}

\bibitem[{{Coughlin} {et~al.}(2019){Coughlin}, {Ro}, \&
  {Quataert}}]{coughlin19b}
{Coughlin}, E.~R., {Ro}, S., \& {Quataert}, E. 2019, \apj, 874, 58,
  \dodoi{10.3847/1538-4357/ab09ec}

\bibitem[{{Cox}(1980)}]{cox80}
{Cox}, J.~P. 1980, {Theory of stellar pulsation}

\bibitem[{{Crowther}(2007)}]{crowther07}
{Crowther}, P.~A. 2007, \araa, 45, 177,
  \dodoi{10.1146/annurev.astro.45.051806.110615}

\bibitem[{{Ensman} \& {Burrows}(1992)}]{ensman92}
{Ensman}, L., \& {Burrows}, A. 1992, \apj, 393, 742, \dodoi{10.1086/171542}

\bibitem[{{Fern{\'a}ndez} {et~al.}(2018){Fern{\'a}ndez}, {Quataert},
  {Kashiyama}, \& {Coughlin}}]{fernandez18}
{Fern{\'a}ndez}, R., {Quataert}, E., {Kashiyama}, K., \& {Coughlin}, E.~R.
  2018, \mnras, 476, 2366, \dodoi{10.1093/mnras/sty306}

\bibitem[{{Filippenko}(1989)}]{filippenko89}
{Filippenko}, A.~V. 1989, \aj, 97, 726, \dodoi{10.1086/115018}

\bibitem[{{Gal-Yam} {et~al.}(2009){Gal-Yam}, {Mazzali}, {Ofek}, {Nugent},
  {Kulkarni}, {Kasliwal}, {Quimby}, {Filippenko}, {Cenko}, {Chornock},
  {Waldman}, {Kasen}, {Sullivan}, {Beshore}, {Drake}, {Thomas}, {Bloom},
  {Poznanski}, {Miller}, {Foley}, {Silverman}, {Arcavi}, {Ellis}, \&
  {Deng}}]{gal-yam09}
{Gal-Yam}, A., {Mazzali}, P., {Ofek}, E.~O., {et~al.} 2009, \nat, 462, 624,
  \dodoi{10.1038/nature08579}

\bibitem[{{Goodman}(1990)}]{goodman90}
{Goodman}, J. 1990, \apj, 358, 214, \dodoi{10.1086/168977}

\bibitem[{{Ivezi{\'c}} {et~al.}(2019){Ivezi{\'c}}, {Kahn}, {Tyson}, {Abel},
  {Acosta}, {Allsman}, {Alonso}, {AlSayyad}, {Anderson}, {Andrew}, {Angel},
  {Angeli}, {Ansari}, {Antilogus}, {Araujo}, {Armstrong}, {Arndt}, {Astier},
  {Aubourg}, {Auza}, {Axelrod}, {Bard}, {Barr}, {Barrau}, {Bartlett}, {Bauer},
  {Bauman}, {Baumont}, {Bechtol}, {Bechtol}, {Becker}, {Becla}, {Beldica},
  {Bellavia}, {Bianco}, {Biswas}, {Blanc}, {Blazek}, {Blandford}, {Bloom},
  {Bogart}, {Bond}, {Booth}, {Borgland}, {Borne}, {Bosch}, {Boutigny},
  {Brackett}, {Bradshaw}, {Brandt}, {Brown}, {Bullock}, {Burchat}, {Burke},
  {Cagnoli}, {Calabrese}, {Callahan}, {Callen}, {Carlin}, {Carlson},
  {Chandrasekharan}, {Charles-Emerson}, {Chesley}, {Cheu}, {Chiang}, {Chiang},
  {Chirino}, {Chow}, {Ciardi}, {Claver}, {Cohen-Tanugi}, {Cockrum}, {Coles},
  {Connolly}, {Cook}, {Cooray}, {Covey}, {Cribbs}, {Cui}, {Cutri}, {Daly},
  {Daniel}, {Daruich}, {Daubard}, {Daues}, {Dawson}, {Delgado}, {Dellapenna},
  {de Peyster}, {de Val-Borro}, {Digel}, {Doherty}, {Dubois},
  {Dubois-Felsmann}, {Durech}, {Economou}, {Eifler}, {Eracleous}, {Emmons},
  {Fausti Neto}, {Ferguson}, {Figueroa}, {Fisher-Levine}, {Focke}, {Foss},
  {Frank}, {Freemon}, {Gangler}, {Gawiser}, {Geary}, {Gee}, {Geha}, {Gessner},
  {Gibson}, {Gilmore}, {Glanzman}, {Glick}, {Goldina}, {Goldstein}, {Goodenow},
  {Graham}, {Gressler}, {Gris}, {Guy}, {Guyonnet}, {Haller}, {Harris},
  {Hascall}, {Haupt}, {Hernandez}, {Herrmann}, {Hileman}, {Hoblitt}, {Hodgson},
  {Hogan}, {Howard}, {Huang}, {Huffer}, {Ingraham}, {Innes}, {Jacoby}, {Jain},
  {Jammes}, {Jee}, {Jenness}, {Jernigan}, {Jevremovi{\'c}}, {Johns}, {Johnson},
  {Johnson}, {Jones}, {Juramy-Gilles}, {Juri{\'c}}, {Kalirai}, {Kallivayalil},
  {Kalmbach}, {Kantor}, {Karst}, {Kasliwal}, {Kelly}, {Kessler}, {Kinnison},
  {Kirkby}, {Knox}, {Kotov}, {Krabbendam}, {Krughoff}, {Kub{\'a}nek},
  {Kuczewski}, {Kulkarni}, {Ku}, {Kurita}, {Lage}, {Lambert}, {Lange},
  {Langton}, {Le Guillou}, {Levine}, {Liang}, {Lim}, {Lintott}, {Long},
  {Lopez}, {Lotz}, {Lupton}, {Lust}, {MacArthur}, {Mahabal}, {Mandelbaum},
  {Markiewicz}, {Marsh}, {Marshall}, {Marshall}, {May}, {McKercher}, {McQueen},
  {Meyers}, {Migliore}, {Miller}, {Mills}, {Miraval}, {Moeyens}, {Moolekamp},
  {Monet}, {Moniez}, {Monkewitz}, {Montgomery}, {Morrison}, {Mueller},
  {Muller}, {Mu{\~n}oz Arancibia}, {Neill}, {Newbry}, {Nief}, {Nomerotski},
  {Nordby}, {O'Connor}, {Oliver}, {Olivier}, {Olsen}, {O'Mullane}, {Ortiz},
  {Osier}, {Owen}, {Pain}, {Palecek}, {Parejko}, {Parsons}, {Pease},
  {Peterson}, {Peterson}, {Petravick}, {Libby Petrick}, {Petry},
  {Pierfederici}, {Pietrowicz}, {Pike}, {Pinto}, {Plante}, {Plate}, {Plutchak},
  {Price}, {Prouza}, {Radeka}, {Rajagopal}, {Rasmussen}, {Regnault}, {Reil},
  {Reiss}, {Reuter}, {Ridgway}, {Riot}, {Ritz}, {Robinson}, {Roby}, {Roodman},
  {Rosing}, {Roucelle}, {Rumore}, {Russo}, {Saha}, {Sassolas}, {Schalk},
  {Schellart}, {Schindler}, {Schmidt}, {Schneider}, {Schneider}, {Schoening},
  {Schumacher}, {Schwamb}, {Sebag}, {Selvy}, {Sembroski}, {Seppala}, {Serio},
  {Serrano}, {Shaw}, {Shipsey}, {Sick}, {Silvestri}, {Slater}, {Smith},
  {Smith}, {Sobhani}, {Soldahl}, {Storrie-Lombardi}, {Stover}, {Strauss},
  {Street}, {Stubbs}, {Sullivan}, {Sweeney}, {Swinbank}, {Szalay}, {Takacs},
  {Tether}, {Thaler}, {Thayer}, {Thomas}, {Thornton}, {Thukral}, {Tice},
  {Trilling}, {Turri}, {Van Berg}, {Vanden Berk}, {Vetter}, {Virieux},
  {Vucina}, {Wahl}, {Walkowicz}, {Walsh}, {Walter}, {Wang}, {Wang}, {Warner},
  {Wiecha}, {Willman}, {Winters}, {Wittman}, {Wolff}, {Wood-Vasey}, {Wu},
  {Xin}, {Yoachim}, \& {Zhan}}]{ivezic19}
{Ivezi{\'c}}, {\v{Z}}., {Kahn}, S.~M., {Tyson}, J.~A., {et~al.} 2019, \apj,
  873, 111, \dodoi{10.3847/1538-4357/ab042c}

\bibitem[{{Kuncarayakti} {et~al.}(2018){Kuncarayakti}, {Maeda}, {Ashall},
  {Prentice}, {Mattila}, {Kankare}, {Fransson}, {Lundqvist}, {Pastorello},
  {Leloudas}, {Anderson}, {Benetti}, {Bersten}, {Cappellaro}, {Cartier},
  {Denneau}, {Della Valle}, {Elias-Rosa}, {Folatelli}, {Fraser}, {Galbany},
  {Gall}, {Gal-Yam}, {Guti{\'e}rrez}, {Hamanowicz}, {Heinze}, {Inserra},
  {Kangas}, {Mazzali}, {Melandri}, {Pignata}, {Rest}, {Reynolds}, {Roy},
  {Smartt}, {Smith}, {Sollerman}, {Somero}, {Stalder}, {Stritzinger}, {Taddia},
  {Tomasella}, {Tonry}, {Weiland}, \& {Young}}]{kuncarayakti18}
{Kuncarayakti}, H., {Maeda}, K., {Ashall}, C.~J., {et~al.} 2018, \apjl, 854,
  L14, \dodoi{10.3847/2041-8213/aaaa1a}

\bibitem[{{Kurf{\"u}rst} {et~al.}(2020){Kurf{\"u}rst}, {Pejcha}, \&
  {Krti{\v{c}}ka}}]{kurfurst20}
{Kurf{\"u}rst}, P., {Pejcha}, O., \& {Krti{\v{c}}ka}, J. 2020, \aap, 642, A214,
  \dodoi{10.1051/0004-6361/202039073}

\bibitem[{{Kushnir} \& {Waxman}(2010)}]{kushnir10}
{Kushnir}, D., \& {Waxman}, E. 2010, \apj, 723, 10,
  \dodoi{10.1088/0004-637X/723/1/10}

\bibitem[{{Kushnir} {et~al.}(2005){Kushnir}, {Waxman}, \&
  {Shvarts}}]{kushnir05}
{Kushnir}, D., {Waxman}, E., \& {Shvarts}, D. 2005, \apj, 634, 407,
  \dodoi{10.1086/496871}

\bibitem[{{Law} {et~al.}(2009){Law}, {Kulkarni}, {Dekany}, {Ofek}, {Quimby},
  {Nugent}, {Surace}, {Grillmair}, {Bloom}, {Kasliwal}, {Bildsten}, {Brown},
  {Cenko}, {Ciardi}, {Croner}, {Djorgovski}, {van Eyken}, {Filippenko}, {Fox},
  {Gal-Yam}, {Hale}, {Hamam}, {Helou}, {Henning}, {Howell}, {Jacobsen},
  {Laher}, {Mattingly}, {McKenna}, {Pickles}, {Poznanski}, {Rahmer}, {Rau},
  {Rosing}, {Shara}, {Smith}, {Starr}, {Sullivan}, {Velur}, {Walters}, \&
  {Zolkower}}]{law09}
{Law}, N.~M., {Kulkarni}, S.~R., {Dekany}, R.~G., {et~al.} 2009, \pasp, 121,
  1395, \dodoi{10.1086/648598}

\bibitem[{LeVeque(2002)}]{leveque02}
LeVeque, R.~J. 2002, Finite Volume Methods for Hyperbolic Problems, Cambridge
  Texts in Applied Mathematics (Cambridge University Press),
  \dodoi{10.1017/CBO9780511791253}

\bibitem[{{Margutti} {et~al.}(2017){Margutti}, {Kamble}, {Milisavljevic},
  {Zapartas}, {de Mink}, {Drout}, {Chornock}, {Risaliti}, {Zauderer},
  {Bietenholz}, {Cantiello}, {Chakraborti}, {Chomiuk}, {Fong}, {Grefenstette},
  {Guidorzi}, {Kirshner}, {Parrent}, {Patnaude}, {Soderberg}, {Gehrels}, \&
  {Harrison}}]{margutti17}
{Margutti}, R., {Kamble}, A., {Milisavljevic}, D., {et~al.} 2017, \apj, 835,
  140, \dodoi{10.3847/1538-4357/835/2/140}

\bibitem[{{Milisavljevic} {et~al.}(2015){Milisavljevic}, {Margutti}, {Kamble},
  {Patnaude}, {Raymond}, {Eldridge}, {Fong}, {Bietenholz}, {Challis},
  {Chornock}, {Drout}, {Fransson}, {Fesen}, {Grindlay}, {Kirshner}, {Lunnan},
  {Mackey}, {Miller}, {Parrent}, {Sanders}, {Soderberg}, \&
  {Zauderer}}]{milisavljevic15}
{Milisavljevic}, D., {Margutti}, R., {Kamble}, A., {et~al.} 2015, \apj, 815,
  120, \dodoi{10.1088/0004-637X/815/2/120}

\bibitem[{{Nakar} \& {Granot}(2007)}]{nakar07}
{Nakar}, E., \& {Granot}, J. 2007, \mnras, 380, 1744,
  \dodoi{10.1111/j.1365-2966.2007.12245.x}

\bibitem[{{Nakar} \& {Piran}(2003)}]{nakar03}
{Nakar}, E., \& {Piran}, T. 2003, \apj, 598, 400, \dodoi{10.1086/378388}

\bibitem[{{Ofek} {et~al.}(2007){Ofek}, {Cameron}, {Kasliwal}, {Gal-Yam}, {Rau},
  {Kulkarni}, {Frail}, {Chandra}, {Cenko}, {Soderberg}, \& {Immler}}]{ofek07}
{Ofek}, E.~O., {Cameron}, P.~B., {Kasliwal}, M.~M., {et~al.} 2007, \apjl, 659,
  L13, \dodoi{10.1086/516749}

\bibitem[{{Ostriker} \& {McKee}(1988)}]{ostriker88}
{Ostriker}, J.~P., \& {McKee}, C.~F. 1988, Reviews of Modern Physics, 60, 1,
  \dodoi{10.1103/RevModPhys.60.1}

\bibitem[{{Quataert} {et~al.}(2019){Quataert}, {Lecoanet}, \&
  {Coughlin}}]{quataert19}
{Quataert}, E., {Lecoanet}, D., \& {Coughlin}, E.~R. 2019, \mnras, 485, L83,
  \dodoi{10.1093/mnrasl/slz031}

\bibitem[{{Quimby} {et~al.}(2007){Quimby}, {Aldering}, {Wheeler},
  {H{\"o}flich}, {Akerlof}, \& {Rykoff}}]{quimby07}
{Quimby}, R.~M., {Aldering}, G., {Wheeler}, J.~C., {et~al.} 2007, \apjl, 668,
  L99, \dodoi{10.1086/522862}

\bibitem[{{Quimby} {et~al.}(2011){Quimby}, {Kulkarni}, {Kasliwal}, {Gal-Yam},
  {Arcavi}, {Sullivan}, {Nugent}, {Thomas}, {Howell}, {Nakar}, {Bildsten},
  {Theissen}, {Law}, {Dekany}, {Rahmer}, {Hale}, {Smith}, {Ofek}, {Zolkower},
  {Velur}, {Walters}, {Henning}, {Bui}, {McKenna}, {Poznanski}, {Cenko}, \&
  {Levitan}}]{quimby11}
{Quimby}, R.~M., {Kulkarni}, S.~R., {Kasliwal}, M.~M., {et~al.} 2011, \nat,
  474, 487, \dodoi{10.1038/nature10095}

\bibitem[{{Ro} {et~al.}(2019){Ro}, {Coughlin}, \& {Quataert}}]{ro19}
{Ro}, S., {Coughlin}, E.~R., \& {Quataert}, E. 2019, \apj, 878, 150,
  \dodoi{10.3847/1538-4357/ab1ea2}

\bibitem[{{Ryu} \& {Vishniac}(1987)}]{ryu87}
{Ryu}, D., \& {Vishniac}, E.~T. 1987, \apj, 313, 820, \dodoi{10.1086/165021}

\bibitem[{{Ryu} \& {Vishniac}(1991)}]{ryu91}
---. 1991, \apj, 368, 411, \dodoi{10.1086/169706}

\bibitem[{{Sanz} {et~al.}(2011){Sanz}, {Bouquet}, \& {Murakami}}]{sanz11}
{Sanz}, J., {Bouquet}, S., \& {Murakami}, M. 2011, \apss, 336, 195,
  \dodoi{10.1007/s10509-010-0563-z}

\bibitem[{{Sanz} {et~al.}(2016){Sanz}, {Bouquet}, {Michaut}, \&
  {Miniere}}]{sanz16}
{Sanz}, J., {Bouquet}, S.~E., {Michaut}, C., \& {Miniere}, J. 2016, Physics of
  Plasmas, 23, 062114, \dodoi{10.1063/1.4953424}

\bibitem[{{Sapir} {et~al.}(2011){Sapir}, {Katz}, \& {Waxman}}]{sapir11}
{Sapir}, N., {Katz}, B., \& {Waxman}, E. 2011, \apj, 742, 36,
  \dodoi{10.1088/0004-637X/742/1/36}

\bibitem[{{Schlegel}(1990)}]{schlegel90}
{Schlegel}, E.~M. 1990, \mnras, 244, 269

\bibitem[{{Sedov}(1959)}]{sedov59}
{Sedov}, L.~I. 1959, {Similarity and Dimensional Methods in Mechanics}
  ({Academic Press, New York, NY})

\bibitem[{{Shiode} \& {Quataert}(2014)}]{shiode14}
{Shiode}, J.~H., \& {Quataert}, E. 2014, \apj, 780, 96,
  \dodoi{10.1088/0004-637X/780/1/96}

\bibitem[{{Smith} {et~al.}(2008){Smith}, {Chornock}, {Li}, {Ganeshalingam},
  {Silverman}, {Foley}, {Filippenko}, \& {Barth}}]{smith08}
{Smith}, N., {Chornock}, R., {Li}, W., {et~al.} 2008, \apj, 686, 467,
  \dodoi{10.1086/591021}

\bibitem[{{Smith} {et~al.}(2007){Smith}, {Li}, {Foley}, {Wheeler}, {Pooley},
  {Chornock}, {Filippenko}, {Silverman}, {Quimby}, {Bloom}, \&
  {Hansen}}]{smith07}
{Smith}, N., {Li}, W., {Foley}, R.~J., {et~al.} 2007, \apj, 666, 1116,
  \dodoi{10.1086/519949}

\bibitem[{{Sollerman} {et~al.}(2020){Sollerman}, {Fransson}, {Barbarino},
  {Fremling}, {Horesh}, {Kool}, {Schulze}, {Sfaradi}, {Yang}, {Bellm},
  {Burruss}, {Cunningham}, {De}, {Drake}, {Golkhou}, {Green}, {Kasliwal},
  {Kulkarni}, {Kupfer}, {Laher}, {Masci}, {Rodriguez}, {Rusholme}, {Williams},
  {Yan}, \& {Zolkower}}]{sollerman20}
{Sollerman}, J., {Fransson}, C., {Barbarino}, C., {et~al.} 2020, \aap, 643,
  A79, \dodoi{10.1051/0004-6361/202038960}

\bibitem[{{Suzuki} {et~al.}(2019){Suzuki}, {Moriya}, \& {Takiwaki}}]{suzuki19}
{Suzuki}, A., {Moriya}, T.~J., \& {Takiwaki}, T. 2019, \apj, 887, 249,
  \dodoi{10.3847/1538-4357/ab5a83}

\bibitem[{{Taylor}(1950)}]{taylor50}
{Taylor}, G. 1950, Proceedings of the Royal Society of London Series A, 201,
  159, \dodoi{10.1098/rspa.1950.0049}

\bibitem[{{To} {et~al.}(2021){To}, {Krause}, {Rozo}, {Wu}, {Gruen}, {Wechsler},
  {Eifler}, {Rykoff}, {Costanzi}, {Becker}, {Bernstein}, {Blazek}, {Bocquet},
  {Bridle}, {Cawthon}, {Choi}, {Crocce}, {Davis}, {DeRose}, {Drlica-Wagner},
  {Elvin-Poole}, {Fang}, {Farahi}, {Friedrich}, {Gatti}, {Gaztanaga},
  {Giannantonio}, {Hartley}, {Hoyle}, {Jarvis}, {MacCrann}, {McClintock},
  {Miranda}, {Pereira}, {Park}, {Porredon}, {Prat}, {Rau}, {Ross}, {Samuroff},
  {S{\'a}nchez}, {Sevilla-Noarbe}, {Sheldon}, {Troxel}, {Varga}, {Vielzeuf},
  {Zhang}, {Zuntz}, {Abbott}, {Aguena}, {Amon}, {Annis}, {Avila}, {Bertin},
  {Bhargava}, {Brooks}, {Burke}, {Carnero Rosell}, {Carrasco Kind},
  {Carretero}, {Chang}, {Conselice}, {da Costa}, {Davis}, {Desai}, {Diehl},
  {Dietrich}, {Everett}, {Evrard}, {Ferrero}, {Flaugher}, {Fosalba}, {Frieman},
  {Garc{\'\i}a-Bellido}, {Gruendl}, {Gutierrez}, {Hinton}, {Hollowood},
  {Honscheid}, {Huterer}, {James}, {Jeltema}, {Kron}, {Kuehn}, {Kuropatkin},
  {Lima}, {Maia}, {Marshall}, {Menanteau}, {Miquel}, {Morgan}, {Muir}, {Myles},
  {Palmese}, {Paz-Chinch{\'o}n}, {Plazas}, {Romer}, {Roodman}, {Sanchez},
  {Santiago}, {Scarpine}, {Serrano}, {Smith}, {Suchyta}, {Swanson}, {Tarle},
  {Thomas}, {Tucker}, {Weller}, {Wester}, {Wilkinson}, \& {DES
  Collaboration}}]{to21}
{To}, C., {Krause}, E., {Rozo}, E., {et~al.} 2021, \prl, 126, 141301,
  \dodoi{10.1103/PhysRevLett.126.141301}

\bibitem[{{Tonry} {et~al.}(2018){Tonry}, {Denneau}, {Heinze}, {Stalder},
  {Smith}, {Smartt}, {Stubbs}, {Weiland}, \& {Rest}}]{tonry18}
{Tonry}, J.~L., {Denneau}, L., {Heinze}, A.~N., {et~al.} 2018, \pasp, 130,
  064505, \dodoi{10.1088/1538-3873/aabadf}

\bibitem[{{Turatto} {et~al.}(1993){Turatto}, {Cappellaro}, {Danziger},
  {Benetti}, {Gouiffes}, \& {della Valle}}]{turatto93}
{Turatto}, M., {Cappellaro}, E., {Danziger}, I.~J., {et~al.} 1993, \mnras, 262,
  128, \dodoi{10.1093/mnras/262.1.128}

\bibitem[{{Wang} \& {Loeb}(2000)}]{wang00}
{Wang}, X., \& {Loeb}, A. 2000, \apj, 535, 788, \dodoi{10.1086/308888}

\bibitem[{{Waxman} \& {Shvarts}(1993)}]{waxman93}
{Waxman}, E., \& {Shvarts}, D. 1993, Physics of Fluids A, 5, 1035,
  \dodoi{10.1063/1.858668}

\bibitem[{{Weaver} {et~al.}(1977){Weaver}, {McCray}, {Castor}, {Shapiro}, \&
  {Moore}}]{weaver77}
{Weaver}, R., {McCray}, R., {Castor}, J., {Shapiro}, P., \& {Moore}, R. 1977,
  \apj, 218, 377, \dodoi{10.1086/155692}

\bibitem[{{Yang} {et~al.}(2020){Yang}, {Sollerman}, {Chen}, {Kool}, {Lunnan},
  {Schulze}, {Strotjohann}, {Horesh}, {Kasliwal}, {Kupfer}, {Mahabal}, {Masci},
  {Nugent}, {Perley}, {Riddle}, {Rusholme}, \& {Sharma}}]{yang20}
{Yang}, S., {Sollerman}, J., {Chen}, T.~W., {et~al.} 2020, arXiv e-prints,
  arXiv:2009.07270.
\newblock \doarXiv{2009.07270}

\end{thebibliography}
\end{document}